\documentclass[a4paper,11pt]{article}
\pdfoutput=1

\usepackage[T1]{fontenc}
\usepackage[dvipsnames]{xcolor}
\usepackage{jheppub}
\usepackage{amsmath}
\usepackage{custom_macros} 
\usepackage{subcaption}

\title{Entanglement C-functions of defects and interfaces in $\mathcal{N}=4$ supersymmetric Yang--Mills theory}

\author[a]{Niko Jokela,}
\author[b]{Jani Kastikainen,}
\author[c,d]{Jos\'e Manuel Pen\'in,}
\author[e]{\\ Ronnie Rodgers,}
\author[a]{and Helime Ruotsalainen}

\affiliation[a]{
    Department of Physics and Helsinki Institute of Physics, P.O. Box 64, FI-00014 University of Helsinki, Finland
}
\affiliation[b]{
    Institute for Theoretical Physics and Astrophysics and W\"urzburg-Dresden Cluster of Excellence ct.qmat, Julius-Maximilians-Universit\"at W\"urzburg, Am Hubland, 97074 W\"urzburg, Germany
}
\affiliation[c]{
    INFN, Sezione di Firenze, Via G. Sansone 1, I-50019 Sesto Fiorentino (Firenze), Italy
}
\affiliation[d]{
    Dipartimento di Fisica e Astronomia, Universitá di Firenze, Via G. Sansone 1,I-50019 Sesto Fiorentino (Firenze), Italy
}
\affiliation[e]{
    Nordita, Stockholm University and KTH Royal Institute of Technology, Hannes Alfv\'ens v\"ag 12, SE-106 91 Stockholm, Sweden
}

\emailAdd{niko.jokela@helsinki.fi}
\emailAdd{jani.kastikainen@uni-wuerzburg.de}
\emailAdd{jmanpen@gmail.com}
\emailAdd{ronnie.rodgers@su.se}
\emailAdd{helime.ruotsalainen@helsinki.fi}

\preprint{   $\begin{array}{rr}
	\text{HIP-2025-27/TH}\\\text{NORDITA 2025-050}\end{array}$
}

\abstract{We consider planar codimension-one defects and interfaces in $\mathcal{N}=4$ supersymmetric Yang--Mills (SYM) theory, 
realized by the D3/D5-brane intersection. Working in the probe limit, where the number of D5-branes is small compared to the number of D3-branes, we obtain analytic results for the holographic entanglement entropy of a ball-shaped region centered on the defect. A defect renormalization group flow is triggered by giving the defect hypermultiplets a mass, which 
corresponds to separating the D3- and D5-branes. Along this flow the entanglement C-function decreases monotonically. We also allow the D5-branes to carry worldvolume flux corresponding to dissolved D3-branes, in which case the setup describes an interface between two copies of $\mathcal{N}=4$ SYM theory with different gauge groups, where an RG flow is triggered by
placing one side of the interface onto the Coulomb branch. Here we again find monotonic behavior of the entanglement C-function, although its interpretation as a measure of effective degrees of freedom is problematic. We investigate possible alternative measures of degrees of freedom.
}

\begin{document} 
\maketitle
\flushbottom
\setcounter{page}{2}

\section{Introduction}
\label{sec:intro}

In quantum field theory (QFT), monotonicity theorems make concrete the notion that degrees of freedom are lost along renormalization group (RG) flows. Each such theorem implies that there is some quantity \(F\) that is larger at the ultraviolet (UV) fixed point than in the infrared (IR), \(F_\mathrm{UV} \geq F_\mathrm{IR}\). Some monotonicity theorems also provide a ``\(C\)-function'' defined along the entire RG flow, that interpolates monotonically from \(F_\mathrm{UV}\) to \(F_\mathrm{IR}\). Established monotonicity theorems are the \(c\)-theorem~\cite{Zamolodchikov:1986gt} in \(d=2\) spacetime dimensions, the \(F\)-theorem in \(d=3\)~\cite{Jafferis:2011zi,Myers:2010tj,Casini:2012ei}, and the \(a\)-theorem in \(d=4\)~\cite{Cardy:1988cwa,Osborn:1989td,Jack:1990eb,Komargodski:2011vj}. Monotonicity theorems are powerful constraints that can be used to rule out CFTs as possible IR fixed points for the RG flow originating from a given UV~\cite{Grover:2012sp}.

In each theorem mentioned above, \(F_\mathrm{UV}\) and \(F_\mathrm{IR}\) are the universal terms in the free energy of the UV and IR conformal field theories (CFTs) on a \(d\)-sphere \sph[d], respectively. When \(d\) is even, \(F\) is also related to the Weyl anomaly, which for an even-dimensional CFT on a curved manifold includes a ``type A'' term proportional to the Euler density~\cite{Deser:1993yx}, with a coefficient proportional to \(F\)~\cite{Casini:2011kv}.\footnote{See ref.~\cite{Giombi:2014xxa} for a definition of \(F\) in arbitrary \(d\).}

\begin{figure}
        \begin{subfigure}{0.5\textwidth}
            \centering
            \includegraphics{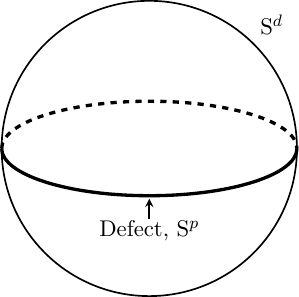}
            \caption{}
            \label{fig:defect_sphere}
        \end{subfigure}\begin{subfigure}{0.5\textwidth}
            \centering
            \includegraphics{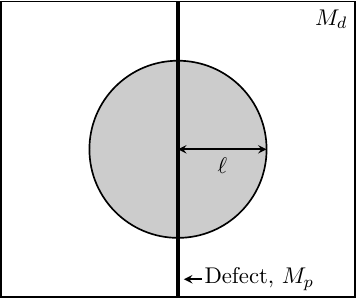}
            \caption{}
            \label{fig:defect_plane}
        \end{subfigure}
        \caption{\textbf{(a):} A \(d\)-dimensional Euclidean CFT on a \(d\)-sphere \(\sph[d]\), with a \(p\)-dimensional conformal defect wrapping a maximal \(\sph[p]\). We denote by \(F_\mathrm{def}\) the defect's contribution to the free energy in this configuration. \textbf{(b):} The CFT in Lorentzian signature, in \(d\)-dimensional Minkowski space \(M_d\). The thick, vertical line represents a planar \(p\)-dimensional defect, i.e. a defect with worldvolume \(M_p\). The relative entropy of a ball centered on the defect, represented by the gray disk in the figure, may be used to define a \(C\)-function~\cite{Casini:2023kyj}.}
\end{figure}

Different monotonicity theorems apply to RG flows localized to defects or boundaries, hereafter collectively referred to as defects, in CFTs. For a \(p\)-dimensional defect there are the \(g\)-theorem for \(p=1\)~\cite{Friedan:2003yc,Cuomo:2021rkm}, the \(b\)-theorem for \(p=2\)~\cite{Jensen:2015swa}, and the defect \(a\)-theorem for \(p=4\)~\cite{Wang:2021mdq}. In each case the monotonic quantity that is larger in the UV than in the IR is \(F_\mathrm{def}\), the universal term in the defect contribution to free energy when the defect wraps a maximal \(p\)-sphere \sph[p] inside \sph[d], as depicted in figure~\ref{fig:defect_sphere}. We will denote \(F_\mathrm{def}\) for the UV and IR fixed points of an RG flow by \(F_\mathrm{def,UV}\) and \(F_\mathrm{def,IR}\), respectively. Thus, each of the theorems mentioned in this paragraph implies \(F_\mathrm{def,UV} \geq F_\mathrm{def,IR}\) for their corresponding values of \(p\). As for bulk RG flows, defect monotonicity theorems are an important tool for constraining possible IR fixed points~\cite{Komargodski:2025jbu}.

Inequalities obeyed by entanglement entropy and related quantum information quantities provide useful tools for proving monotonicity theorems. Strong subadditivity of entanglement entropy was key to the proof of the \(F\)-theorem~\cite{Casini:2012ei}, and has also been used to establish alternate, entropic versions of the \(c\)-theorem and \(a\)-theorem~\cite{Casini:2004bw,Casini:2017vbe}. The relationship between entanglement entropy and monotonicity arises because at the UV and IR fixed points, the universal coefficient in the entanglement entropy of a ball-shaped region is proportional to \(F_\mathrm{UV}\) and \(F_\mathrm{IR}\), respectively, in the absence of a defect~\cite{Casini:2011kv}.

For defect RG flows the situation is more complicated. The universal coefficient in a defect's contribution to entanglement entropy is not generally monotonic under RG flows~\cite{Kobayashi:2018lil}. This is because for defects of codimension \(d-p \geq 2\), this coefficient in the entanglement entropy is not proportional to \(F_\mathrm{def}\). For example, when \(p\) is even \(F_\mathrm{def}\) is proportional to the coefficient of the \(p\)-dimensional Euler density in the defect's contribution to the Weyl anomaly, similarly to the bulk \(F\) in even \(d\). However, a \(p\)-dimensional defect can have additional terms in its contribution to the Weyl anomaly compared to a \(p\)-dimensional CFT~\cite{Graham:1999pm,Henningson:1999xi,Schwimmer:2008yh,Herzog:2015ioa,Herzog:2017kkj,FarajiAstaneh:2021skf,Chalabi:2021jud}, and the coefficients of some of these additional terms contribute to the entanglement entropy~\cite{Jensen:2018rxu}.

In general, to isolate \(F_\mathrm{def}\) from an entropy one must instead compute a relative entropy~\cite{Kobayashi:2018lil}. Casini, Salazar Landea, and Torroba (CST) considered the relative entropy of a ball centered on a planar defect as depicted in figure~\ref{fig:defect_plane}, showing that it may be used to define a \(C\)-function that decreases monotonically along defect RG flows for defects of dimension \(p\leq 4\)~\cite{Casini:2023kyj}. Unfortunately, the relative entropy is a difficult quantity to compute, so explicit examples of this \(C\)-function are hard to find. The situation simplifies somewhat for codimension-one defects, for which CST's \(C\)-function may be obtained from entanglement entropy alone. However, entanglement entropy is still not easy to compute in interacting QFTs.

A useful tool is holography~\cite{Maldacena:1997re,Gubser:1998bc,Witten:1998qj}, also known as the AdS/CFT correspondence, which simplifies the computation of entanglement entropy in a holographic QFT to the problem of finding the area of a minimal surface in the dual asymptotically locally-anti-de Sitter (\ads) spacetime~\cite{Ryu:2006bv,Ryu:2006ef,Casini:2011kv,Lewkowycz:2013nqa}. For defects, prior to CST's monotonicity proof, holography was used to explore the possible monotonicity of entanglement entropy under defect RG flows~\cite{Estes:2014hka,Kumar:2017vjv,Kobayashi:2018lil,Rodgers:2018mvq,Estes:2018tnu}. These holographic results helped to provide evidence that for defects the RG central charge is not simply the universal term in the entanglement entropy~\cite{Kobayashi:2018lil,Jensen:2018rxu}. A general approach for investigating defects and their RG flows in bottom-up holographic models has recently been proposed~\cite{Nakayama:2025hqo}.

Defects are typically holographically dual to extended objects such as D-branes. Since asymptotically locally-\ads\ gravitational solutions containing branes are often challenging to find, particularly when the dual defect breaks conformal symmetry, many of the holographic results mentioned above make use of probe branes~\cite{Karch:2002sh,Karch:2005ms}, where the backreaction of the branes on the metric and other gravity fields is neglected. A variety of methods exist for computing probe brane contributions to entanglement entropy~\cite{Jensen:2013lxa,Karch:2014ufa,Chang:2013mca,Jones:2015twa,Jokela:2024cxb}.

In this article, we use holography to compute entanglement entropy of a ball of radius \(\ell\), centered on a planar codimension-one defect in four-dimensional \(\cN=4\) supersymmetric Yang-Mills (SYM) theory with gauge group \(\SU(N_3)\) and 't Hooft coupling \(\la\). The defect hosts \(N_5\)  three-dimensional hypermultiplets. When the hypermultiplets are massless, this defect QFT is conformal~\cite{DeWolfe:2001pq,Erdmenger:2002ex}. We will give the hypermultiplets a non-zero mass, breaking conformal symmetry and triggering a defect RG flow. The holographic dual description consists of \(N_5\) D5-branes in the \(\ads[5] \times \sph[5]\) background of type IIB supergravity that arises as the near-horizon geometry of \(N_3\gg1\) D3-branes~\cite{Karch:2000gx}, with the hypermultiplet mass dual to a slipping mode on \sph[5]. We will work in the probe limit, \(N_3 \gg \sqrt{\la} \, N_5\).

Since the defect we consider is codimension-one, CST's \(C\)-function may be obtained purely from entanglement entropy. Concretely, denoting the defect's contribution to the entanglement entropy by \(S_\mathrm{def}(\ell)\), the \(C\)-function is
\begin{equation} \label{eq:C_function_definition}
    C(\ell) = (\ell \, \p_\ell-1)\, S_\mathrm{def}(\ell)\, .
\end{equation}
We will evaluate this \(C\)-function using our holographic entanglement entropy results, confirming that it indeed decreases monotonically with \(\ell\). The precise definition of \(S_\mathrm{def}\) is reviewed in section~\ref{sec:C_functions}.

In addition, we will allow for the D5-branes to carry charge corresponding to \(n_3\) dissolved D3-branes, in which case the D5-branes are no longer dual to a defect, but rather an interface between two copies of \(\cN=4\) SYM theory on half of Minkowski space, with different rank gauge groups on either side~\cite{DeWolfe:2001pq,Skenderis:2002vf}. When \(n_3\neq 0\) there are no hypermultiplets or other degrees of freedom localized to the interface in the perturbative description available at weak coupling~\cite{Gaiotto:2008sa,Ipsen:2019jne}; however, in the S-dual formulation, the interface supports explicit bifundamental hypermultiplets. It is still possible to break conformal symmetry by sourcing the slipping mode for the D5-branes on the \sph[5], holographically dual to putting the copy of \(\cN=4\) SYM theory on one side of the interface onto the Coulomb branch~\cite{Arean:2006vg,Arean:2007nh}. In this case, because the conformal symmetry is broken in the bulk of \(\cN=4\) SYM theory, CST's proof of monotonicity of \(C(\ell)\) defined in equation~\eqref{eq:C_function_definition} does not apply. 
Indeed, for such combined bulk and defect flows, there is generally no reason to expect a universal RG monotone.
Nevertheless, we find that \(C(\ell)\) is monotonic even with \(n_3 \neq 0\). While it diverges to negative values in the IR, we interpret this monotonicity as physically tracking the decoupling of the interface degrees of freedom (manifest in the S-dual frame) as they acquire mass.
We will also explore other candidate \(C\)-functions obtained by taking derivatives of the entanglement entropy, that interpolate between finite limits between the UV and IR, showing that they are generally not monotonic functions of \(\ell\). Denoting by \(S(\ell)\) the entanglement entropy of a ball-shaped region of radius \(\ell\), the best candidate that we find is 
\begin{equation}
    \AtLM(\ell) \equiv  -\frac{1}{2}\,\ell\,\partial_\ell\,(\ell\,\partial_{\ell} - 2)(\ell \, \partial_\ell - 1)\,S(\ell) \,,
\end{equation}
inspired by the \(C\)-functions of Liu and Mezei~\cite{Liu:2012eea,Liu:2013una}. We find that \(\AtLM\) is smaller in the IR than in the UV, with sensible limits in both cases, although it is not always monotonic in between.

Entanglement entropy for interfaces has been a topic of interest lately, with for example much recent work on holographic entanglement entropy and effective central charges for interfaces between two-dimensional CFTs~\cite{Gutperle:2015hcv,Gutperle:2015kmw,Karch:2021qhd,Karch:2022vot,Karch:2023evr,Karch:2024udk,Afxonidis:2024gne,Gutperle:2024yiz,Afxonidis:2025jph}. Holographic entanglement entropy has also recently been computed for a family of conformal interfaces in \(\cN=4\) SYM theory~\cite{Uhlemann:2023oea}. These are different from the interfaces we study in this article, since their brane intersection description includes NS5-branes. When defect conformal symmetry is preserved, the contribution to entanglement entropy from the interface that we consider has recently been related to the light crossing time~\cite{Harvey:2025ttz}. Also known as causal depth, the light crossing time is a singularity in correlation functions of a holographic defect/interface/boundary CFT due to the presence of one or more branes in the dual gravitational description~\cite{Reeves:2021sab}.

The structure of this article is as follows. Section~\ref{sec:d3_d5} contains a review of the features of the D3/probe D5 system needed for our entanglement entropy calculations, which appear in section~\ref{sec:entanglement_calculations}. Section~\ref{sec:C_functions} contains the results for \(C\)-functions. We close with conclusions and outlook for the future in section~\ref{sec:discussion}. There are several appendices with additional calculations. In appendix \ref{app:free_energy} we compute \(F_\mathrm{def,UV}\) holographically for the D3/D5 system in the probe limit. Appendix~\ref{app:integrals} contains details of the evaluation of integrals that appear in the entanglement entropy calculation. Appendix~\ref{sec:coulomb} contains details of the holographic computation of entanglement entropy on the Coulomb branch, originally performed in ref.~\cite{Chalabi:2020tlw}, which will be useful for our discussion.

\section{Review of the D3/probe D5 system}
\label{sec:d3_d5}

In this section we briefly describe the probe D5-brane embeddings in \(\ads[5] \times \sph[5]\) holographically dual to codimension-one defects in \(\cN=4\) SYM theory. Readers familiar with these embeddings, which have previously been described in refs.~\cite{Skenderis:2002vf,Arean:2006vg,Arean:2007nh}, may nevertheless not wish to skip this section, as we use it to establish notation that will appear throughout this article.

We write the metric and four-form flux of \(\ads[5] \times \sph[5]\), arising as the near-horizon limit of the geometry sourced by \(N_3\) coincident D3-branes, as
\begin{subequations} \label{eq:ads5xs5}
\begin{align}
    \diff s^2 &= \frac{L^2}{z^2} \le(
        -\diff t^2 + \diff x^2 + \diff  r^2 + r^2 \diff \f^2 
    \ri)
    + \frac{L^2}{z^2} \diff z^2\label{eq:ads5xs5_metric} \\ 
    &\qquad\qquad\qquad+ L^2 \le(\diff \q^2 + \cos^2 \q \diff s_{\sph[2]_{A}}^2 + \sin^2 \q  \diff s_{\sph[2]_{B}}^2 \ri) ,
    \nonumber \\
    C_4 &= \frac{L^4}{z^4} r  \diff t \wedge \diff x \wedge \diff r \wedge \diff \f  + \dots \; ,
    \label{eq:ads5xs5_c4}
\end{align}
\end{subequations}
where \(\diff s_{\sph[2]_{A,B}}^2\) are the metrics of two unit, round two-spheres \(\sph[2]_{A}\) and \(\sph[2]_{B}\), and the dots in \(C_4\) denote a term with legs on \(\sph[2]_{A}\) and \(\sph[2]_{B}\) needed to make \(F_5=\diff C_4\) self-dual. The curvature radius \(L\) of the \ads[5] and \sph[5] factors is related to the number of D3-branes, the closed string coupling \(g_s\), and the Regge parameter \(\a'\) by \(L^4 = 4 \pi  g_s N_3 \a'^2\). In the limit of large \(N_3\) followed by large `t Hooft coupling \(\la = 4\pi g_s N_3\), type IIB supergravity in this background is holographically dual to \(\cN=4\) SYM theory with gauge group \(\SU(N_3)\)~\cite{Maldacena:1997re}.

\begin{table}[t]
    \begin{center}
    \begin{tabular}{r | c c c c c c c c c c}
        & \(t\) & \(x\) & \(r\) & \(\f\) & \(z\) & \multicolumn{2}{|c|}{\(\sph[2]_{A}\)}  &  \multicolumn{2}{|c|}{\(\sph[2]_{B}\)} & \(\q\)
        \\ \hline
        \(N_3\) D3 & \(\times\) & \(\times\) & \(\times\) & \(\times\) & \(\cdot\) & \(\cdot\) & \(\cdot\) & \(\cdot\) & \(\cdot\) & \(\cdot\)
        \\
        \(N_5\) D5 & \(\times\) & \(\cdot\) & \(\times\) & \(\times\) & \(\times\) & \(\times\) & \(\times\) & \(\cdot\) & \(\cdot\) & \(\cdot\)
        \\
        \(n_3\) D3 & \(\times\) & \(>\) & \(\times\) & \(\times\) & \(\cdot\) & \(\cdot\) & \(\cdot\) & \(\cdot\) & \(\cdot\) & \(\cdot\)
    \end{tabular}
        \caption{The D-brane intersection corresponding to a codimension-one defect or interface in \(\cN=4\) SYM theory~\cite{Karch:2000gx}, as depicted in figure~\ref{fig:brane_intersection}. The crosses indicate directions spanned by the branes. The \(>\) symbol in the table indicates that the \(n_3\) D3-branes are only present on one side of the D5-branes. In the limit of strong coupling and large \(N_3 \), the \(N_3\) D3-branes are replaced by the \(\ads[5] \times \sph[5]\) geometry in equation~\eqref{eq:ads5xs5}, in which the D5-branes appear as probes. In this limit, the \(n_3\) D3-branes are not additional probes added to the system, but are rather present as flux of the D5-branes' worldvolume gauge field as in equation~\eqref{eq:q_def}.}
    \label{tab:brane_configuration}
    \end{center}
\end{table}

Now we introduce \(N_5\) D5-branes spanning \((t,x,r,\f,\sph[2]_{A})\), as summarized in table~\ref{tab:brane_configuration}. We work in a probe limit, in which the D5-branes do not backreact on the \(\ads[5]\times\sph[5]\) geometry in equation~\eqref{eq:ads5xs5}. This limit is valid provided \(\sqrt{\la} \, N_5 \ll N_3\)~\cite{Karch:2000gx,Karch:2002sh}. We make the following ansatz for the worldvolume fields on the D5-branes,
\begin{equation} \label{eq:ansatz}
    x=x(z) \, ,
    \qquad
    \q = \q(z) \, ,
    \qquad
    \mathcal{F} = \frac{q L^2}{2 \pi \a'}\, \mathrm{vol}(\sph[2]_{A}) \, ,
\end{equation}
where \(\mathcal{F}\) is the field strength of the D5-branes' worldvolume gauge field and \(\mathrm{vol}(\sph[2]_{A})\) is the volume form on \(\sph[2]_{A}\). When the constant parameter \(q\) is non-zero, the D5-branes carry \(n_3\) units of D3-brane charge, where \(n_3\) is given by the integral of the worldvolume field strength over the wrapped \(\sph[2]_{A}\),
\begin{equation}\label{eq:q_def}
    n_3 = \frac{N_5}{2\pi} \biggl|\, \int_{\sph[2]_{A}} \mathcal{F} \, \biggr|= \frac{\sqrt{\la}}{\pi} N_5 |q| \, .
\end{equation}
Thus, when \(q\) is of order unity then \(n_3\) is large because \(\la\) is large. For validity of the probe limit we require that \(n_3 \ll N_3\), so that the backreaction of the \(n_3\) D3-branes is negligible compared to that of the \(N_3\) D3-branes sourcing the geometry.

The D5-brane action \(I_\mathrm{D5}\) evaluated on the ansatz in equation~\eqref{eq:ansatz} is
\begin{align} \label{eq:action_ansatz}
    I_\mathrm{D5} &= - N_5 T_5 \int \diff^6 \xi \, \sqrt{- \det(g+ 2 \pi \a' \mathcal{F})} + 2 \pi \a' N_5 T_5 \int \cF \wedge P[C_4] 
    \nonumber \\
    &= - \frac{\sqrt{\la} \, N_3 N_5}{2\pi^3} \int \diff t \diff r \diff z \,
    \frac{r}{z^4} \le[  \sqrt{(q^2 + \cos^4 \q)(1 + x'^2 + z^2 \q'^2)} - q x'\ri]  ,
\end{align}
where \(T_5\) is the D5-brane tension, \(g\) is the induced metric on the worldvolume of the D5-branes, and \(P[C_4]\) is the pullback of \(C_4\) to the worldvolume. Primes denote derivatives with respect to \(z\). The Euler--Lagrange equations for \(\q\) and \(x\) following from this action admit solutions of the form
\begin{equation}
    \sin \q = \m z \,, \qquad x = \frac{q z}{\sqrt{1 - \m^2 z^2}} \, ,
    \label{eq:D5_embedding}
\end{equation}
where \(\m\) is an integration constant and we have used translational symmetry in the \(x\) direction to set \(x(z=0) = 0\). This family of solutions is holographically dual to the codimension-one defect described in section~\ref{sec:intro} and in greater detail below. The defect or interface is located at \(x=0\).

\begin{figure}
    \begin{center}
    \includegraphics{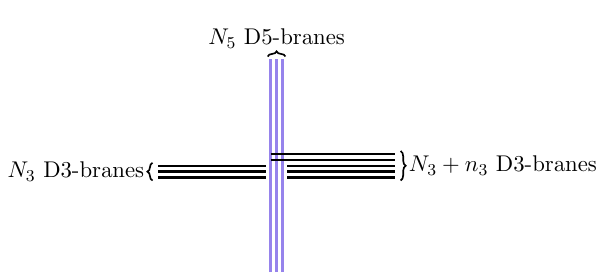}
    \end{center}
    \caption{The D3/D5 intersection giving rise to the codimension-one defect or interface in \(\cN=4\) SYM theory that we consider. The horizontal and vertical lines represent stacks of coincident D3 and D5-branes, respectively. Some number \(n_3\) of the D3-branes may end on the D5-branes, in which case the defect forms an interface between two copies of \(\cN=4\) SYM theory with different rank gauge groups. An RG flow may be triggered by separating the D5-branes and any D3-branes that end on them from the rest of the D3-branes.}
    \label{fig:brane_intersection}
\end{figure}

\paragraph{Field theory description.} The QFT holographically dual to the  D3/probe D5 system arises as the low-energy description of the brane intersection in figure~\ref{fig:brane_intersection}.  The description of the dual QFT depends on whether $n_3 = 0$ or $n_3>0$. We will discuss both cases. For the description of the \(n_3 > 0\) case, we will assume that \(q > 0\). The description of the \(q < 0\) case is the same up to an \(x \to - x\) reflection. Our results for entanglement entropy and \(C\)-functions given in later sections are valid for either sign of \(q\).

When \(n_3 = 0\), the dual QFT consists of $\mathcal{N} = 4$ SYM theory on Minkowski space $M_4$ with coordinates \((t,x,r,\f)\), coupled to a codimension-one planar defect at $x = 0$, on which are supported \(N_5\) three-dimensional hypermultiplets. When the hypermultiplets are massless, the theory is invariant under the defect conformal group \(\SO(3,2)\)~\cite{Karch:2000ct,DeWolfe:2001pq,Erdmenger:2002ex}. One can break conformal symmetry and trigger a defect RG flow by giving the hypermultiplets a non-zero mass \(m\), so that in the IR the hypermultiplet decouples and the defect flows to a trivial defect. In the brane intersection description, the hypermultiplets arise from strings stretched between D3-branes and the D5-branes. The hypermultiplet mass is then equal to the string tension multiplied by the minimum length of such strings~\cite{Karch:2002sh}, and is thus proportional to the integration constant \(\m\) in the D5-brane embedding \eqref{eq:D5_embedding} with $q = 0$,
\begin{equation}
    m = \frac{\sqrt{\la}}{2\pi}\, \mu \, .
    \label{eq:m_mu_relation}
\end{equation}

When \(n_3 > 0\), there are no hypermultiplets at \(x=0\). Instead, the hypersurface at \(x=0\) forms an interface between two copies of \(\cN=4\) SYM theory. The gauge groups of the copies at negative and positive \(x\) are \(\SU(N_3)\) and \(\SU(N_3 + n_3)\), respectively. Let us denote the $\mathcal{N} = 4$ SYM theory with gauge group $\SU(N_3)$ at $x<0$ by CFT$_-$ and the theory with gauge group $\SU(N_3+n_3)$ by CFT$_+$. When $n_3 > 1$ , three of the six adjoint-valued scalars of \(\cN = 4\) SYM theory, which we denote $\{\Phi_i\}_{i=1,\ldots, 6}$, have singular boundary conditions on the $x>0$ side of the interface~\cite{Gaiotto:2008sa}. These boundary conditions take the block form
\begin{equation}
    \Phi_{1,2,3}^+ =\frac{1}{x}\begin{pmatrix}
    0_{N_3\times N_3} & 0_{N_3\times n_3}\\
    0_{n_3\times N_3} & D_{1,2,3}
    \end{pmatrix} + \ldots
    \,,\qquad x \rightarrow 0^+\,,
    \label{eq:Phi_plus_expansion}
\end{equation}
where $\{D_i\}_{i = 1,2,3}$ are elements of the Lie algebra $\mathfrak{su}(n_3)$ that form an $\mathfrak{su}(2)$ subalgebra, \(0_{M \times N}\) denotes an \(M \times N\) matrix of zeroes, and the ellipsis denotes terms that are finite in the limit \(x \to 0^+\). The singular boundary conditions lead to non-trivial one-point functions for \(\F_{1,2,3}^+\)~\cite{Nagasaki:2012re} and break the gauge group of CFT$_+$ to $\SU(N_3)\times \mathrm{U}(n_3)$.

The boundary conditions in equation~\eqref{eq:Phi_plus_expansion} preserve the boundary conformal group \(\SO(3,2)\). In this case a mass scale \(m\) may be introduced by modifying the boundary conditions of CFT$_+$ at spatial infinity. Instead of assuming that $\Phi_{4,5,6}$ vanish asymptotically far from the interface, we can take an \(n_3 \times n_3\) block of $\Phi_{4,5,6}^+$ to asymptote to a constant matrix proportional to \(m\) as $x\rightarrow \infty$, so that $\SU(N_3)\times \mathrm{U}(n_3)$ is preserved and that at large \(x\) there is a vacuum expectation value $\langle \Phi_{4,5,6}\rangle \propto m$ ~\cite{Arean:2006vg}. In other words, CFT$_+$ is put on the Coulomb branch asymptotically. In addition, this gives \(n_3\) of the vector multiplets a mass $m$ for $x>0$. Thus scale symmetry is broken, triggering a non-trivial RG flow of CFT$_+$~\cite{Arean:2006vg,Arean:2007nh}. In the IR, field content of CFT$_+$ reduces to that of \(\cN=4\) SYM theory with gauge group \(\SU(N_3) \times \mathrm{U}(n_3)\).

Equation~\eqref{eq:m_mu_relation} is true for the interface, i.e. for $n_3>0$, even though $m$ is a mass of a four-dimensional vector multiplet rather than defect-localized hypermultiplets as was the case for \(n_3 = 0\). The interpretation is that the Coulomb branch of CFT$_+$ amounts to separating $n_3$ D3-branes --- which we will refer to as
\(\mathrm{D3}'\) --- from the stack of \(N_3\) D3-branes in a direction orthogonal to both the $N_3$ D3-branes (that we call D3) and the $N_5$ D5-branes. The mass of the vector multiplet is equal to the mass of strings stretching between the two stacks of $N_3$ and $n_3$ D3-branes (the D3--\(\mathrm{D3}'\) strings). However, in the limit $ 1 \ll n_3 \ll N_3$, the $n_3$ D3-branes are dissolved into the D5-branes and are best thought of as a solitonic configuration on the D5-brane worldvolume in the manner of ref.~\cite{Callan:1997kz}, corresponding to the non-trivial embedding $x = x(z)$ in \eqref{eq:D5_embedding}. Therefore the D3--\(\mathrm{D3}'\) strings attached to the dissolved \(\mathrm{D3}'\)-branes are identified with the D3--D5 strings and \eqref{eq:m_mu_relation} holds. 

Notice that the interface arising as the dual description of the D3/D5 system when $n_3>0$ is fundamentally different from the defect arising when $n_3 = 0$, because there are no additional manifest degrees of freedom localized at the interface \cite{Mikhaylov:2014aoa}; the interface is described purely in terms of the boundary conditions~\eqref{eq:Phi_plus_expansion}~\cite{Gaiotto:2008sa}. We note, however, that these degrees of freedom are only absent in the perturbative Lagrangian. In the S-dual NS5-brane frame, the interface is described by three-dimensional bifundamental hypermultiplets coupled to the gauge fields on the two sides of the interface~\cite{Gaiotto:2008ak}. The disappearance of defect degrees of freedom when going from $n_3 = 0$ to $n_3>0$ in the original frame seems puzzling, but it can be understood, for example, by the following argument~\cite{Ipsen:2019jne}.

Consider a codimension-one defect in \(\cN=4\) SYM theory, with gauge group $\SU(N_3+n_3)$ on both sides of the defect, corresponding to a stack of $N_3+n_3$ D3-branes intersecting a stack of $N_5$ D5-branes. We separate $n_3$ D3-branes --- which we label \(\mathrm{D3}''\) --- from the $x < 0 $ stack, putting CFT$_-$ onto the Coulomb branch and the defect theory onto the Higgs branch, in which the defect hypermultiplets pick up a mass proportional to the separation of the left \(\mathrm{D3}''\)-branes from the stack~\cite{Arean:2006vg}. In the limit of large separation, the D3--\(\mathrm{D3}''\) strings completely decouple, leaving an $\SU(N_3)$ gauge theory at \(x<0\). In addition, the defect hypermultiplets become infinitely heavy, leaving no degrees of freedom localized to the interface.

\section{Probe brane defect entanglement entropy}
\label{sec:entanglement_calculations}

In this section we use holography to compute the entanglement entropy \(S\) of a ball-shaped region of radius \(\ell\) in the defect or interface QFT dual to the D3/probe D5 system reviewed in section~\ref{sec:d3_d5}. To make the computation tractable we will consider only the most symmetric case, for which the entangling region is centered on the defect, as in figure~\ref{fig:defect_plane}.

\begin{figure}
\begin{center}\includegraphics{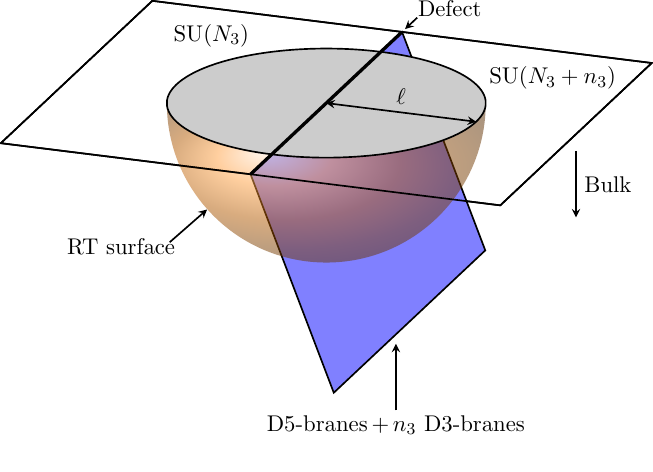}\end{center}
    \caption{Cartoon of the holographic description of the D3/probe D5 system. The box at the top represents the conformal boundary of \ads[5], with the bulk of \ads[5] extending below. The blue parallelogram represents the stack of coincident probe D5-branes, which intersect the boundary along a flat hypersurface, corresponding to the location of the defect in the dual QFT. If the D5-branes carry dissolved D3-brane charge, then the rank of the gauge group either side of the interface is different. The gray disk represents the ball-shaped region, centered on the defect, for which we will compute the entanglement entropy. Holographically, the entanglement entropy is proportional to the area of the RT surface.}
    \label{fig:entangling_region}
\end{figure}

The holographic setup is depicted in figure~\ref{fig:entangling_region}. The parallelogram at the top of the figure represents the boundary of \ads[5] where we think of the dual QFT as ``living'', and therefore corresponds to figure~\ref{fig:defect_plane}. The thick black line bisecting the boundary is the defect, while the gray disk is the entangling region. Below the boundary is the bulk of \ads[5]. The defect is dual to a stack of D5-branes in the bulk. The rank of the gauge group on either side of the defect can be different, which corresponds to the D5-branes carrying D3-brane charge. The entanglement entropy is proportional to the area of the Ryu--Takayanagi (RT) surface, the minimal area bulk surface homologous to the entangling region and sharing the same boundary~\cite{Ryu:2006bv,Ryu:2006ef}.

Our aim is to compute the contribution $S_{\text{def}}(\ell)$ of the defect or the interface to the entanglement entropy, arising holographically from the presence of the D5-branes. We will do so in the probe limit, computing the leading contribution to $S_{\text{def}}(\ell)$ when \(\sqrt{\la}\, N_5\) and \(n_3\) are both much less than \(N_3\). We will use techniques developed in refs.~\cite{Jensen:2013lxa,Karch:2014ufa} to obtain $S_{\text{def}}(\ell)$ in the probe limit without explicitly computing the backreaction of the D5-branes on the surrounding \(\ads[5]\times \sph[5]\) spacetime.

In section~\ref{subsec:EE_definitions} we review the definition of \(S_\mathrm{def}\). Then, in section~\ref{sec:conformal} we will apply the probe brane techniques compute \(S_\mathrm{def}\) for \(m=0\), preserving defect conformal symmetry. For this case $S_{\text{def}}$ was computed in ref.~\cite{Estes:2014hka} by applying the RT prescription in the geometry corresponding to backreacting D5-branes~\cite{Gomis:2006cu,DHoker:2007zhm}. Our result matches the \(\sqrt{\la}\, N_5 \ll N_3\) and \(n_3 \ll N_3\) limit of ref.~\cite{Estes:2014hka}'s result, providing a consistency check of our probe calculations. Then in section~\ref{sec:flow_ee} we will compute the \(S_\mathrm{def}\) for \(m \neq 0\), for which no backreacted geometry is known, with the exception of a backreacted solution corresponding to \(n_3 =0\) which uses smeared D5-branes~\cite{Jokela:2019tsb}.

\subsection{Entanglement entropies of defects and interfaces}\label{subsec:EE_definitions}

We are interested in the entanglement entropy of a ball-shaped region of radius $\ell$ in $\mathcal{N} = 4$ SYM theory when there is either a defect or an interface present of the type described at the end of section~\ref{sec:d3_d5}. In this subsection we review what is meant by the contribution of a defect or boundary to this entanglement entropy.

\paragraph{Entropy of defects.} First, consider the entanglement entropy of a ball-shaped region of radius $\ell$ centered at $x = r = 0$ in the ground state\footnote{To quantize a theory on Minkowski space, one must impose appropriate boundary conditions at spatial infinity. Here we assume standard conditions such that all one-point functions in the absence of the defect are vanishing. The conditions at infinity will play an important role in our discussion of interfaces.} of $\mathcal{N} = 4$ SYM theory coupled to the defect described in the previous section on Minkowski space $M_4$. Let $\rho$ be the reduced density matrix of the ball in the ground state of the system when the defect hypermultiplet has mass $m$. The entanglement entropy in this ground state is defined as the von Neumann entropy
\begin{equation}
    S(\ell) \equiv -\tr{(\rho\log{\rho})}\,,
    \label{eq:von_Neumann_rho}
\end{equation}
which can be decomposed as
\begin{equation}
    S(\ell) = S^{(N_3)}(\ell)+S_{\text{def}}(\ell)\,,
    \label{eq:ball_entropy}
\end{equation}
where $S^{(N_3)}(\ell)$ is the entropy of the ball in the $\SO(4,2)$-invariant vacuum of $\mathcal{N}=4$ SYM theory with gauge group $\SU(N_3)$ with no defect present and $S_{\text{def}}(\ell)$ encodes all contributions arising due to the defect. Both \(S^{(N_3)}\) and \(S_\mathrm{def}\) are UV divergent, and so must be regulated by the introduction of a short-distance cutoff \(\e\). The former has the divergence structure~\cite{Ryu:2006bv,Ryu:2006ef}
\begin{equation}
    S^{(N_3)}(\ell) = p_2\,\frac{\ell^2}{\epsilon^2} - A_{\text{UV}}^{(N_3)}\log{\frac{\ell}{\epsilon}} + O(\epsilon^0)\,,
    \label{eq:UV_fixed_point_entropy}
\end{equation}
with constant coefficients \(p_2\) and \(A_\mathrm{UV}^{(N_3)}\). The coefficient \(p_2\) and the \(O(\e^0)\) terms are scheme-dependent, in the sense that they are not invariant under a multiplicative rescaling of the cutoff \(\e\), while the coefficient of the logarithm \(A_\mathrm{UV}^{(N_3)}\) is scheme-independent and contains physical information --- it is proportional to the type A Weyl anomaly coefficient of \(\cN=4\) SYM theory~\cite{Casini:2011kv}.

The defect contribution to the entanglement entropy has divergence structure~\cite{Casini:2023kyj}
\begin{equation}
    S_{\text{def}}(\ell) = p_1\,\frac{\ell}{\epsilon} - P_{\text{def}}(a)+ O(\e)\,,
    \label{eq:Sdef_expansion}
\end{equation}
with a scheme-dependent coefficient \(p_1\), and a scheme-independent coefficient \(P_\mathrm{def}(a)\) which is a function of the dimensionless radius of the entangling region, $a \propto m\ell$ with $m$ being the mass of the defect hypermultiplets.\footnote{In order for \(P_\mathrm{def}\) to be scheme independent we must define \(S_\mathrm{def}\) as the difference between entanglement entropies in \(\cN=4\) SYM theory with and without the defect, with the same UV cutoff \(\e\) in both cases. Otherwise \(P_\mathrm{def}\) would be contaminated by the scheme-dependence of the \(O(\e^0)\) term in equation~\eqref{eq:UV_fixed_point_entropy}. See for example the discussion in ref.~\cite{Jensen:2013lxa}.} At the UV fixed point $a = 0$, the finite term is equal to an $\ell$-independent constant $P_{\text{def}}(0) = F_{\text{def,UV}}$, which by conformal invariance, coincides with the contribution of the defect to the free energy of the CFT on a maximal $\sph[3] \subset \sph[4]$~\cite{Casini:2011kv}. For $a> 0$, $P_{\text{def}}$ is not related to any free energy.

\paragraph{Entropy of interfaces.} Now consider the interface described in section~\ref{sec:d3_d5} between two $\mathcal{N} = 4$ SYM theories with different gauge groups $\SU(N_3)$ and \(\SU(N_3 + n_3)\), which we label as CFT\(_-\) and CFT\(_+\), respectively. We would like to compute the entanglement entropy of a ball-shaped region in the ground state of the system when it is quantized with boundary conditions \eqref{eq:Phi_plus_expansion} imposed at the interface and, schematically, $\Phi_{4,5,6}^{-}\rightarrow 0$ as $x\rightarrow -\infty$ and $\Phi_{4,5,6}^+\rightarrow m$ as $x \rightarrow \infty$. This ground state can be prepared by a Euclidean path integral with these boundary conditions over a semi-infinite interval in Euclidean time. We denote the entanglement entropy of a ball-shaped region of radius $\ell$ centered at the interface $x = 0$ in this ground state by $S(\ell)$. Following ref.~\cite{Estes:2014hka}, we decompose this as
\begin{equation}
    S(\ell) =\frac{S^{(N_3)}(\ell)+S^{(N_3+n_3)}(\ell)}{2} + S_{\mathrm{def}}(\ell)\,,
    \label{eq:interface_entropy}
\end{equation}
where $S^{(N)}(\ell)$ is the entanglement entropy \eqref{eq:UV_fixed_point_entropy} of a ball of radius $\ell$ in the $\SO(4,2)$-invariant vacuum of $\mathcal{N} = 4$ SYM theory with gauge group $\SU(N)$ on all of Minkowski space, with no defect or interface present. The factor of one half in equation~\eqref{eq:interface_entropy} takes into account that only half of the ball lies on each side of the interface. The term $S_{\mathrm{def}}(\ell)$ captures all contributions of the interface to the entropy (we use the same symbol \(S_\mathrm{def}\) as for the contribution of a defect to entanglement entropy, as many of the equations we will write later apply to both the defect and interface entanglement entropies) and it is expected to have the same divergence structure as the defect contribution~\eqref{eq:Sdef_expansion} with $P_{\text{def}}(a)$ a function of the radius $a \propto m\ell$. Recall that in this case $m$ is not the mass of hypermultiplets localized to a defect (the interface does not support any degrees of freedom \cite{Mikhaylov:2014aoa}), but is the one-point function of some of the adjoint scalar fields of \(\cN=4\) SYM theory, or equivalently, is the mass of some four-dimensional vector multiplets, arising due to spontaneous symmetry breaking. This is reviewed in section~\ref{sec:d3_d5}.

There is another natural way of separating contributions to \(S(\ell)\). Since CFT\(_+\) is on the Coulomb branch, \(S_\mathrm{def}\) as defined by equation~\eqref{eq:interface_entropy} includes UV-finite terms that have a four-dimensional origin. Concretely, let \(S_{\hspace{1pt}\text{Coul}}^{(N,n)}\) denote the entanglement entropy of a ball of radius \(\ell\) in \(\cN=4\) SYM theory with gauge group \(\SU(N + n)\), on the Coulomb branch where the gauge group is spontaneously broken to \(\SU(N) \times \mathrm{U}(n)\), causing some vector multiplets to gain a mass \(m\). This entanglement entropy takes the form
\begin{equation}
    S_{\hspace{1pt}\text{Coul}}^{(N,n)}(\ell) =S^{(N+n)}(\ell) + P_{\text{Coul}}^{(N,n)}(a) + O(\e) \, ,
    \label{eq:Coulomb_branch_entropy_expansion}
\end{equation}
where the first term on the right-hand side is the entanglement entropy in the conformal vacuum, while $P_{\text{Coul}}^{(N,n)}(a)$ is a UV finite function of the dimensionless radius \(a \propto m \ell\) which vanishes at the fixed point $P_{\text{Coul}}^{(N,n)}(0) = 0$. Returning to the interface, since the theory at $x>0$ is on the Coulomb branch, it is natural to decompose the entanglement entropy as
\begin{equation}
    S(\ell) =\frac{S^{(N_3)}(\ell)+S^{(N_3,n_3)}_{\hspace{1pt}\text{Coul}}(\ell)}{2} + \widetilde{S}_{\mathrm{def}}(\ell)\,.
    \label{eq:interface_entropy_2}
\end{equation}
Comparing to equation~\eqref{eq:interface_entropy} and using equation~\eqref{eq:Coulomb_branch_entropy_expansion}, we see that \(\widetilde{S}_\mathrm{def}(\ell) = S_\mathrm{def}(\ell) - \frac{1}{2}P^{(N,n)}_\mathrm{Coul}(a)\). This has the divergence structure
\begin{equation}
    \widetilde{S}_{\mathrm{def}}(\ell) = p_1\,\frac{\ell}{\epsilon} - \widetilde{P}_{\text{def}}(a) + O(\e) \, ,
    \label{eq:Ftilde_exp}
\end{equation}
where $\widetilde{P}_{\text{def}}(a) \equiv P_{\text{def}}(a) -\frac{1}{2}\,P_{\text{Coul}}^{(N,n)}(a)$ differs from $P_{\text{def}}(a) $ by a finite term. Thus $\widetilde{S}_{\mathrm{def}}(\ell)$ is an intrinsically three-dimensional contribution to the entropy due to the interface since all four-dimensional terms involving the Coulomb branch mass scale $m$ have been subtracted.

\paragraph{Holographic calculation.} Holographically, the defect and interface that we consider are described by the same D3/probe D5-brane system of section~\ref{sec:d3_d5}. The entanglement entropies can be computed using the RT formula as an area of a minimal surface $\mathcal{A}$ anchored on the sphere $r = \ell$ on the conformal boundary $z = 0$ \cite{Ryu:2006bv,Ryu:2006ef}
\begin{equation}
    S(\ell) = \frac{\text{Area}\,(\mathcal{A})}{4G_{\text{N}}}\,.
\end{equation}
The contribution of the defect or the interface $S_{\mathrm{def}}(\ell)$ to the entropy arises from the backreaction of the D5-branes on the geometry which changes the area of the minimal surface. In the probe limit, the area has an expansion in powers of $\sqrt{\la} \, N_5 / N_3 \ll 1$ which translates to the expansion
\begin{equation}
    S_{\mathrm{def}}(\ell) = S_1(\ell)+S_2(\ell) + \ldots\,,
    \label{eq:Sdef_S1}
\end{equation}
where \(S_n \propto N_3^2\, (\sqrt{\la} \, N_5/N_3)^n\). The purpose of the next sections is to compute $S_1(\ell)$. For the zeroth order contributions without a brane present, the RT formula gives the result \cite{Ryu:2006bv,Ryu:2006ef}
\begin{equation} \label{eq:sym_entropy}
    S^{(N)}(\ell) = N^2 \le[\frac{\ell^2}{\e^2} - \log \le(\frac{\ell}{\e}\ri) \ri] + O(\epsilon^0)\,,
\end{equation}
where \(\e\) is a short-distance cutoff. Comparing with equation~\eqref{eq:UV_fixed_point_entropy} we obtain $A_{\text{UV}}^{(N)} = N^2$, as expected for the type A Weyl anomaly coefficient at large \(N\)~\cite{Henningson:1998gx}. This is the entanglement entropy of the $\SO(4,2)$-invariant vacuum of $\mathcal{N} = 4$ SYM theory at large-$N$ and strong coupling.

\subsection{Entanglement entropy of conformal defects}
\label{sec:conformal}

In this section we compute the contribution to entanglement entropy of the conformal defect or interface, i.e. with \(m=0\), corresponding to \(\m=0\) in the D5-brane embeddings described in section~\ref{sec:d3_d5}. Thus, from equation~\eqref{eq:D5_embedding} we have that \(\q = 0\) for all \(z\), so the D5-branes wrap a maximal \sph[2] for all values of the radial coordinate, while the solution for \(x\) becomes
\begin{equation} \label{eq:d5_conformal_solution}
    x = q z \, .
\end{equation}
Such D5-branes have an \(\ads[4]\times\sph[2]\) worldvolume, so they are dual to a conformal defect for $q = 0$ and to a conformal interface for $q\neq 0$. Our aim is to compute the leading-order contribution of the D5-branes to the holographic entanglement entropy in the probe limit. The challenge faced in any holographic entanglement entropy calculation involving probe branes is that, as discussed above, the entanglement entropy is proportional to the area of a minimal surface. In the probe limit we neglect the backreaction of the D5-branes on the metric, so the area of the minimal surface is unchanged.

One possible approach to obtain the entanglement entropy contribution of probe branes is to compute the linearized backreaction of the branes~\cite{Chang:2013mca,Jones:2015twa}. However, when the probe branes preserve defect conformal symmetry there is a simpler method~\cite{Jensen:2013lxa}, based on the observation by Casini, Huerta, and Myers (CHM) that the entanglement entropy of a ball in a CFT on \(d\)-dimensional Minkowski space is equal to the thermal entropy of that CFT on \(\mathbb{R} \times \mathbb{H}_{d-1}\)~\cite{Casini:2011kv}, where \(\mathbb{H}_{d-1}\) is \((d-1)\)-dimensional hyperbolic space. Ref.~\cite{Jensen:2013lxa} used their method to compute the entanglement entropy for the D3/probe D5-brane system for \(m=q=0\). The same result for the \(m=q=0\) entanglement entropy has also been obtained from the linearized backreaction~\cite{Chang:2013mca}. In this section we extend the calculation of ref.~\cite{Jensen:2013lxa} to non-zero \(q\). The holographic entanglement entropy for \(m=0\) has also been computed outside the probe limit using the full backreaction of the D5-branes~\cite{Estes:2014hka}. The result that we will find agrees with the probe limit of the latter calculation.

The required thermal entropy can be computed in holography by performing a bulk coordinate transformation to put \ads[5] in \(\mathbb{R} \times \mathbb{H}_3\) slicing. Starting from the metric and four-form in equation~\eqref{eq:ads5xs5}, we first Wick rotate to Euclidean time \(t_\mathrm{E}\), by defining \(t = - i t_\mathrm{E}\), and then define new coordinates \((\t,\z,v,\xi)\) through~\cite{Casini:2011kv}
\begin{equation}\begin{aligned} \label{eq:chm_map}
    t_\mathrm{E} &= \Omega^{-1} \ell \sqrt{\z^2 - 1} \sin \t \, , \qquad &
    x &= \Omega^{-1} \ell \, \z \sinh v \cos \xi \, ,
    \\
    z &= \Omega^{-1} \ell \, , &
    r &= \Omega^{-1} \ell \, \z \sinh v \sin \xi \, ,
\end{aligned}\end{equation}
where \(\Omega = \z \cosh v + \sqrt{\z^2 - 1} \cos \t\). The new coordinates take values in the ranges \(\t \in [0,2\pi]\), \(\z \in [1,\infty)\), \(v \in [0,\infty)\), and \(\xi \in [0,2\pi]\). The Euclidean \(\ads[5] \times \sph[5]\) metric becomes
\begin{multline}
    \diff s^2 =L^2 \biggl[  f(\z) \diff \t^2 + \frac{\diff \z^2}{f(\z)} + \z^2 \diff v^2 + \z^2 \sinh^2 v  \le( \diff \xi^2 + \sin^2 \xi \diff \f^2 \ri)
    \\
    + \diff \q^2 + \cos^2 \q \diff s_{\sph[2]_{A}}^2 + \sin^2 \q \diff s_{\sph[2]_{B}}^2
    \biggr] ,
    \label{eq:chm_metric}
\end{multline}
where \(f(\z) = \z^2 -1\). The four-form potential in these coordinates may be taken to be
\begin{equation} \label{eq:chm_c4_v1}
    C_4 = -i L^4 (\z^4 - 1) \sinh^2 v \, \sin \xi \, \diff \t \wedge \diff v \wedge \diff \xi \wedge \diff \f + \dots \; ,
\end{equation}
which differs from the direct coordinate transformation of~\eqref{eq:ads5xs5_c4} by a gauge transformation that makes \(C_4\) vanish at \(\z=1\). This gauge transformation is necessary for the probe brane method to yield the correct result for the entanglement entropy~\cite{Kumar:2017vjv}. 

The metric in equation~\eqref{eq:chm_metric} has a Euclidean horizon at \(\z=1\) where the circle wound by \(\t\) degenerates. The horizon has a dimensionless Hawking temperature
\begin{equation}
    T = \frac{1}{2\pi}  \, ,
\end{equation}
and the holographic entanglement entropy \(S\) is equal to the Bekenstein--Hawking entropy of this horizon. To compute the contribution of a probe brane to this entropy, it is convenient to express the entropy as the temperature derivative of the free energy $F$,
\begin{equation} \label{eq:entropy_from_free_energy}
    S = - \le.\frac{\p F}{\p T} \ri|_{T = 1/2\pi} \, .
\end{equation}
The free energy is given holographically by \(F = T I^\star\), where \(I^\star\) is the Euclidean action for the gravitational system (type IIB supergravity plus probe branes), with the star denoting that the action is to be evaluated on-shell.

To compute the temperature derivative in equation~\eqref{eq:entropy_from_free_energy}, we need to be able to vary \(T\). The metric~\eqref{eq:chm_metric} and four-form~\eqref{eq:chm_c4_v1} remain a solution to the type IIB supergravity equations of motion with \(f(\z)\) replaced by~\cite{Emparan:1999gf,Jensen:2013lxa}
\begin{equation} \label{eq:chm_metric_function}
    f(\z) =  \z^2 - 1 - \frac{\z_H^4 - \z_H^2}{\z^2} \, ,
\end{equation}
which shifts the location of the horizon to \(\z=\z_H\). The Hawking temperature of this horizon becomes
\begin{equation} \label{eq:chm_temperature}
    T = \frac{2 \z_H^2 - 1}{2\pi \z_H} \, .
\end{equation}
We also perform a gauge transformation on \(C_4\) so that it still vanishes at the horizon~\cite{Kumar:2017vjv}
\begin{equation} \label{eq:chm_c4}
    C_4 = -i L^4 (\z^4 - \z_H^4) \sinh^2 v \, \sin \xi \, \diff \t \wedge \diff v \wedge \diff \xi \wedge \diff \f + \dots \; .
\end{equation}

The on-shell action is a sum of two pieces,
\begin{equation}
    I^\star = I_\mathrm{bulk}^\star + I_\mathrm{D5}^\star \, ,
\end{equation}
where \(I_\mathrm{bulk}\) is the type IIB supergravity bulk action, while \(I_\mathrm{D5}\) is the Euclidean D5-brane action. In the probe limit, \(I_\mathrm{bulk} \propto N_3^2\) while \(I_\mathrm{D5} \propto \sqrt{\la} \, N_3 N_5 \ll I_\mathrm{bulk}\). The defect or interface contribution $S_1$ to the entanglement entropy, as defined in equation~\eqref{eq:Sdef_S1}, arises from the D5-brane action \eqref{eq:action_ansatz},
\begin{equation} \label{eq:S1_chm}
    S_1 = - \le.\frac{\p (T I_\mathrm{D5}^\star)}{\p T} \ri|_{T = 1/2\pi} \, ,
\end{equation}
which means that our task is to evaluate this derivative.

After making the coordinate transformation in equation~\eqref{eq:chm_map}, the solution in equation~\eqref{eq:d5_conformal_solution} becomes
\begin{equation} \label{eq:d5_conformal_solution_chm}
    \cos\xi = \frac{q}{\z \sinh v} \, .
\end{equation}
Note that when \(T \neq 1/2\pi\), the expression in equation~\eqref{eq:d5_conformal_solution_chm} is no longer a solution to the equations of motion. However, the variation of the embedding due to a small shift in temperature does not affect the value of the on-shell action, since by definition the action is stationary on solutions to the equations of motion. Thus, to evaluate the right-hand side of the  equation~\eqref{eq:S1_chm} it is sufficient to use the \(T=1/2\pi\) solution~\eqref{eq:d5_conformal_solution_chm}. Substituting this solution into the D5-brane action, one finds
\begin{align}
    I^\star_\mathrm{D5}(n) = \frac{2\sqrt{\la}\, N_3 N_5 }{\pi T} \int_{\z_H}^\infty \diff \z \, \int_{v_\mathrm{min}(\z)}^{v_\mathrm{max}(\z)} \diff v\, \sin \xi \sinh v\,  \cL \, ,
    \label{eq:d5_conformal_chm_action}
\end{align}
where the factor of \(1/T\) arises from the integral over \(\t\in[0,1/T]\), and
\begin{equation}
    \cL = \z^2 \sqrt{1+q^2}\sqrt{1 + \sinh^2 v\le[(\p_v \xi)^2 + \z^2 f(\z) (\p_\z \xi)^2 \ri]}
         - q(\z^4 - \z_H^4) \sinh v \, \p_\z \xi \, .
\end{equation}
The lower limit \(v_\mathrm{min}(\z)\) on the \(v\) integral in equation~\eqref{eq:d5_conformal_chm_action} is fixed by the requirement that we integrate only over the worldvolume of the D5-branes, corresponding to \(\cos \xi \leq 1\) in equation~\eqref{eq:d5_conformal_solution_chm},
\begin{eqnarray} \label{eq:d5_conformal_vmin}
    v_\mathrm{min}(\z) = \sinh^{-1} \le(\frac{q}{\z}\ri) \, .
\end{eqnarray}
The upper limit \(v_\mathrm{max}(\z)\) arises because we regulate UV divergences in the entanglement entropy by imposing a near-boundary cutoff \(z \geq \e\) for some small positive \(\e\) in the Poincar\'e coordinates of equation~\eqref{eq:ads5xs5_metric}.\footnote{Crucially, this is the same cutoff used to obtain equation~\eqref{eq:sym_entropy} for \(S^{(N)}\).} From the coordinate transformation~\eqref{eq:chm_map}, this implies that~\cite{Casini:2011kv}
\begin{equation} \label{eq:vmax}
    v_\mathrm{max}(\z) = \log\le(\frac{2\ell}{\z\e}\ri) \, .
\end{equation}

Applying the formula~\eqref{eq:S1_chm} to the action~\eqref{eq:d5_conformal_chm_action}, we arrive at an integral expression for the defect or interface contribution \(S_1\) to entanglement entropy, which can be split into two pieces,
\begin{equation} \label{eq:S1_conformal_decomposition}
    S_1 = \Sbrane + \Shor \, .
\end{equation}
The first term, \(\Sbrane\), arises from taking the derivative with respect to \(T\) of the explicit factors of \(\z_H\) appearing in \(\cL\)
\begin{align}\label{eq:Sbulk_conformal}
    \Sbrane &\equiv \frac{2\sqrt{\la}\, N_3 N_5 }{\pi} \frac{\p \z_H}{\p T} \le. \int_{\z_H}^\infty \diff \z \, \int_{v_\mathrm{min}(\z)}^{v_\mathrm{max}(\z)} \diff v\, \sin \xi \sinh v\,  \frac{\p \cL}{\p \z_H} \ri|_{T=1/2\pi}
    \nonumber \\
    &= - \frac{2 q^2\sqrt{\la} \,  N_3 N_5}{\pi} \int_{\z_H}^\infty \diff \z \, \int_{v_\mathrm{min}(\z)}^{v_\mathrm{max}(\z)} \diff v \frac{\sinh v}{\z^2}
    \\
    &= \frac{\sqrt{\la}\, N_3 N_5}{\pi} \le[-q^2 \frac{\ell}{\e} + q^2 \sqrt{1+ q^2} + q \sinh^{-1} q \ri]+ O(\e) \, .
    \nonumber
\end{align}
The second term in equation~\eqref{eq:S1_conformal_decomposition}, \(\Shor\), arises from the derivative of the lower limit of the \(\z\) integral,
\begin{equation}\begin{aligned} \label{eq:Shor_conformal}
    \Shor &\equiv
    -\le. \frac{2\sqrt{\la}\, N_3 N_5 }{\pi}  \frac{\p \z_H}{\p T}  \, \int_{v_\mathrm{min}(\z)}^{v_\mathrm{max}(\z)} \diff v\, \sin \xi \sinh v\,  \cL |_{\z = \z_H} \ri|_{T=1/2\pi}
    \\
    &= \frac{2 (1+q^2)\sqrt{\la} \, N_3 N_5}{3\pi} \int_{v_\mathrm{min}(1)}^{v_\mathrm{max}(\z)} \diff v \, \sinh v
    \\
    &= (1+q^2) \frac{2 \sqrt{\la} \, N_3 N_5}{3\pi} \le[ \frac{\ell}{\e} - \sqrt{1 + q^2} \ri] + O(\e) \, .
\end{aligned} \end{equation}

Summing the two contributions \eqref{eq:Sbulk_conformal} and \eqref{eq:Shor_conformal}, we find that the defect or interface contribution to the entanglement entropy is, in the probe limit
\begin{equation} \label{eq:d5_conformal_entanglement}
    S_1 = 
    \frac{\sqrt{\la} \, N_3 N_5}{3\pi}\le[ (2-q^2)\frac{\ell}{\e} - (2-q^2)\sqrt{1 + q^2} + 3 q \sinh^{-1} q\ri] + O(\e)\, .
\end{equation}
This result provides a consistency check of the probe formalism, as the finite part of the holographic entanglement entropy for the this defect or interface with \(m=0\) and arbitrary \(N_5\) and \(n_3\) was computed in ref.~\cite{Estes:2014hka} using the RT prescription in the appropriate backreacted supergravity solutions~\cite{Gomis:2006cu,DHoker:2007zhm}. The finite term in equation~\eqref{eq:d5_conformal_entanglement} matches the \(\sqrt{\la}\, N_5 \ll N_3\) and \(n_3 \ll N_3\) limit of the result of ref.~\cite{Estes:2014hka}.\footnote{As a further check of equation~\eqref{eq:d5_conformal_entanglement}, we can set \(q=0\) to find
\[
    S_1 = \frac{2\sqrt{\la} \, N_3 N_5}{3\pi}\le( \frac{\ell}{\e} - 1\ri) + O(\e)\, ,
\]
which is in agreement with equation~(3.51) of ref.~\cite{Jensen:2013lxa} (see also ref.~\cite{Chang:2013mca}).} As expected~\cite{Kobayashi:2018lil}, the finite term in equation~\eqref{eq:d5_conformal_entanglement} is \(-F_\mathrm{def}\), where \(F_\mathrm{def}\) is the contribution of the defect or interface to the free energy on a sphere; as we show in appendix~\ref{app:free_energy},
\begin{equation}\label{eq:free-energy}
    F_{\text{def}} =\frac{\sqrt{\la} \, N_3 N_5}{3 \pi} \le[(2-q^2)\sqrt{1+q^2} - 3 q \sinh^{-1} q\ri]
\end{equation}
to leading order in the probe limit.

\subsection{Entanglement entropy of RG flows}
\label{sec:flow_ee}

In this subsection, we compute the probe contribution to entanglement entropy when \(m \neq 0\), i.e. when the defect hypermultiplets have non-zero mass when \(n_3 = 0\), or when the theory on the right-hand side of the defect is asymptotically on the Coulomb branch for \(n_3 \neq 0\). In either case this breaks conformal symmetry, so we can no longer interpret the entanglement entropy as a thermal entropy on \(\mathbb{R} \times \mathbb{H}_{3}\). Instead, we will use the method of ref.~\cite{Karch:2014ufa}, which arises from the application of generalized gravitational entropy~\cite{Lewkowycz:2013nqa} to probe branes. We briefly summarise the essentials of this method needed for our calculation below.

\paragraph{Method.}

Consider the probe D5-branes embedded in the spacetime in equation~\eqref{eq:chm_metric}, with \(f(\z)\) as in equation~\eqref{eq:chm_metric_function} with arbitrary \(\z_H\). Let \([I_\mathrm{D5}^\star]_{2\pi}\) denote the on-shell action of the probe D5-branes, with the integral over \(\t\) restricted to \(\t \in 2\pi\). Then the probe D5-branes' contribution to the entanglement entropy of a ball, as defined in equation~\eqref{eq:Sdef_S1}, is~\cite{Karch:2014ufa}
\begin{equation} \label{eq:generalised_gravitational_entropy}
    S_1 = - \frac{1}{3}\lim_{\z_H \to 1} \p_{\z_H} [I_\mathrm{D5}^\star(\z_H)]_{2\pi} \, .
\end{equation}
For \(m=0\) this is equivalent to equation~\eqref{eq:S1_chm}. Let us write
\begin{equation}
    [I_\mathrm{D5}^\star(\z_H)]_{2\pi} = \int_{\z_H}^\infty \diff \z  \int_0^{2\pi} \diff \t \, \cL(\z_H; X) \, ,
\end{equation}
where the Lagrangian density \(\cL\) depends explicitly on \(\z_H\) and on the embedding scalars, collectively denoted by \(X = (X^1,X^2) \equiv (\q,\xi)\). Since the action is to be evaluated on-shell, \(\cL\) also has implicit dependence on \(\z_H\) through the form of the solutions for the embedding scalars \(X\). The entanglement entropy obtained from equation~\eqref{eq:generalised_gravitational_entropy} then splits into a sum of three terms
\begin{equation} \label{eq:S1_split}
    S_1 = \Sbrane + \Shor + \Svar\,.
\end{equation}

The first term in equation~\eqref{eq:S1_split}, \(\Sbrane\), is the contribution from differentiating the Lagrangian density with respect to its explicit dependence on \(\z_H\), keeping the embedding \(X\) fixed,
\begin{equation} \label{eq:Sbrane_def}
    \Sbrane = - \frac{1}{3} \lim_{\z_H \to 1} \int_1^\infty \diff \z \int_0^{2\pi} \diff \t  \le(\frac{\p \cL}{\p \z_H}\ri)_X \,.
\end{equation}
The second term in \eqref{eq:S1_split}, \(\Shor\), is the boundary term at \(\z = \z_H\), coming from the \(\z_H\)-derivative acting on the lower limit of the \(\z\) integral,
\begin{equation} \label{eq:Shor_def}
    \Shor = \frac{1}{3} \lim_{\z_H \to 1}\int_0^{2\pi} \diff \t \le.\cL\ri|_{\z=\z_H}\,.
\end{equation}
Finally, \(\Svar\) arises from the implicit dependence of \([I^\star]_{2\pi}\) on \(\z_H\) due to its dependence on the embedding \(X\). Due to stationarity of the action when evaluated on solutions to the equations of motion, \(\Svar\) is also a boundary term at \(\z = \z_H\),
\begin{equation} \label{eq:Svar_def}
    \Svar = \frac{1}{3} \lim_{\z_H \to 1} \int_0^{2\pi} \diff \t \le.\le(\frac{\p \cL}{\p (\p_\z X^n)}\ri)_{\z_H} \frac{\p X^n}{\p \z_H} \ri|_{\z = \z_H}\, .
\end{equation}
Further details on the decomposition in equation~\eqref{eq:S1_split} may be found in refs.~\cite{Karch:2014ufa,Kumar:2017vjv,Chalabi:2020tlw}.

The first two contributions to equation~\eqref{eq:S1_split}, \(\Sbrane\) and \(\Shor\), depend only on the form of the embedding scalars when \(\z_H = 1\). This may be obtained from the embedding in Poincar\'e coordinates, described in section~\ref{sec:d3_d5}, by applying the coordinate transformation in equation~\eqref{eq:chm_map}. Explicitly, in Poincar\'e coordinates the non-zero \(q\) embedding derived in section~\ref{sec:d3_d5} is
\begin{equation} \label{eq:massive_solution_poincare}
    \sin \q = \m z \, ,
    \qquad
    x = \frac{q z}{\sqrt{1 - \m^2 z^2}} \, .
\end{equation}
In the coordinates used in equation~\eqref{eq:chm_metric}, the embedding is specified by the angular coordinates \(\q\) and \(\xi\). Applying the transformation in equation~\eqref{eq:chm_map} to equation~\eqref{eq:massive_solution_poincare}, we obtain
\begin{equation} \label{eq:massive_solution_chm}
    \sin \q = \frac{\m\ell}{\z \cosh v + \sqrt{\z^2 - 1} \cos \t} \, ,
    \qquad
    \cos \xi = \frac{q}{\z \sinh v \, \cos\q} \, .
\end{equation}
This form of the embedding may be substituted into equations~\eqref{eq:Sbrane_def} and \eqref{eq:Shor_def}. On the other hand, in order to evaluate \(\p X^n/\p \z_H\) in \(\Svar\) we need to determine the form of the embedding close to \(\z = \z_H\), slightly away from \(\z_H = 1\). This may be achieved using the series solution approach laid out in ref.~\cite{Chalabi:2020tlw}.

\paragraph{Calculation.}

We will now evaluate the entanglement entropy contributions~\(\Sbrane\), \(\Shor\), and \(\Svar\) described above. The Euclidean D5-brane action evaluated on the ansatz in which the only non-trivial  worldvolume scalars are \((\q,\xi)\) depending on \((\z,v,\t)\), and with a worldvolume gauge field \(\cF\) as in equation~\eqref{eq:ansatz}, is
\begin{equation} \label{eq:d5_chm_action}
    [I_\mathrm{D5}^\star]_{2\pi} = \frac{\sqrt{\la}\, N_3 N_5}{\pi^2} \int_{\cR} \diff v \diff \z \diff \t  \sinh v \sin\xi
    \,
    \le[\z^2 \sqrt{q^2  + \cos^4 \q} \,\sqrt{\Sigma} - q (\z^4 - \z_H^4) \sinh v \, \p_\z \xi\ri]  ,
\end{equation}
where\footnote{Our convention for the antisymmetrization of indices includes normalization by a factor of half,
\[
    \p_{[a} \q \, \p_{b]} \xi = \frac{1}{2} (\p_a \q \, \p_b \xi - \p_b \q \, \p_a \xi) \, .
\]}
\begin{equation}\begin{aligned}
    \Sigma &= 1 + \p_a \q \, \p^a \q
    + \z^2 \sinh^2 v \, \p_a \xi \, \p^a \xi
    - 2 \z^2 \sinh^2 v \, \p_{[a} \, \q \p_{b]} \xi \, \p^{[a} \, \q \p^{b]} \xi\, ,
\end{aligned}\end{equation}
with \(a,b\) indices raised by the metric
\begin{equation}
    \diff s^2 = f(\z) \diff \t^2 + \frac{1}{f(\z)} \diff \z^2 + \z^2  \diff v^2 \, .
\end{equation}
For \(\z_H=1\), the integration region \(\cR\) is defined by
\begin{equation}
    0 \leq \t \leq 2\pi \, ,
    \qquad
    \z \geq 1\, ,
    \qquad
    0 \leq v \leq v_\mathrm{max}(\z) \equiv \log \le(\frac{2\ell}{\z\e}\ri)\ ,
    \qquad
    \cos \xi \leq 1 \, ,
\end{equation}
where \(\e \ll 1\) (and thus \(v_\mathrm{max}(\z) \gg 1\)) is the same near-boundary cutoff as was used in section~\ref{sec:conformal}. Most of the challenge in computing \(\Sbrane\) comes from the condition \(\cos\xi \leq 1\), which from equation~\eqref{eq:massive_solution_chm} implies that
\begin{equation} \label{eq:complicated_boundary}
    \frac{q}{\z \sinh v \sqrt{1 - \frac{a^2}{(\z \cosh v + \sqrt{\z^2 - 1} \cos \t)^2}}} \leq 1 \, .
\end{equation}
In principle, when defining \(\cR\) we should also require \(\sin \q \leq 1\), implying that
\begin{equation}
    \frac{a}{\z \cosh v + \sqrt{\z^2 - 1} \cos \t} \leq 1 \, .
\end{equation}
However, except at \(q=0\) this is always less strict than the inequality in equation~\eqref{eq:complicated_boundary}.

In order to simplify the results below, we define the combinations
\begin{equation} \label{eq:a_sigma_def}
    a = \frac{2\pi}{\sqrt{\la}} m \ell \, ,
    \qquad
    \s = \frac{\sqrt{
			1 + a^2 + q^2 - \sqrt{(1-a^2 +q^2)^2 + 4 a^2 q^2}
		}}{\sqrt{2}} \, ,
\end{equation}
which arise naturally from the holographic calculation. The parameter \(a\) is a dimensionless radius of the entangling region, measured in units of \(\m = 2\pi m/\sqrt{\la}\), while \(\s\) is the maximum value of \(\m z\) at the intersection of the RT surface \(x^2 + r^2 + z^2 = \ell^2\) with the D5-branes at \(x = q z/\sqrt{1 -\m^2 z^2}\).

\begin{figure}
    \begin{center}
    \includegraphics[]{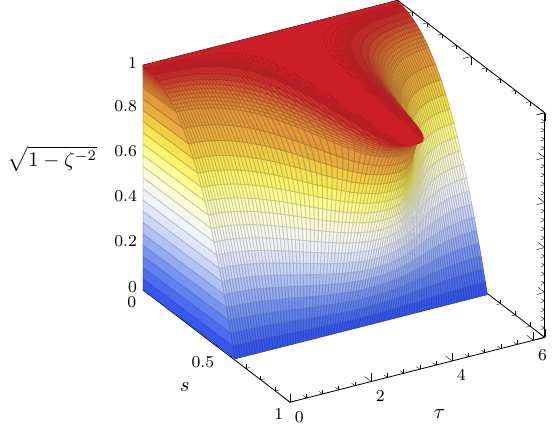}
    \caption{The shaded region shows the typical form of the integration region for \(\Sbrane\) in the \((\t,\z,s)\) coordinates used in equation~\eqref{eq:Sbrane_integral}. This plot shows the exact integration region for \(q=a=1\), but the shape of the integration region is qualitatively similar for other non-zero values of \(q\) and \(a\).}
    \label{fig:integration_region}
    \end{center}
\end{figure}

Differentiating equation~\eqref{eq:d5_chm_action} with respect to the explicit factors of \(\z_H\) appearing in the integrand, we obtain an integral for \(\Sbrane\) defined in equation~\eqref{eq:Sbrane_def}. The resulting integrand is cumbersome, so we will not write it explicitly. Instead we will write the expression obtained after one changes integration variable from \(v\) to \(s \equiv \sin \q\) using equation~\eqref{eq:massive_solution_chm}, which gives a significantly simpler integrand,
\begin{equation}\begin{aligned} \label{eq:Sbrane_integral}
    \Sbrane =
    \frac{\sqrt{\la} \, N_3 N_5}{3\pi^2 a} \! \int_{\cR} \! \frac{\diff s \diff \z  \diff \t}{(1-s^2)^{3/2}} \! \biggl[&
        s^2 \le(q^2 + (1-s^2)^2 \ri) \le(\frac{\cos(2\t)}{\z^2 - 1} - \sin^2 \t  + \frac{2 a \cos \t}{s \sqrt{\z^2 - 1}}\ri)
        \\
        & + a^2 \le(\frac{q^2}{s^2} + (1-s^2)^2 \ri)
        - 4 q^2 a \frac{s^3 \cos\t + s \sqrt{\z^2 -1}}{s^2 \z^3 (1-s^2)^{3/2} \sqrt{\z^2 - 1}}
    \biggr] \, .
\end{aligned}\end{equation}
In these coordinates, the integration region \(\cR\) is defined by \(0 \leq \t \leq 2\pi\), \(\z \geq 1\), \(\m\e \leq s \leq 1\), and
\begin{equation} \label{eq:s_max}
    q^2 s^2 \leq (1-s^2) \le(a - s \z - s \sqrt{\z^2 - 1} \, \cos \t \ri)\le(a + s \z - s \sqrt{\z^2 - 1} \, \cos \t \ri) ,
\end{equation}
the latter of which follows from equation~\eqref{eq:complicated_boundary}. The typical form of \(\cR\) is shown in figure~\ref{fig:integration_region}, where we see that equation~\eqref{eq:s_max} gives a quartic equation for a \((\t,\z)\)-dependent maximum value of \(s\). This makes equation~\eqref{eq:Sbrane_integral} challenging to evaluate directly, since the roots of this quartic are extremely cumbersome. To make progress we follow the strategy of ref.~\cite{Karch:2014ufa} of changing integration variables back to the Poincar\'e coordinates used in equation~\eqref{eq:ads5xs5_metric}, which simplifies the integration region at the cost of making the integrand more complicated.

The necessary change of variables to Poincar\'e coordinates is obtained by inverting the transformation in equation~\eqref{eq:chm_map}, restricted to the worldvolume of the D5-branes, i.e. at \(s \equiv \sin\q=\m z\) and \(x = q z/\sqrt{1-\m^2 z^2}\). For notational simplicity we will drop the subscript ``E'' from the Euclidean time, and rescale both \(t\) and \(r\) by factors of \(\ell\) to make them dimensionless. The resulting change of variables is
\begin{equation}\begin{aligned} \label{eq:to_poincare}
    \z &= \frac{1}{2 a s(1-s^2)}\sqrt{(1-s^2)^2\le[(R^2 + a^2)^2 - 4 a^2 r^2\ri] + 2 q^2 s^2 (1-s^2) (R^2 - a^2) + q^4 s^4} \, ,
    \\
    \cos \t &= \frac{(1-s^2)(a^2 - R^2) - q^2 s^2}{
        \sqrt{
            (1-s^2)^2 \le[(R^2 - a^2)^2 + 4 a^2 t^2 \ri] + 2 q^2 s^2 (1-s^2) (R^2 - a^2) + q^4 s^4
        }
    } \, ,
\end{aligned}\end{equation}
with \(R^2 = r^2 + s^2 + t^2\). In this coordinate system the integration region \(\cR\), which corresponds to the worldvolume of the D5-branes with a near-boundary cutoff \(z \geq \e\), is defined by \(s \in [\m \e,1]\), \(t \in (-\infty,\infty)\), and \(r \in [0,\infty)\).

The details of the evaluation of the integral for \(\Sbrane\) in Poincar\'e coordinates \((s,t,r)\) are given in appendix~\ref{app:integrals}. The result is
\begin{equation}\begin{aligned} \label{eq:Sbrane_result}
    \Sbrane = \frac{\sqrt{\la} \, N_3 N_5}{3\pi} \biggl[
            - 3 q^2 \frac{a}{\m\e} +& \frac{2 a^2 + 2 q^2 - 1}{4a} \sin^{-1} \s + 3 q \sinh^{-1}\le(\frac{q}{\sqrt{1-\s^2}}\ri)
        \\
            &+ \sqrt{1-\s^2} \le(3(1-2\s^2)\frac{a^3}{\s^3} - \frac{7 - 12 \s^2}{2} \frac{a}{\s}+ \frac{3 \s}{4a} \ri)
    \biggr] \, .
\end{aligned}\end{equation}
There is a subtlety associated to the evaluation of \(\Sbrane\) in Poincar\'e coordinates, to do with the fact that the integrand in equation~\eqref{eq:Sbrane_integral} is singular at \(\z=1\). We comment on this subtlety in section~\ref{sec:boundary_term}.

The boundary term at \(\z=1\) defined in equation~\eqref{eq:Shor_def} is more straightforward to compute. Evaluating the integrand of equation~\eqref{eq:d5_chm_action} on the solution~\eqref{eq:massive_solution_chm} at \(\z=1\), one finds
\begin{equation}\label{eq:Shor_text}
    \Shor =\frac{2\sqrt{\la} \, N_3 N_5}{3 \pi} \int_{v_\mathrm{min}(1)}^{v_\mathrm{max}(1)} \diff v \, \tanh v \, \frac{q^2 \cosh^4 v + (\cosh^2 v - a^2)^2}{(\cosh^2 v - a^2)^{3/2}}\, .
\end{equation}
The lower limit of the integral over \(v\) is determined by equation~\eqref{eq:complicated_boundary} evaluated at \(\z=1\), \(v_\mathrm{min} = \sinh^{-1}(q/\sqrt{1-\s^2})\), while the upper limit comes from equation~\eqref{eq:vmax}. The integrand in equation \eqref{eq:Shor_text} may be simplified by rewriting it in terms of \(s \equiv \sin \q = a \sech v\),
\begin{subequations}\begin{align}
    \Shor &= \frac{2 \sqrt{\la} \, N_3 N_5}{3 \pi} a \int_{\m\e}^{\s} \diff s \, \frac{q^2+(1-s^2)^2}{s^2(1-s^2)^{3/2}}
    \\
    &= \frac{2\sqrt{\la} \, N_3 N_5}{3\pi} a \le[\frac{1+q^2}{\m\e} + \frac{\s^2 - 1 + q^2 (2 \s^2 - 1)}{\s \sqrt{1 - \s^2}} - \sin^{-1} \s\ri]  .
    \label{eq:Shor_result}
\end{align}\end{subequations}
The lower limit on the integral comes from \(v_\mathrm{max}(1)\), and arises because the embedding is \(s = \m z\) and we are imposing the UV cutoff \(z \geq \e\). The upper limit \(\s\), which was defined in equation~\eqref{eq:a_sigma_def}, comes from the fact that we are integrating over the surface at \(\z=1\), which in Poincar\'e coordinates corresponds to the RT surface, \(r^2 + x^2 + z^2 = \ell^2\). Since \(r\) is real, this implies that \(x^2 + z^2 \leq \ell^2\). Using the D5-brane embedding in equation~\eqref{eq:massive_solution_poincare} to express \(x\) in terms of \(z\), and then writing \(z = \m^{-1} s\), we find that \(x^2 + z^2 \leq \ell^2\) implies that
\begin{equation} \label{eq:s_max_sigma}
    \frac{q^2 s^2}{1 - s^2} + s^2 \leq a^2 
    \quad
    \Rightarrow
    \quad
    s \leq \s \, .
\end{equation}

The second boundary term at \(\z=1\), defined by equation~\eqref{eq:Svar_def}, is
\begin{subequations}
\begin{align}
    \Svar &= -\frac{\sqrt{\la} \, N_3 N_5}{3\pi^2} a^2\int_0^{2\pi} \diff \t \int_{v_\mathrm{min}}^{v_\mathrm{max}} \diff v \, \cos^2 \t \, \tanh v \, \frac{q^2 + (1-a^2 \sech^2 v)^2}{(\cosh^2 v - a^2)^{3/2}}
    \label{eq:Svar_integral}
    \\
    &= -\frac{2 \sqrt{\la} \, N_3 N_5}{3\pi a} \int_{\m\e}^\s \diff s \,  \frac{ s^2 \le[ q^2 + (1-s^2)^2 \ri]}{(1-s^2)^{3/2}}
    \label{eq:Svar_integral_s}
    \\
    &= \frac{\sqrt{\la} \, N_3 N_5}{24 \pi a} \le[\s (1-2\s^2) \sqrt{1-\s^2} - \frac{8 q^2 \s}{\sqrt{1-\s^2}} - (1 - 8 q^2) \sin^{-1}\s\ri] ,
    \label{eq:Svar_result}
\end{align}
\end{subequations}
where we drop terms that vanish in the limit \(\e\to 0\).
The derivation of the integral in equation~\eqref{eq:Svar_integral} using methods from ref.~\cite{Chalabi:2020tlw} is given in appendix~\ref{sec:Svar}. Equation~\eqref{eq:Svar_integral_s} is obtained from~\eqref{eq:Svar_integral} by the substitution \(s = a \sech v\).

We note that \(\Svar = 0\) at \(m=0\), consistent with the fact that this boundary term is generically absent for brane embeddings preserving defect conformal symmetry~\cite{Chalabi:2020tlw}. This follows from the definitions in equation~\eqref{eq:a_sigma_def}, which imply that \(m=0\) (for fixed \(\ell\)) corresponds to \(a=0\), and that for small \(a\) one has \(\s \approx a/\sqrt{1+q^2}\). Using this, one finds from equation~\eqref{eq:Svar_result} that \(\Svar\) vanishes as \(a \to 0\).

\paragraph{Result for the entanglement entropy.} Summing the contributions in equations~\eqref{eq:Sbrane_result}, \eqref{eq:Shor_result}, and~\eqref{eq:Svar_result}, we obtain our result for the leading order contribution of the defect or interface to the entanglement entropy in the probe limit,\footnote{As a reminder, as defined in equation~\eqref{eq:a_sigma_def} \( a = \m \ell = \frac{2\pi}{\sqrt{\la}} m \ell\) is a dimensionless parameter proportional to the radius of the entangling region in units of the mass of the hypermultiplet fields (for \(n_3 =0\)) or vector multiplets (for \(n_3 \neq 0\)), and
\[
    \s = \frac{\sqrt{
			1 + a^2 + q^2 - \sqrt{(1-a^2 +q^2)^2 + 4 a^2 q^2}
		}}{\sqrt{2}} \, .
\]}
\begin{equation}\begin{aligned} \label{eq:ee_full}
    S_1 = \frac{\sqrt{\la} \, N_3 N_5}{3\pi} \biggl[
        (2-q^2)&\frac{a}{\m\e}
        - \frac{3(1+4a^2-4q^2)}{8a} \sin^{-1}\s
        + 3 q \sinh^{-1}\le(\frac{q}{\sqrt{1-\s^2}} \ri)
        \\&
        + \sqrt{1-\s^2} \le(
            (1-2\s^2) \frac{a^3}{\s^3} + \frac{4 \s^2 - 9}{2} \frac{a}{\s} + \frac{15 - 2\s^2}{8} \frac{\s}{a}    
        \ri)
    \biggr]\, ,
\end{aligned}\end{equation}
This is one of the key results of this article. As a cross-check, the result in equation~\eqref{eq:ee_full} reduces to the conformal result in equation~\eqref{eq:d5_conformal_entanglement} in the limit \(m \to 0\) with \(\ell\) held fixed, as discussed further below.

When \(q=0\), the result for the entanglement entropy simplifies greatly,
\begin{equation} \label{eq:delta_ee_q0}
    \le. S_1\ri|_{q=0} = \frac{\sqrt{\la} \, N_3 N_5}{24\pi} \times \begin{cases}
        \dfrac{16 a}{\m\e} - 3 (1 + 4 a^2) \dfrac{\sin^{-1} a}{a} - (13+2a^2) \sqrt{1-a^2} \, ,
        &
        a \leq 1 \, ,
        \\[1em]
        \dfrac{16 a}{\m\e}- \dfrac{3\pi}{2} (1 + 4 a^2) \ , & a > 1 \, .
    \end{cases}
\end{equation}
We note that \(S_1|_{q=0}\) is continuous at \(a=1\), as are its first three derivatives with respect to \(a\). On the other hand, \(\p_a^4 \le. S_1\ri|_{q=0}\) is discontinuous at \(a=1\). The piecewise form of this result arises because at \(q=0\) one has that \(\s = \mathrm{min}(a,1)\), which can be understood holographically as follows.

\begin{figure}
    \centering
    \includegraphics[width=0.95\linewidth]{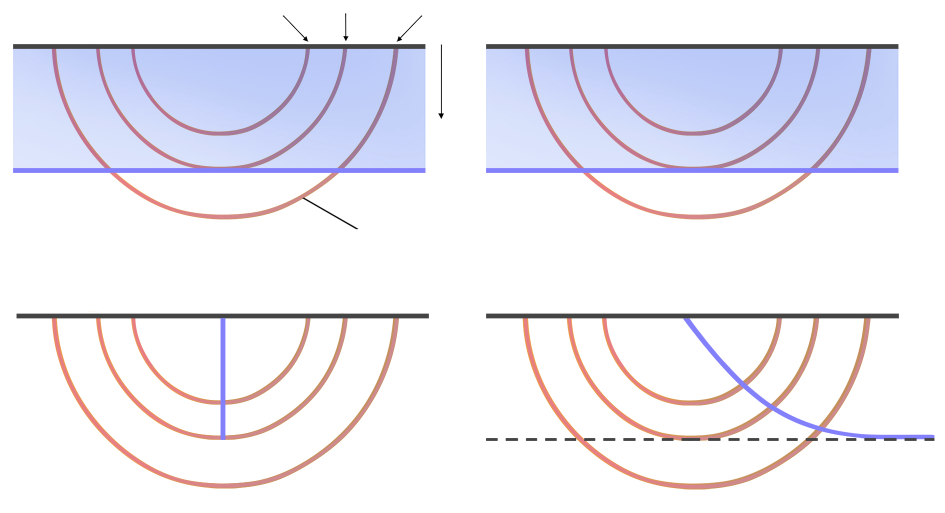}
    \put(-325,-14){$q = 0$}
    \put(-120,-14){$q > 0$}
    \put(-314,220){$a<1$}
    \put(-274,220){$a=1$}
    \put(-234,220){$a>1$}
    \put(-215,180){$z$}
    \put(-325,138){$z=1/\m$}
    \put(-325,22){$z=1/\m$}
    \put(-45,20){$z=1/\m$}
    \put(-330,65){D5}
    \put(-400,160){D5}
    \put(-190,160){D5}
    \put(-123,65){D5}
    \put(-402,205){boundary $(x,z=0)$}
    \put(-402,90){boundary $(r,z=0)$}
    \put(-190,205){boundary $(x,z=0)$}
    \put(-190,90){boundary $(r,z=0)$}
    \put(-265,110){RT-surface}
    \caption{Cross-sections of \ads[5] for D3/probe D5 system, with RT surfaces. The top panels show the projection onto the $(x,z)$ plane, where $x$ is the boundary direction orthogonal to the defect and $z$ is the bulk coordinate. The D5-branes correspond to the shaded region. The bottom panels show the projection onto the $(r,z)$ plane, where $r$ is a radial polar coordinate on the plane of the defect. The D5-branes are represented by the vertical, purple lines. \textbf{(Left)} When $q=0$, the D5-branes extend to $z=1/\mu$ straight into the bulk, orthogonal to the boundary in the $r$ direction and the RT-surfaces. The maximal value \(\s\) of \(\m z\) on the intersection of the D5-branes with the RT surface is \(\s = 1\) for all \(a \geq 1\).  \textbf{(Right)} For $q > 0$ the D5-branes bend and extend to infinity in a direction parallel to the boundary. As a result, \(\s\) grows smoothly to its maximal value of \(\s=1\) as \(a \to \infty\).}
    \label{fig:q}
\end{figure}
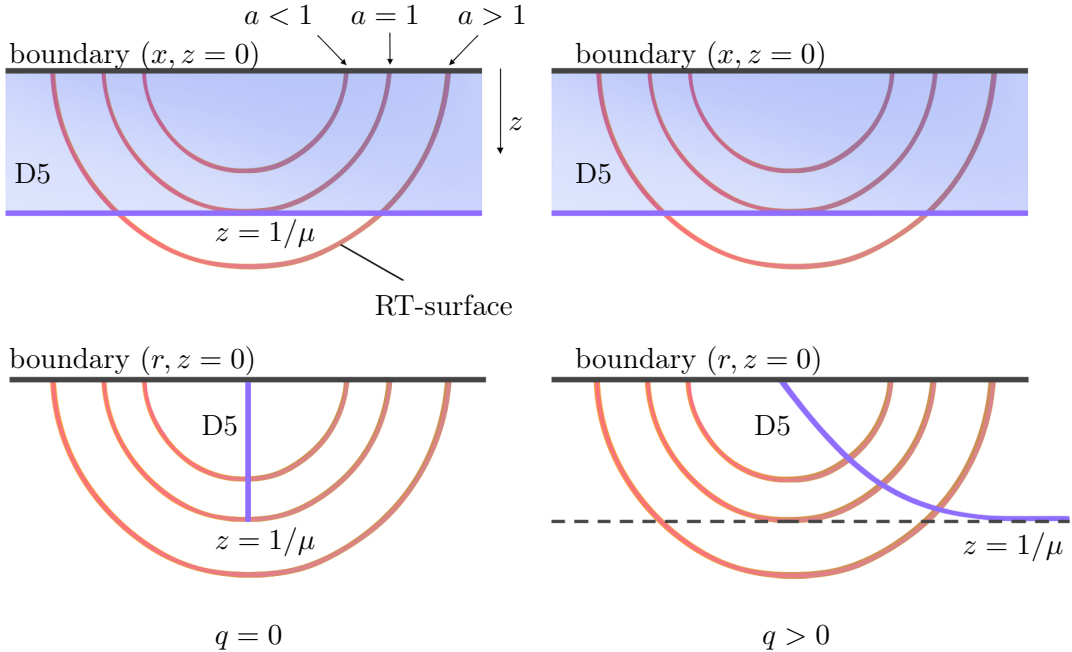

Recall from the discussion around equation~\eqref{eq:s_max_sigma} that the quantity \(\s\) is the maximal value of \(s = \m z\) on the intersection between the D5-branes and the RT surface. As depicted on the left-hand side of figure~\ref{fig:q}, when \(q=0\) and \(m \neq 0\), the probe D5-branes have a maximal extent into the bulk of \ads[5]. When the dimensionless radius \(a\) of the entangling region is small, \(\s\) is set by the extent of the RT surface into the bulk. As the size of the entangling region grows, the RT surface probes further into the bulk, eventually probing further than the maximal extent of the D5-branes. This occurs exactly at \(a=1\), at which point \(\s\) saturates to its maximal value \(\s=1\). In contrast, when \(q \neq 0\) the D5-branes extend toward infinity in the bulk, in one of the directions parallel to the boundary, as depicted on the right-hand side of figure~\ref{fig:q}. Consequently, for \(q \neq 0\) we have that \(\s\) grows smoothly towards \(\s \to1\) as \(a \to \infty\), which is why the result in equation~\eqref{eq:ee_full} for \(q \neq 0\) does not take a piecewise form. 

Expanding equation~\eqref{eq:ee_full} for small radius \(a \ll 1\), we find
\begin{equation}\begin{aligned} \label{eq:ee_uv_expansion}
    S_1 = \frac{\sqrt{\la} \, N_3 N_5}{3\pi} \biggl[
        (2-q^2)\frac{\ell}{\e} - (2-q^2)\sqrt{1 + q^2} + 3 q \sinh^{-1} q
        - \sqrt{1+q^2} \, a^2
    \biggr] + O(a^4)\, ,
\end{aligned}\end{equation}
where in the term proportional to \(\e^{-1}\) we have replaced \(a\) with \(\ell = a/\m\). The terms at order \(O(a^0)\) in the small-\(a\) expansion match the \(m=0\) result in equation~\eqref{eq:d5_conformal_entanglement}. This makes sense since by dimensional analysis, the limit of small \(a\) is equivalent to the limit of small \(m\) with fixed \(\ell\). If we define \(\Delta S_1\) as the entanglement entropy minus this UV contribution, \(\D S_1 \equiv S_1 - \le.S_1\ri|_{m=0}\), then we have that at leading order in small \(a\)
\begin{equation}
    \Delta S_1 = - \sqrt{1+q^2}\frac{\sqrt{\la} \, N_3 N_5}{3\pi}  \, a^2 + O(a^4) \, .
\end{equation}
It may be possible to compute this leading term in \(\Delta S_1\) using the RT prescription in the back-reacted geometry obtained starting from the \(m=0\) geometry of refs.~\cite{Gomis:2006cu,DHoker:2007zhm} and turning on perturbatively small \(m\). This would provide another check of our probe calculation, although would likely be technically challenging. Furthermore, the leading $O(a^2)$ term is controlled by conformal perturbation theory which may be accessible using field theory methods.

In the opposite limit of large radius, expanding \(\Delta S_1\) for \(a \gg 1\) and using equation~\eqref{eq:q_def} to replace \(q\) with \(n_3\), we find
\begin{equation} \label{eq:Delta_S_large_a}
    \Delta S_1 = - \frac{n_3 N_3}{3} a^2 + \frac{\sqrt{\la} \, N_3 N_5}{4} a + n_3 N_3 \log a +  O(a^0) \, .
\end{equation}
The \( O(a)\) term is independent of \(n_3\), and describes the leading-order contribution of D5-branes without dissolved D3-brane charge to \(\Delta S_1\). The \( O(a^2)\) and \( O(\log a)\) terms are proportional to \(n_3\) and have a natural interpretation; they are exactly one half of the leading and first subleading terms at large radius in the entanglement entropy contribution from \(n_3\) probe D3-branes spanning the \((t,x,r,\f)\) directions~\cite{Chalabi:2020tlw}, \(P^{(N_3,n_3)}_\mathrm{Coul}(a)\) as defined equation~\eqref{eq:Coulomb_branch_entropy_expansion}. These terms thus arise in our result for the entanglement entropy for \(n_3 \neq 0\) because the D5-branes contain dissolved D3-brane charge, with the factor of one half arising because the dissolved D3-branes are only semi-infinite in the \(x\) direction as indicated in table~\ref{tab:brane_configuration}. See appendix~\ref{sec:coulomb} for the translation of the results of ref.~\cite{Chalabi:2020tlw} into our notation, with \(P^{(N_3,n_3)}_\mathrm{Coul}(a)\) given in equation~\eqref{eq:S_coul}.

\subsection{On the evaluation of worldvolume integrals}
\label{sec:boundary_term}

In this subsection we comment on a subtlety that arises in the computation of the integral for \(\Sbrane\) via transformation to Poincar\'e coordinates. The integrand in equation~\eqref{eq:Sbrane_integral} contains a term which diverges as \((\z-1)^{-1}\) as \(\z\to1\),
\begin{equation} \label{eq:Sbrane_singular}
    \Sbrane =
    \frac{\sqrt{\la} \, N_3 N_5}{3\pi^2 a} \int_{\cR} \diff s \diff \z \diff \t \, \frac{1}{(1-s^2)^{3/2}}\biggl[
        s^2 \le(q^2 + (1-s^2)^2 \ri) \frac{\cos(2\t)}{\z^2 - 1} + \ldots
    \biggr] \, ,
\end{equation}
where the dots denote terms of the integrand which are subleading as \(\z \to 1\). The singular term in equation~\eqref{eq:Sbrane_singular} means that the integral is not absolutely convergent, making the order of integration important; in order to obtain a finite result for the entanglement entropy, we must perform the integral over \(\t \in [0,2\pi]\) to eliminate this singular term before performing the integral over \(\z\). This singular term also means that we must be extremely cautious when performing a change of variables. Integrands containing terms proportional to \(\cos(2\t)/(\z^2-1)\) occur generically in entanglement entropy calculations for probe branes with non-\ads\ worldvolumes~\cite{Karch:2014ufa,Kumar:2017vjv,Chalabi:2020tlw,Jokela:2024cxb}, so the comments we make here have general implications for probe brane entanglement entropy calculations.

\begin{figure}
    \begin{center}
    \begin{subfigure}{0.5\textwidth}
        \centering
        \hspace{0.45cm}
        \includegraphics{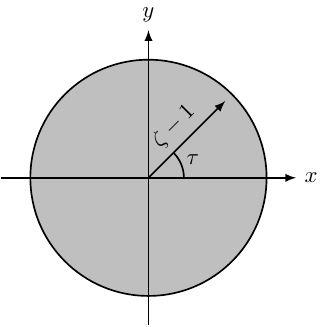}
        \caption{}
        \label{fig:disk_shaded}
    \end{subfigure}\begin{subfigure}{0.5\textwidth}
        \centering
        \hspace{0.45cm}
        \includegraphics{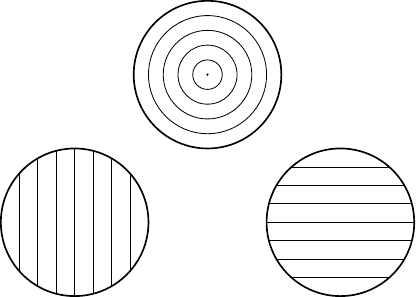}
        \caption{}
        \label{fig:disk_slicings}
    \end{subfigure}
    \caption{\textbf{(a):} The gray disk is the integration region for the integral in equation~\eqref{eq:model_integral}. \textbf{(b):} The integral in equation~\eqref{eq:model_integral} corresponds to slicing the disk up into circles, as depicted at the top, and integrating the contribution from each circle. The integral in Cartesian coordinates \((x,y)\) in equation~\eqref{eq:model_integral_cartesian} corresponds instead to integrating the contributions of vertical lines, as depicted at the bottom left. If we swap the order of the \(x\) and \(y\) integrals then we instead integrate the contributions of horizontal lines, as depicted at the bottom right. Because the integral is not absolutely convergent Fubini's theorem does not apply, so each of these options defines a different integral, evaluating to a different result.}
    \label{fig:disk}
    \end{center}
\end{figure}

To see that the singular term in equation~\eqref{eq:Sbrane_singular} can cause issues when changing variables, it may be useful to consider a simpler example, the integral
\begin{equation} \label{eq:model_integral}
    A = \int_1^2 \diff \z \int_0^{2\pi} \diff \t \frac{\cos(2\t)}{\z-1} = 0\, ,
\end{equation}
which we might think of as an integral over the unit disk \(|\z-1| \leq 1\) as depicted in figure~\ref{fig:disk_shaded}. If we swap the order of the integrals then we do not obtain \(A=0\) since the integral over \(\z\) diverges logarithmically near \(\z=1\). This happens because the integral is not absolutely convergent, so Fubini's theorem does not apply; we must think of the integral in equation~\eqref{eq:model_integral} as an iterated integral first over \(\t\), then over \(\z\), not as a double integral over the unit disk.

Suppose we ignore this fact, and try to transform the integral to Cartesian coordinates,
\begin{equation}
    x = (\z-1) \cos \t \, , \qquad y = (\z-1) \sin\t \, .
\end{equation}
Then if we integrate over \(y\) first we find
\begin{equation} \label{eq:model_integral_cartesian}
    A^{(\mathrm{Cart})} = \int_{-1}^1 \diff x \int_{-\sqrt{1-x^2}}^{\sqrt{1-x^2}} \diff y \frac{x^2 - y^2}{(x^2 + y^2)^2} = \pi \, ,
\end{equation}
where we have added the superscript (Cart) to indicate that the operations we have just performed define a different integral. Indeed, comparing equations~\eqref{eq:model_integral} and~\eqref{eq:model_integral_cartesian} we have that \(A \neq A^\mathrm{(Cart)}\). Since the integrand in equation~\eqref{eq:model_integral_cartesian} is odd under the interchange \(x \leftrightarrow y\), if we had decided to integrate over \(x\) first then we would have found \(A^\mathrm{(Cart)} = -\pi\) instead. Each of these integrals corresponds to slicing up the unit disk in a different way, as depicted in figure~\ref{fig:disk_slicings}.

Returning now to the evaluation of \(\Sbrane\), the above discussion makes clear that if we want to safely transform to Poincar\'e coordinates then we should first subtract the singular term in equation~\eqref{eq:Sbrane_singular} from the integrand before performing the change of variables. We could then transform the subtracted integral to Poincar\'e coordinates. However, to complete the evaluation of \(\Sbrane\) we would need to evaluate the integral of the piece we subtract,
\begin{equation} \label{eq:singular_integral}
    \frac{\sqrt{\la} \, N_3 N_5}{3\pi^2 a}\int_{\cR} \diff s \diff \z \diff \t \,
        \frac{s^2}{(1-s^2)^{3/2}} \le(q^2 + (1-s^2)^2 \ri) \frac{\cos(2\t)}{\z^2 - 1} \,  .
\end{equation}
If we perform the integral over \(\t\) before the integral over \(\z\) then this integral is finite but not zero; as can be seen in figure~\ref{fig:integration_region}, \(\t\) ranges over \([0,2\pi]\) at \(\z=1\) for all \(s\), but there are values of \(\z\) and \(s\) for which the range of \(\t\) is more limited, yielding a non-zero integral of \(\cos(2\t)\). In fact, the complicated form of the integration region means that we have been unable to directly evaluate the integral in equation~\eqref{eq:singular_integral}.

Instead, in appendix~\ref{app:integrals} we take a more naive approach, closing our eyes to the fact that the integrand is singular and directly converting integral in equation~\eqref{eq:Sbrane_integral} to Poincar\'e coordinates, \((s,\z,\t) \to (s,t,r)\). In fact, the singular integrand means that the value of the integral \emph{does} change when we perform this change of variables. Let us denote the result of the integral in Poincar\'e coordinates, integrating first over \(r\), then \(t\), then \(s\), by \(\SbranePoinc\). We have strong reason to believe that the relation 
\begin{equation} \label{eq:S_brane_poincare_difference}
    \Sbrane = \SbranePoinc - \Svar\,,
\end{equation}
holds in general for probe brane entanglement entropy calculations, where $\Svar$ is defined in equation~\eqref{eq:Svar_def}. In appendix~\ref{app:integrals} we compute \(\SbranePoinc\) analytically. Combined with the value of \(\Svar\) given in equation~\eqref{eq:Svar_result}, we use equation~\eqref{eq:S_brane_poincare_difference} to obtain the result for \(\Sbrane\) quoted in equation~\eqref{eq:Sbrane_result}.

Why do we believe that equation~\eqref{eq:S_brane_poincare_difference} is true? The first piece of evidence comes from an earlier probe brane calculation. For \(n_3\) probe D3-branes spanning the \((t,x,r,\f)\) directions, holographically dual to \(\cN=4\) SYM theory on the Coulomb branch, it is possible to compute \(\Sbrane\) analytically in the CHM coordinates used in equation~\eqref{eq:chm_metric}~\cite{Chalabi:2020tlw}, and in Poincar\'e coordinates, with the results differing by \(\Svar\) for the D3-branes as shown in appendix~\ref{sec:coulomb}.

Further evidence for equation~\eqref{eq:S_brane_poincare_difference} comes from probe D7-branes in \(\ads[5] \times \sph[5]\) and probe D6-branes on \(\ads[4] \times \mathbb{C}\mathrm{P}^3\). In ref.~\cite{Jokela:2024cxb} the contribution of these branes to entanglement entropy in the dual QFT were computed using the probe methods described above, and compared to the appropriate limit of the entanglement entropy computed using the RT prescription in the backreacted geometry sourced by smeared flavor branes. The two results were found to agree provided the contribution of \(\Svar\) was dropped from the probe brane entanglement entropy. If equation~\eqref{eq:S_brane_poincare_difference} holds then this explains why, since \(\Sbrane\) was computed in Poincar\'e coordinates in ref.~\cite{Jokela:2024cxb}\footnote{Ref.~\cite{Jokela:2024cxb} also performed the integral over \(r\) before \(t\).} and equation~\eqref{eq:S_brane_poincare_difference} implies that the probe contribution to entanglement entropy is
\begin{equation}
    \Sbrane + \Shor + \Svar = \SbranePoinc + \Shor \, .
\end{equation}
Hence, computing only \(\SbranePoinc\) and dropping \(\Svar\) gives the same result for the entanglement entropy as including all three contributions to equation~\eqref{eq:S1_split}.

\begin{figure}
    \begin{center}
        \includegraphics{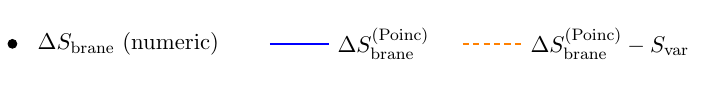}
        \includegraphics{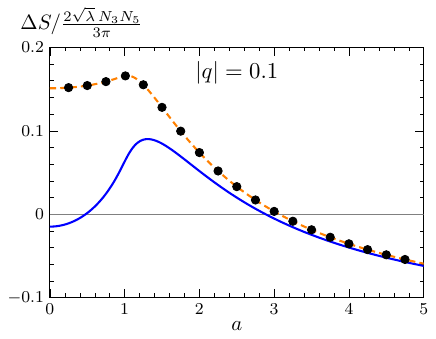}
        \hfill
        \includegraphics{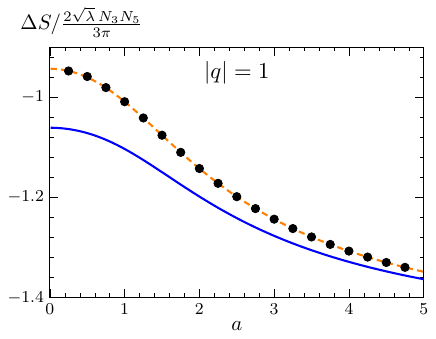}
        \caption{Comparison of results for \(\Delta \Sbrane\), defined in equation~\eqref{eq:Delta_Sbrane}, as functions of \(a\) for two different values of \(q\). The black dots show numerical results computed in the CHM coordinates \((\z,\t,s)\). The solid blue curves show the analytic result of the integral obtained by transforming to Poincar\'e coordinates \((t,r,s)\). These two results disagree because of the singular integrand. The difference between them is equal to \(\Svar\), as shown by the dashed orange curves.}
        \label{fig:Sbrane_comparison}
    \end{center}
\end{figure}

Finally, in figure~\ref{fig:Sbrane_comparison} we show numerical evidence that equation~\eqref{eq:S_brane_poincare_difference} holds for the D3/probe D5 system that we consider in this work. The black points in the figure show results for 
\begin{equation} \label{eq:Delta_Sbrane}
    \Delta \Sbrane \equiv \Sbrane - \le.\Sbrane\ri|_{m=0} \, ,
\end{equation}
computed by numerically evaluating the integral in equation~\eqref{eq:Sbrane_integral} in the CHM coordinates \((\t,\z,s)\). The subtraction of the \(m=0\) result is performed to eliminate the UV-divergent term in \(\Sbrane\). The solid blue curves in the figure show \(\Delta \SbranePoinc\), defined in the same way as \(\Delta \Sbrane\) but using our analytic result for \(\SbranePoinc\) computed in appendix~\ref{app:integrals}. Clearly \(\Delta \Sbrane\) and \(\Delta \SbranePoinc\) are different, showing that the results of the integrals in the two coordinate systems are different due to the singular term in the integrand. The dashed orange curves show \(\Delta \SbranePoinc - \Svar\), which agrees well with the numerical results for \(\Delta \Sbrane\), as expected from equation~\eqref{eq:S_brane_poincare_difference}.\footnote{Recall that \(\Svar|_{m=0} = 0\), so that \(\Delta \Svar \equiv \Svar - \le.\Svar\ri|_{m=0} = \Svar\).}

Based on this evidence, we assume equation~\eqref{eq:S_brane_poincare_difference} is true. This allows us to obtain the analytic result for \(\Sbrane\) in equation~\eqref{eq:Sbrane_result}. However, it is not obvious to us how to derive the former equation. A proof of equation~\eqref{eq:S_brane_poincare_difference} would be highly desirable in order to fully complete our probe calculation. Towards this, we note that it is presumably not a coincidence that if we define the functional

\begin{equation}
    \cI[h(\z,\t)] =
    \frac{\sqrt{\la} \, N_3 N_5}{3\pi^2 a} \int_{\cR} \diff s \diff \z \diff \t \, \frac{s^2 \le(q^2 + (1-s^2)^2 \ri)}{(1-s^2)^{3/2}} h(\z,\t) \, ,
\end{equation}
then the singular term in equation~\eqref{eq:Sbrane_singular} is \(\cI\le[\cos(2\t)/(\z^2 - 1)\ri]\), while comparing to equation~\eqref{eq:Svar_integral_s} we have that \(\Svar = - \cI\le[\d(\z-1) \ri]\), where \(\d\) is the Dirac delta function.

\section{Computation of holographic \texorpdfstring{\(C\)}{C}-functions}
\label{sec:C_functions}

In this section, we review $C$-functions that can be used to characterize the RG flow triggered by the presence of a non-conformal defect or an interface. We compute them in the holographic example of the previous section and study their monotonicity. For a defect RG flow, a monotonic $C$-function interpolating between constant values in the UV and the IR can be defined as in ref.~\cite{Casini:2023kyj}. In the interface case, \(n_3 > 0\), we find that this \(C\)-function is monotonic, even though it does not have to be, but it diverges to $-\infty$ in the IR. Bad IR behavior of this \(C\)-function is unsurprising, because the RG flow in the case of the interface is actually four-dimensional. We show that finite IR values can be obtained by considering \(C\)-functions adapted to four-dimensions, which we refer to as ``$A$-functions''. We will define and analyze several such \(A\)-functions inspired by the work of refs.~\cite{Liu:2012eea,Liu:2013una,Casini:2017vbe}.

\subsection{Definitions of defect and interface \texorpdfstring{\(C\)}{C}-functions}\label{subsec:C_A_function_definitions}

\paragraph{Defect \texorpdfstring{\(C\)}{C}-functions.} A $C$-function characterizing RG flows localized to defects was proposed in ref.~\cite{Casini:2023kyj}. In this setup, a $d$-dimensional CFT on Minkowski space \(M_d\) with action $I_{\text{CFT}}$ is supplemented with a $p$-dimensional planar defect spanning \(M_p\), with action
\begin{equation}
    I_{\text{def}} = I_{\text{def,UV}} + m\int_{M_p} \mathrm{d}^p x\,\mathcal{O}(x)\,,
\end{equation}
where the operator $\mathcal{O}$ is relevant from the defect point of view, with scaling dimension $\Delta < p$ under the defect conformal group. We assume that the action $I_{\text{def,UV}}$ is localized on $M_p$ and preserves defect conformal symmetry, which is then broken by the relevant deformation \(m \cO\). This induces a defect RG flow that can be characterized by the CST $C$-function~\cite{Casini:2023kyj}
\begin{equation} \label{eq:relative_entropy_C_function}
    C_\mathrm{CST}(\ell) \equiv (p-2)\, S_\mathrm{rel}(\ell) - \ell\, S_\mathrm{rel}'(\ell)\,,
\end{equation}
where $S_\mathrm{rel}(\ell)$ is the relative entropy of reduced density matrices $\rho$ and $\rho_0 = \rho\vert_{m = 0}$ of a ball-shaped entangling region of radius $\ell$, centered at the defect in the ground states of the CFT with a conformal defect $I_{\text{CFT}}+I_{\text{def,UV}}$ and the deformed theory $I_{\text{QFT}} = I_{\text{CFT}}+I_{\text{def}}$ respectively. Explicitly
\begin{equation}
    S_\mathrm{rel}(\ell) \equiv S(\rho\Vert\rho_0) = \tr(\r\, K_0) - \tr(\r_0 K_0) - (S(\ell)-S(\ell)\vert_{m = 0})\,,
\end{equation}
where $K_0 = -\log{\rho_0}$ is the modular Hamiltonian of \(\r_0\), $S(\ell)$ is the von Neumann entropy \eqref{eq:von_Neumann_rho} of $\rho$ and the traces are assumed to be computed on the null Cauchy surface of the domain of dependence of the ball \cite{Casini:2023kyj}.

Using positivity of the relative entropy and the quantum null energy condition, it is proven in ref.~\cite{Casini:2023kyj} that $C(\ell)$ is a monotonically decreasing function of the radius $\ell$, in other words, it satisfies the inequality
\begin{equation}
    C'_\mathrm{CST}(\ell) = (p-3)\, S_\mathrm{rel}'(\ell)-\ell \, S_\mathrm{rel}''(\ell)  \leq 0\,.
    \label{eq:relative_entropy_inequality}
\end{equation}
The inequality \eqref{eq:relative_entropy_inequality} generalizes earlier results for \(p=1\)~\cite{Casini:2016fgb,Casini:2022bsu}, and for \(p=2\) with codimension \(d-p < 2\)~\cite{Casini:2016udt,Casini:2018nym}.

Unfortunately, the relative entropy is a difficult quantity to compute, so explicit examples of this \(C\)-function are hard to find. The situation simplifies somewhat for codimension-one defects, for which the modular Hamiltonian contribution cancels from equation~\eqref{eq:relative_entropy_C_function}, so that one may obtain the \(C\)-function from entanglement entropy alone. Concretely, for \(p = d-1\) equation~\eqref{eq:relative_entropy_C_function} becomes~\cite{Casini:2023kyj}
\begin{equation} \label{eq:ee_C_function}
    C_\mathrm{CST}(\ell) = \ell \, \Delta S'(\ell) - (d-3)\, \Delta S(\ell)\ ,
\end{equation}
where $\Delta S(\ell) \equiv S(\ell)-S_{0}(\ell) $ is the UV subtracted entanglement entropy of the deformed theory.

The case of a planar codimension-one defect in $\mathcal{N} = 4$ SYM theory defined above corresponds to $d = 4$ and $p = 3$ in which case we obtain
\begin{equation}
    C_\mathrm{CST}(\ell) = (\ell\,\partial_{\ell} -1)\,\Delta S(\ell) \,.
    \label{eq:c_function_Delta_S}
\end{equation}
As explained in section \ref{subsec:EE_definitions}, the entropy in the vacuum of the deformed theory has the divergence structure \cite{Casini:2023kyj}
\begin{equation}
    S(\ell) = p_2\,\frac{\ell^2}{\epsilon^2}+ p_1\,\frac{\ell}{\epsilon} - A_{\text{UV}}^{(N)}\log{\frac{\ell}{\epsilon}}  - P_{\text{def}}(a)+ \ldots\,,\quad \epsilon\rightarrow 0\,,
\end{equation}
where $a \propto m\ell$ is the dimensionless radius of the entangling region.\footnote{Recall that for our D3/probe D5 system, we define \(a = 2\pi m \ell/\sqrt{\la}\) in equation~\eqref{eq:a_sigma_def}.} The divergence structure of the entropy $S_{0}(\ell)$ in the UV fixed-point theory is the same except that the finite term is equal to an $\ell$-independent constant $P_{\text{def}}(0) = F_{\text{def,UV}}$. Therefore the UV subtracted entropy does not contain any divergences and is given by
\begin{equation}
    \Delta S(\ell)  = F_{\text{def,UV}}-P_{\text{def}}(a)  + \ldots\,,\quad \epsilon\rightarrow 0\,.
\end{equation}
We see that the $C$-function \eqref{eq:c_function_Delta_S} is UV finite in the $\epsilon\rightarrow 0$ limit and given by
\begin{equation}\label{eq:CST_simplified}
    C_\mathrm{CST}(a) = -(a\,\partial_{a} -1)\,P_{\text{def}}(a) - F_{\text{def,UV}} \,.
\end{equation}
At the UV fixed point $a = 0$ this vanishes, $C_\mathrm{CST}(0) = 0$, and the theorem $C'_\mathrm{CST}(a)\leq 0$ of ref.~\cite{Casini:2023kyj} then implies that $C_\mathrm{CST}(a)$ decreases monotonically as a function of $a$ to negative values.
Note, however, that measures of defect degrees of freedom can be negative, since they count degrees of freedom relative to the bulk theory. For example, this is shown for two-dimensional defects in~\cite{Nozaki:2012qd,Jensen:2015swa}. In our context, the physical significance lies in the monotonic decrease, which tracks the decoupling of the interface degrees of freedom along the RG flow.

In this case, though, there is a simple way to modify the \(C\)-function in equation~\eqref{eq:CST_simplified} to obtain something positive in the UV, and therefore potentially non-negative in the IR. To do so, we define
\begin{equation}
     C(a) \equiv (\ell\,\partial_{\ell} -1)\,S_{\text{def}}(\ell)\,,
     \label{eq:CST_modified_1}
\end{equation}
where $S_{\text{def}}(\ell)$ is the contribution of the defect to the entropy as defined in \eqref{eq:ball_entropy}. Note that $S_{0}(\ell)$ still contains contributions from the conformal defect so that $\Delta S \neq S_{\text{def}}(\ell)$ is not equal to the defect contribution. Substituting the UV expansion \eqref{eq:Sdef_expansion}, it is simply
\begin{equation}
    C(a) = -(a\,\partial_{a} -1)\,P_{\text{def}}(a)\,,
    \label{eq:CST_modified}
\end{equation}
which differs from $C_\mathrm{CST}(a)$ defined in equation~\eqref{eq:CST_simplified} only by a constant shift and thus also decreases monotonically, $C'(a)\leq 0$ by the theorem of ref.~\cite{Casini:2023kyj}. This \(C\)-function matches with the defect free energy at the UV fixed-point $C(0) = F_{\text{def,UV}}$.

\paragraph{Interface \texorpdfstring{\(C\)}{C}-functions.} $C$-functions characterizing RG flows in the presence of interfaces are less studied. We will still attempt a similar approach as for defect RG flows and construct $C$-functions from the entanglement entropy of a ball-shaped subregion centered at the interface. The interface we consider is presented in section~\ref{sec:d3_d5}: $\mathcal{N} = 4$ SYM theory with gauge groups $\SU(N_3)$ and $\SU(N_3+n_3)$ on $x<0$ and $x>0$ respectively. The RG flow is induced by a non-trivial boundary condition at spatial infinity $x\rightarrow \infty$ which gives a vacuum expectation value for the adjoint scalars $\Phi_{4,5,6}$ inducing a mass on the four-dimensional gauge fields on the right side of the interface. Unlike for defects, the RG flow is four-dimensional and there are no degrees of freedom localized at the interface that can undergo an RG flow. There are no quantities that are proven to be monotonic for such RG flows, so we will consider several candidate \(C\)-functions.

The entanglement entropy $S(\ell)$ of the ball in this state is defined in section~\ref{subsec:EE_definitions}. It decomposes into fixed-point and interface contributions as in equation~\eqref{eq:interface_entropy}. There are two avenues to defining candidate $C$-functions: defining intrinsically three-dimensional $C$-functions that match with $F_{\text{def,UV}}$ in the UV, and four-dimensional functions that capture the type A anomaly coefficient $A_{\text{UV}}$ in the UV. We refer to the latter as ``$A$-functions'', in order to draw a distinction. We will begin our discussion with the former --- three-dimensional candidates defined using the interface contribution to the entropy.

We define an interface $C$-function $C(a)$ using the same expression \eqref{eq:CST_modified_1} as in the defect case, but where $S_{\text{def}}(\ell)$ stands for the interface contribution to the entanglement entropy as defined in \eqref{eq:interface_entropy}. For defects, $C(a)$ is a monotonically decreasing function along defect RG flows \cite{Casini:2023kyj}. When applied to the interface here, the parameter $a\propto m\ell$ is the ball radius in the units of the four-dimensional mass, or equivalently, in the units of the asymptotic value of the adjoint scalar one-point functions. Thus these $C$-functions are quantifying a four-dimensional RG flow on a half-space, so the theorem implying monotonicity of $C(a)$ does not apply~\cite{Casini:2023kyj}. Regardless, we will discover below that it is monotonically decreasing in our holographic example.

Since $S_{\text{def}}(a)$ also contains four-dimensional contributions as discussed in section~\ref{subsec:EE_definitions}, it may be natural to consider a slight modification $\widetilde{C}(a)$ of $C(a)$ where in place of \(S_\mathrm{def}(a)\) we use $\widetilde{S}_{\text{def}}(a)$ defined in equation~\eqref{eq:interface_entropy_2}. The modified $C$-function is defined as
\begin{equation}
    \widetilde{C}(a) \equiv  (\ell\,\partial_{\ell} -1)\,\widetilde{S}_{\text{def}}(\ell) = -(a\,\partial_{a} -1)\,\widetilde{P}_{\text{def}}(a) \, ,
    \label{eq:Ctilde_interface}
\end{equation}
where the second equality  follows from equation~\eqref{eq:Ftilde_exp}. Recall that $\widetilde{S}_{\text{def}}(a)$ is equal to \(S_\mathrm{def}(a)\) minus one half of the Coulomb branch entanglement entropy, and correspondingly $\widetilde{P}_{\text{def}}(a) = P_{\text{def}}(a) -\frac{1}{2}\,P_{\text{Coul}}^{(N,n)}(a)$ where $P_{\text{Coul}}^{(N,n)}(a)$ is the finite term in the entanglement entropy~\eqref{eq:Coulomb_branch_entropy_expansion} of the Coulomb branch vacuum on all of Minkowski space. At the UV fixed point, both $C$-functions give $\widetilde{C}(0) = C(0) = F_{\text{def,UV}}$ since $P_{\text{Coul}}^{(N,n)}(0) = 0$.

The two $C$-functions $C(a)$ and $\widetilde{C}(a)$ are motivated by three-dimensional defects, but because the flow on the right-side of the interface is four-dimensional, it is natural to also consider quantities that are adapted to four dimensions. Indeed, it is known that defect \(C\)-functions can increase under bulk RG flows~\cite{Green:2007wr,Sato:2020upl,Shachar:2024ubf}. As mentioned above, in order to draw a distinction between \(C\)-functions adapted to three and four dimensions, we will refer to the latter as ``\(A\)-functions''. We will consider a several possible such \(A\)-functions.

First, we will consider the Casini--Test\'e--Torroba (CTT) $A$-function defined as \cite{Casini:2017vbe}
\begin{equation}
    A_{\text{CTT}}(a) = (\ell\,\partial_{\ell} - 2)\,\Delta S(\ell) = (\ell\,\partial_{\ell} - 2)\,\Delta S_{\text{def}}(\ell)\,,
    \label{eq:A_CTT}
\end{equation}
where $\Delta S(\ell) = S(\ell) - S(\ell)\vert_{m = 0}  $ is the UV subtracted entropy in the presence of the interface, $ \Delta S(\ell) =\Delta S_{\text{def}}(\ell) \equiv S_{\text{def}}(\ell) - S_{\text{def}}(\ell)\vert_{m = 0}$ by equation~\eqref{eq:interface_entropy}. The UV subtraction in equation~\eqref{eq:A_CTT} is needed to remove the logarithmic and linear divergences from the entanglement entropy (if we did not perform the subtraction, the differential operator would cancel the $\ell^2\slash \epsilon^2$ divergence, but not the others). When there is no interface present and $S(\ell)$ is the entropy in the ground state of a Poincar\'e-invariant theory, \(A_\mathrm{CTT}\) is monotonically decreasing~\cite{Casini:2017vbe}. However, monotonicity is not guaranteed in the presence of an interface.

The \(A\)-function defined in equation~\eqref{eq:A_CTT} vanishes at the UV fixed point, $A_{\text{CTT}}(0) = 0 $, where it does not provide a viable measure of the amount of degrees of freedom. To fix this issue, motivated by the work of Liu and Mezei~\cite{Liu:2012eea,Liu:2013una}, we define another candidate $A$-function as
\begin{equation}
    A_{\text{LM}}(a) = \frac{1}{2}\,\ell\,\partial_\ell\,(\ell\,\partial_{\ell} - 2)(S(\ell)-S_{\text{def}}(\ell)\vert_{m = 0})\,,
    \label{eq:A_LM}
\end{equation}
where the differential operator removes the four-dimensional quadratic and logarithmic UV divergences, while the subtraction of the UV interface contribution is needed to remove the three-dimensional linear divergence arising from the interface. Without the UV subtraction, \(A_\mathrm{LM}\) matches the $A$-function of refs.~\cite{Liu:2012eea,Liu:2013una} which was proposed for four-dimensional RG flows in Poincar\'e-invariant theories.

Using the decomposition of the entanglement entropy in equation~\eqref{eq:interface_entropy} and the entanglement entropy of \(\cN=4\) SYM in equation~\eqref{eq:UV_fixed_point_entropy}, equation~\eqref{eq:A_LM} becomes
\begin{equation}    \label{eq:A_LM2}
    A_{\text{LM}}(a) = \frac{A_{\text{UV}}^{(N_3)}+A_{\text{UV}}^{(N_3+n_3)}}{2}+\frac{1}{2}\,a\,\partial_a\,(a\,\partial_{a} - 2)\,\Delta S_{\text{def}}(a)\,,
\end{equation}
where $A_{\text{UV}}^{(N)} = N^2$ is the type A Weyl anomaly coefficient of \(\SU(N)\) \(\cN=4\) SYM theory at large \(N\)~\cite{Henningson:1998gx}. We will see that the contribution of \(\Delta S_\mathrm{def}\) to equation~\eqref{eq:A_LM2} vanishes at \(a=0\), so that at the UV fixed point \(A_\mathrm{LM}(a)\) evaluates to the the average of the four-dimensional type A Weyl anomaly coefficients of the theories on either side of the interface, 
\begin{equation}
    A_{\text{LM}}(0) = \frac{A_{\text{UV}}^{(N_3)}+A_{\text{UV}}^{(N_3+n_3)}}{2}\,.
    \label{eq:A_LM_0}
\end{equation}
This value is positive and is known to be a good measure of degrees of freedom at fixed points. The requirement that $A_{\text{LM}}(0)\vert_{n_3 = 0} = A_{\text{UV}}^{(N_3)}$ fixes the sign and the factor of one half in the definition~\eqref{eq:A_LM}. 

Instead of subtracting the $\ell\slash \epsilon$ divergence of $S(\ell)$ with the fixed point value as in equation~\eqref{eq:A_LM}, we can remove it by acting on the entanglement entropy with an additional differential operator. This is achieved in the $A$-function
\begin{equation}
    \AtLM(a) \equiv  -\frac{1}{2}\,\ell\,\partial_\ell\,(\ell\,\partial_{\ell} - 2)(\ell \, \partial_\ell - 1)\,S(\ell) \,.
    \label{eq:ALM_tilde}
\end{equation}
From equations~\eqref{eq:UV_fixed_point_entropy} and~\eqref{eq:interface_entropy}, this gives for our system
\begin{equation}
    \AtLM(a) = \frac{A_{\text{UV}}^{(N_3)}+A_{\text{UV}}^{(N_3+n_3)}}{2}-\frac{1}{2}\,\ell\,\partial_\ell\,(\ell\,\partial_{\ell} - 2)(\ell \, \partial_\ell - 1)\,S_{\text{def}}(\ell) \,,
    \label{eq:ALM_tilde_alt}
\end{equation}
which at the UV fixed-point matches with the average \eqref{eq:A_LM_0} and satisfies $\AtLM(0)\vert_{n_3 = 0} = A_{\text{UV}}^{(N_3)}$. This fixes the factor of $-1\slash 2$ in the definition \eqref{eq:ALM_tilde}.

\subsection{Holographic defect \texorpdfstring{\(C\)}{C}-functions}
\label{sec:C_function}

We will now analyze $C$-functions adapted to three-dimensional RG flows defined in the previous subsection by computing them using the holographic entanglement entropy found in section \ref{sec:flow_ee}. We will first consider the defect $C$-function \eqref{eq:CST_modified_1} which at leading order in the probe limit takes the form
\begin{equation}\label{eq:C_function_definition_probe}
    C(a) = (a\,\partial_a-1)\,S_1\,,
\end{equation}
where $S_1$ is given by \eqref{eq:ee_full}.

\begin{figure}[t]
    \begin{center}
        \includegraphics{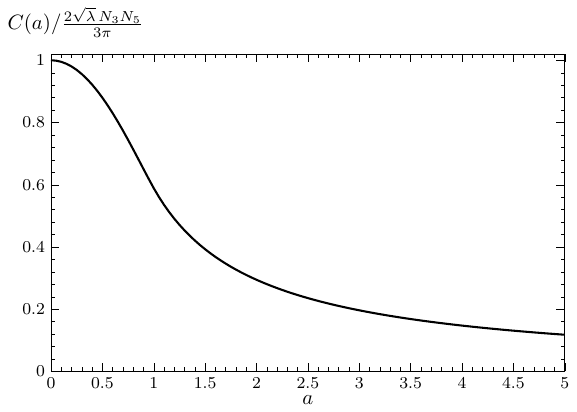}
        \caption{The entropic \(C\)-function~\eqref{eq:q0_C} for \(q=0\), corresponding to equal rank gauge groups on either side of the defect, plotted as a function of the dimensionless radius \(a\) of the entangling region defined in equation~\eqref{eq:a_sigma_def}. In the UV, \(a \to 0\), the \(C\)-function approaches \(F_\mathrm{def,UV} = 2 \sqrt{\la}\, N_3 N_5/(3\pi)\), contribution of the conformal defect to the free energy on a sphere. The \(C\)-function vanishes in the IR, \(a \to \infty\).}
        \label{fig:q0_c}
    \end{center}
\end{figure}

For \(q=0\) this \(C\)-function evaluates to
\begin{equation} \label{eq:q0_C}
    \le.C(a)\ri|_{q=0} = \frac{\sqrt{\la} \, N_3 N_5}{4\pi}  \begin{cases}
        \dfrac{\sin^{-1} a}{a} + \dfrac{5-2a^2}{3} \sqrt{1-a^2}
        \,, \quad & a \leq 1 \, ,
        \\[0.5em]
        \dfrac{\pi}{2a} \, , \quad& a >1  \, .
    \end{cases}
\end{equation}
This is plotted in figure~\ref{fig:q0_c}. We see that \(C(a)\) is a monotonically decreasing function of \(a\), consistent with the theorem of ref.~\cite{Casini:2023kyj}. In the UV limit \(a \to 0\), the \(C\)-function is
\begin{equation}\label{eq:defect_C_UV}
    \le.C(0) \ri|_{q=0} = \frac{2 \sqrt{\la} \, N_3 N_5}{3\pi} \, ,
\end{equation}
which is equal to \(F_\mathrm{def}\) \eqref{eq:free-energy} for the \(q=0\) conformal defect. The \(C\)-function vanishes in the IR limit \(a \to \infty\), which makes sense since the hypermultiplet mass gaps out all of the defect degrees of freedom. Clearly, $C(a)\vert_{q = 0}$ is a positive monotonically decreasing function, and therefore provides a viable measure of the number of defect degrees of freedom along the whole RG flow.

\begin{figure}
    \begin{center}
    \includegraphics{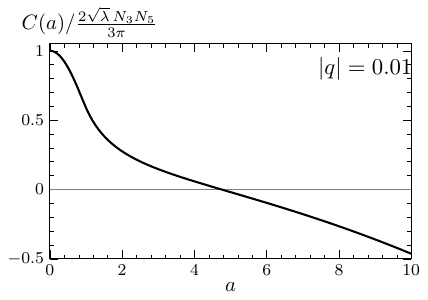}
    \includegraphics{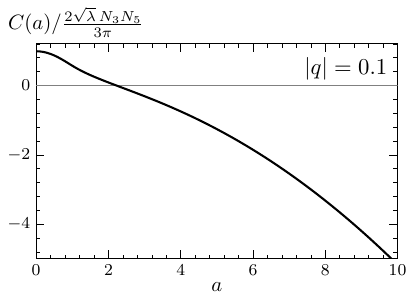}
    \includegraphics{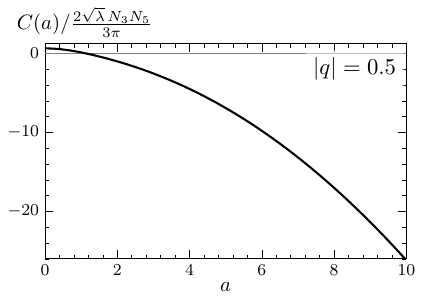}
    \includegraphics{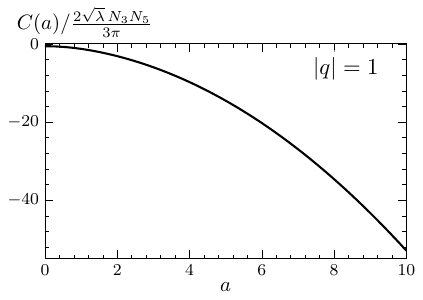}
    \caption{The \(C\)-function defined in equation~\eqref{eq:C_function_definition_probe} for sample non-zero values of \(q\). In each case, the \(C\)-function decreases monotonically with increasing \(a\). However, for \(q \neq 0\) the \(C\)-function does not approach a constant in the IR limit \(a\to\infty\), but rather diverges quadratically.}
    \label{fig:C_function_full}
    \end{center}
\end{figure}

If we compute the \(C\)-function defined in equation~\eqref{eq:C_function_definition_probe} for \(q \neq 0\) (an interface), we find
\begin{equation}\begin{aligned} \label{eq:C_function_full}
    C(a) = \frac{\sqrt{\la} \, N_3 N_5}{3\pi} \biggl[&
			3 \frac{(1 - 4q^2)}{4a} \sin^{-1} \s
			-3 q \sinh^{-1} \le(\frac{ q}{\sqrt{1-\s^2}}\ri)
			\\& \hspace{2cm}
			- \le( \frac{a^3}{\s^3} + (\s^2 - 6) \frac{a}{\s} + \frac{15-2\s^2}{4} \frac{\s}{a}\ri) \sqrt{1-\s^2}
		\biggr] \, .
\end{aligned}\end{equation}
We plot this \(C\)-function for sample values of \(q\) in figure~\ref{fig:C_function_full}. Note that \(C\) is an even function of \(q\), so it is sufficient to plot it only for positive \(q\). We find that \(C(a)\) decreases monotonically with increasing \(a\) for every value of \(q\) that we have checked. A priori there is no reason for \(C(a)\) to be monotonic for the interface $q\neq 0$, because, as already explained in section~\ref{subsec:C_A_function_definitions}, the proof of the monotonicity theorem in ref.~\cite{Casini:2023kyj} assumes a three-dimensional RG flow localized on the defect while here the RG flow is four-dimensional. However, it is interesting to note that \(C(a)\) is monotonic for our interface.

In the UV limit \(a \to 0\) we find
\begin{equation}
    C(0) = \frac{\sqrt{\la} \, N_3 N_5}{3\pi} \le[(2-q^2)\sqrt{1+q^2} - 3 q \sinh^{-1} q \ri] = F_\mathrm{def,UV} \, .
\end{equation}
We note that \(C(0)\) is negative for \(|q| \gtrsim 0.797\), suggesting that $F_\mathrm{def,UV}$does not count the UV degrees of freedom of the interface. In addition, for any non-zero \(q\) the \(C\)-function does not asymptote to a constant as \(a \to \infty\), but rather diverges due to the \(O(a^2)\) and \( O(\log a)\) terms in equation~\eqref{eq:Delta_S_large_a},
\begin{equation}
        C(a) = - \frac{n_3 N_3}{3} a^2 - n_3 N_3 \log a + O(1) \,,\quad a\rightarrow \infty\, .
        \label{eq:C_q_asymptotics}
\end{equation}
Therefore $C(a)$ also does not provide a viable measure of the number of degrees of freedom in the IR when $q\neq 0$. 

\begin{figure}
    \centering
    \includegraphics{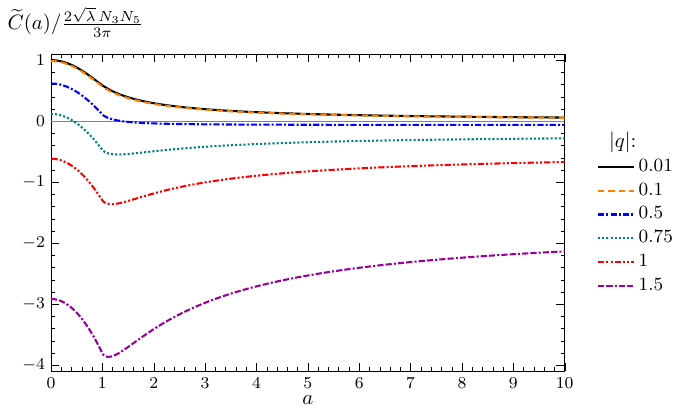}
    \caption{The putative \(C\)-function \(\widetilde{C}\) obtained if we subtract half of the Coulomb branch entanglement entropy from \(S_1\), as in equation~\eqref{eq:ctilde}. Unlike the CST \(C\)-function plotted in figure~\ref{fig:C_function_full}, \(\widetilde{C}\) interpolates between finite values in the UV \(a\to 0\) and IR \(a \to \infty\) limits. However, it is not a monotonic function of \(a\) for sufficiently large \(|q|\). In addition, for sufficiently large \(|q| \gtrsim 0.9397\), \(\widetilde{C}\) is larger in the IR than in the UV.}
    \label{fig:ctilde}
\end{figure}

As explained below equation~\eqref{eq:Delta_S_large_a}, the \(O(a^2)\) and \(O(\log a)\) terms in \(S_1\) that cause \(C(a)\) to diverge as $a\rightarrow \infty$ have a four-dimensional origin; they are one-half of the finite term $P_\mathrm{Coul}^{(N,n)}$ in the Coulomb branch entanglement entropy $S_{\hspace{1pt}\text{Coul}}^{(N,n)}(\ell)$ \eqref{eq:Coulomb_branch_entropy_expansion} of a ball-shaped region \cite{Chalabi:2020tlw}. Therefore a finite value in the IR can be obtained by utilizing the modified $C$-function $\widetilde{C}(a) $ defined in equation~\eqref{eq:Ctilde_interface}, where the Coulomb branch contribution to the entanglement entropy is subtracted. Explicitly,
\begin{equation}\label{eq:Ctilde_rewritten}
    \widetilde{C}(a) = (a\,\partial_a-1)\,\Bigl(S_1-\tfrac{1}{2}\,P_\mathrm{Coul}^{(N,n)}\Bigr) \ ,
\end{equation}
Using the result for $P_\mathrm{Coul}^{(N,n)}$ in the probe limit computed in ref.~\cite{Chalabi:2020tlw} (see appendix~\ref{sec:coulomb} for a translation of their results into our notation), we find that
\begin{equation}
    \widetilde{C}(a)  = C(a) + \frac{\sqrt{\la} \, N_3 N_5}{3\pi} \frac{q}{a} \le[3 a \cosh^{-1} a + (a^2 - 4)\sqrt{a^2 - 1} \ri] \Theta(a-1) \, ,
    \label{eq:ctilde}
\end{equation}
where \(\Q\) is the Heaviside step function. This is plotted for sample values of \(q\) in figure~\ref{fig:ctilde}. It interpolates (non-monotonically for sufficiently large \(q\)) between the UV and IR limits
\begin{equation}
    \widetilde{C}(a=0) = \frac{\sqrt{\la} \, N_3 N_5}{3\pi} \le(\frac{2 + q^2 -q^4}{\sqrt{1+q^2}} -3 q \sinh^{-1} q \ri)  ,
    \qquad
    \widetilde{C}(a \to \infty) = -\frac{\sqrt{\la} \, N_3 N_5}{3\pi} |q|^3 \, . 
\end{equation}
As expected, \(\widetilde{C}(a)\) takes a finite value in the IR, unlike $C(a)$. However, the value is always negative (for small \(q\), \(\widetilde{C}(a)\) becomes negative at very large \(a\), and so this is not always visible in figure~\ref{fig:ctilde}). For \(|q| \lesssim 0.9397\) we have that \(\widetilde{C}\) is larger in the UV than in the IR, \(\widetilde{C}(a=0) > \widetilde{C}(a \to \infty)\). This situation is reversed for \(|q| \gtrsim 0.9397\), where \(\widetilde{C}\) is larger in the IR than in the UV. Due to these issues, \(\widetilde{C}(a)\) does not provide a viable count of degrees of freedom along the interface RG flow.

\subsection{Holographic interface \texorpdfstring{\(A\)}{A}-functions}
\label{sec:liu_mezei}

\begin{figure}
    \begin{center}
    \includegraphics{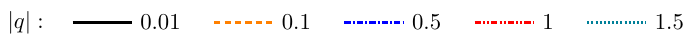}
    \begin{subfigure}{0.5\textwidth}
        \includegraphics{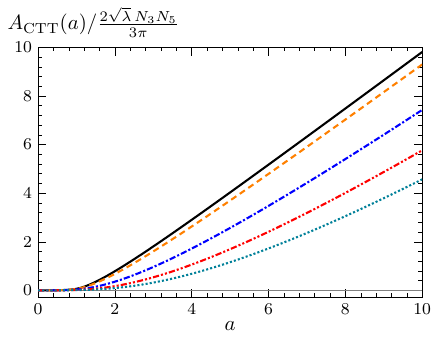}
        \caption{The CTT \(A\)-function of equation~\eqref{eq:holographic_ACTT}.}
        \label{fig:Actt}
    \end{subfigure}\begin{subfigure}{0.5\textwidth}
        \includegraphics{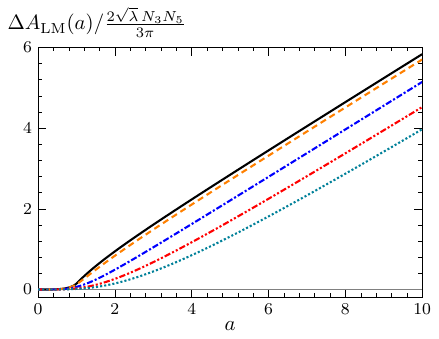}
        \caption{The LM \(A\)-function of equation~\eqref{eq:Delta_Alm_def}.}
        \label{fig:Alm}
    \end{subfigure}
    \end{center}
    \caption{Two putative \(A\)-functions, for sample values of \(q\). The left-hand panel shows \(A_\mathrm{CTT}\) defined in equation~\eqref{eq:holographic_ACTT}, while the right-hand panel shows \(\Delta A_\mathrm{LM}\) defined in equation~\eqref{eq:Delta_Alm_def} as one half of the logarithmic derivative of \(A_\mathrm{CTT}\) with respect to \(a\). Both of these candidate \(A\)-functions monotonically increase along the RG flows that we consider, so do not provide good measures of the change in the number of degrees of freedom.}
    \label{fig:Alm_C4d}
\end{figure}

We will now consider the $A$-functions defined in section \ref{subsec:C_A_function_definitions} that are adapted to four-dimensional RG flows and are therefore more appropriate in the interface case $q\neq 0$, $n_3>0$. We will first consider the $A$-function $A_{\text{CTT}}(a)$ defined in equation~\eqref{eq:A_CTT}, which in the probe limit becomes
\begin{equation} \label{eq:holographic_ACTT}
    A_{\text{CTT}}(a) = (a\,\partial_a -2)\,\Delta S_1 \, ,
\end{equation}
where $\Delta S_1 \equiv S_1 - S_1\vert_{m = 0}$. Using the expression for $S_1$ given in equation~\eqref{eq:ee_full}, we find
\begin{equation} \begin{aligned}
    A_{\text{CTT}}(a) \frac{\sqrt{\la} \, N_3 N_5}{\pi} \biggl[&
        \frac{3 + 4 a^2 - 12 q^2}{8a} \sin^{-1} \s + 2 q \sinh^{-1}q - 2 q \sinh^{-1} \le( \frac{q}{\sqrt{1-\s^2}} \ri)
        \\
    & + \frac{2 (q^2 - 2)}{3} \sqrt{1 + q^2}
    + \frac{23 - 16 q^4 - 25 \s^2 + 2 \s^4 + 52 q^2 + 8 \s^2 q^2}{24 \sqrt{1 + q^2 - \s^2}}
    \biggr] \, .
\end{aligned}\end{equation}
We plot \(A_{\text{CTT}}(a)\) for sample values of \(q\) in figure~\ref{fig:Actt}. The UV and IR limits  of this \(A\)-function are
\begin{equation}\begin{aligned}
    a \to 0: && A_{\text{CTT}}(a) &= \frac{\sqrt{\la } \, N_3 N_5}{30 \pi (1 +q^2)^{3/2}} a^4 + O(a^6) \,,
    \\
    a \to \infty: && A_{\text{CTT}}(a) &= \frac{\sqrt{\la} \, N_3 N_5}{4} a + O (\log a) \,.
\end{aligned}\end{equation}
For all values of \(q\), we find that \(A_\mathrm{CTT}(a)\) is a positive function that increases monotonically from zero in the IR to $+\infty$ in the IR, and is therefore not a good candidate to count degrees of freedom. For four-dimensional RG flows driven by a relevant deformation in the absence of an interface, $A_{\text{CTT}}(a) $ is proven to be monotonically decreasing function \cite{Casini:2017vbe}. The monotonic growth observed here is not in tension with this theorem, because the RG flow in our case occurs only on a half-space, breaking the assumed Lorentz invariance in the direction orthogonal to the interface \cite{Casini:2012ei,Casini:2017vbe}.

Next, we consider the function $A_{\text{LM}}(a)$ defined in equation~\eqref{eq:A_LM}. In order to isolate the probe contribution to this \(A\)-function we must first subtract off the constant UV piece, defining
\begin{equation}
    A_{\text{LM}}(a) = \frac{A_{\text{UV}}^{(N_3)}+A_{\text{UV}}^{(N_3+n_3)}}{2} + \D A_{\text{LM}}(a) \, ,
\end{equation}
so that, at leading order in the probe limit,
\begin{equation}\label{eq:Delta_Alm_def}
    \Delta A_{\text{LM}}(a) = \frac{1}{2}\,a\,\partial_a\,(a\,\partial_{a} - 2)\,\Delta S_1 \, .
\end{equation}
Substituting the expression for \(S_1\) in equation~\eqref{eq:ee_full}, we find
\begin{equation}\label{eq:A}
    \Delta A_{\text{LM}}(a) = \frac{\sqrt{\la} \, N_3 N_5}{16 \pi} \le[\frac{12 q^2 + 4 a^2 - 3}{a} \sin^{-1} \s + \frac{3 -12 q^2 - 5 \s^2 + 2 \s^4}{\sqrt{1 + q^2 - \s^2}} \ri] .
\end{equation}
In the UV and IR limits, \(\Delta A_\mathrm{LM}(a)\) behaves as
\begin{equation}\begin{aligned}
    a \to 0: && \Delta A_{\text{LM}}(a) &= \frac{\sqrt{\la } \, N_3 N_5}{15 \pi (1 +q^2)^{3/2}}\, a^4 + O(a^6) \,,
    \\
    a \to \infty: && \Delta A_{\text{LM}}(a) &= \frac{\sqrt{\la} \, N_3 N_5}{8}\, a - \frac{\sqrt{\la} \, N_3 N_5 q}{\pi} + O (a^{-1}) \,.
\end{aligned}\end{equation}
Therefore the full function $A_{\text{LM}}(a)$ diverges to $+\infty$ in the IR $a\rightarrow \infty$ and so also does not provide a viable measure of degrees of freedom along the flow We plot \(\Delta A_\mathrm{LM}(a)\) in figure~\ref{fig:Alm} for sample values of \(q\), where we see that in each case \(\Delta A_{\text{LM}}(a)\) is a monotonically increasing function of \(a\), similar to \(A_\mathrm{CTT}(a)\).

Finally, to overcome the issue of divergence in the IR, we consider the function \(\AtLM\) defined in equation~\eqref{eq:ALM_tilde_alt}. We will again separate the probe part, 
\begin{equation}
    \AtLM(a) = \frac{A_{\text{UV}}^{(N_3)}+A_{\text{UV}}^{(N_3+n_3)}}{2} + \D \AtLM(a) \, ,
\end{equation}
so that at leading order in the probe limit we have that
\begin{equation} \label{eq:clm_definition}
    \Delta \AtLM (a) = -\frac{1}{2}\,a\, \p_a (a \, \p_a - 1)(a \, \p_a - 2)\, S_1 \, .
\end{equation}
Now, substituting the expression for \(S_1\) in equation~\eqref{eq:ee_full}, we have
\begin{equation} \label{eq:clm}
    \D \AtLM(a) = - \frac{\sqrt{\la} N_3 N_5}{2\pi} \le[
        \frac{3(1-4q^2)}{4a} \sin^{-1}\s - \le((\s^2-3)\frac{a}{\s} + \frac{15 - 2\s^2}{4} \frac{\s}{a}\ri) \sqrt{1-\s^2}
    \ri] .
\end{equation}

By construction, for all values of \(q\) we have that \(\D\AtLM\) vanishes in the UV, i.e. \(\D\AtLM(a=0) = 0\), so that the UV value of $\widetilde{A}_{\text{LM}}(0)$ is the average of type-A Weyl anomaly coefficients $A_{\text{UV}}^{(N)} = N^2$ of the theories on the two sides of the interface,
\begin{equation}
    \widetilde{A}_{\text{LM}}(0) = \frac{N_3^2 + (N_3+n_3)^2}{2} = N_3^2\le[1 + \frac{n_3}{N_3} +O(n_3^2/N_3^2)\ri]\,,
    \label{eq:UV_value_average}
\end{equation}
where $n_3\slash N_3 \ll 1$ in our D3/probe D5 system. In the IR we find
\begin{equation} \label{eq:clm_ir}
    \lim_{a \to \infty} \D \AtLM(a) = - \frac{\sqrt{\la} \, N_3 N_5}{\pi} |q| = - n_3 N_3 \, .
\end{equation}
\begin{figure}
    \begin{center}
        \includegraphics{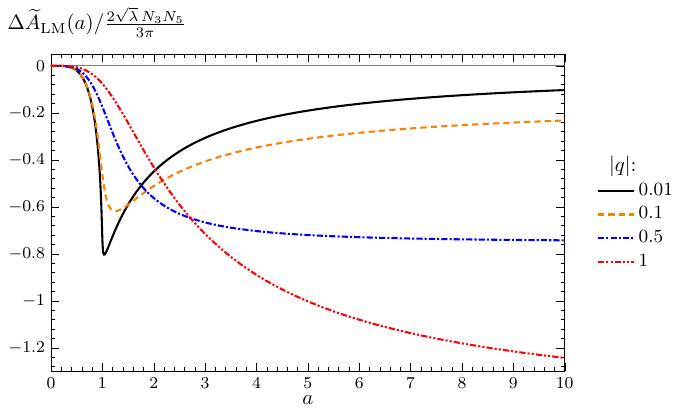}
    \caption{The putative \(C\)-function \(\AtLM\) inspired by refs.~\cite{Liu:2012eea,Liu:2013una}, defined in equation~\eqref{eq:A_LM}. We plot the difference \(\D\AtLM\) obtained after subtracting the UV contribution, as in equation~\eqref{eq:clm_definition}, for sample values of \(q\). For all \(q\) we find that \(\AtLM\) decreases from \(\D\AtLM(0)=0\) at small \(a\), while at large \(a\) we find that \(\D \AtLM\) saturates to a value less than zero, proportional to \(n_3\) given in equation~\eqref{eq:clm_ir}. Thus, \(\AtLM\) as we have defined it is finite in both the UV and IR limits, and smaller in the IR that in the UV. However, for \(q \leq 1/2\) \(\AtLM\) is not a monotonic function of \(a\).}
    \label{fig:clm}
    \end{center}
\end{figure}
Thus, up to leading order in the probe limit, the full \(\AtLM\) function behaves in the IR as
\begin{equation}\label{eq:correct_IR_value}
    \lim_{a \to \infty} \AtLM(a) = N_3^2 \le[1 + O(n_3^2/N_3^2) \ri],
\end{equation}
where the linear $n_3\slash N_3$ term in the average of the Weyl anomaly coefficients \eqref{eq:UV_value_average} has canceled against the contribution \eqref{eq:clm_ir} coming from the interface (the D5-branes). The result \(\widetilde{A}_\mathrm{LM}(a \to \infty) = N_3^2\) is the large \(N_3\) limit of the type A Weyl anomaly coefficient for \(\SU(N_3)\) \(\cN=4\) SYM theory without an interface, which is the expected result in the IR in the probe limit \(n_3 \ll N_3\). One would need to go beyond the probe limit by computing backreaction to see the contribution of the \(\mathrm{U}(n_3)\) factor on one side of the interface, which should contribute a term of order \(n_3^2\). See the paragraph below equation~\eqref{eq:Phi_plus_expansion} for a discussion of the IR of this RG flow.

The form of \(\D \AtLM(a)\) is plotted for sample values of \(q\) in figure~\ref{fig:clm}, which shows that although \(\AtLM(a)\) interpolates between sensible limits in the UV and IR, it is not a monotonic function \(a\), in general. Indeed, we find that our result for \(\D \AtLM(a)\) is a monotonically decreasing function of \(a\) for \(q \geq 1/2\). On the other hand, for \(q < 1/2\) our \(\D \AtLM(a)\) exhibits a sharp global minimum at an intermediate value of \(a\). This can be seen most explicitly in the limit of \(q=0\), where equation~\eqref{eq:clm} becomes
\begin{equation} \label{eq:q0_clm}
    \le. \D \AtLM(a)\ri|_{q=0} = -\frac{\sqrt{\la} \, N_3 N_5}{8\pi}\begin{cases}
      \dfrac{3\sin^{-1} a}{a} - \le(2a^2+3\ri) \sqrt{1-a^2} \, ,
      & a \leq 1 \, ,
      \\
      \dfrac{3\pi}{2a} \, , & a > 1\, ,
    \end{cases}
\end{equation}
which is a decreasing or increasing function of \(a\) for \(a<1\) or \(a>1\), respectively, with a cusp at \(a=1\). Plainly, for \(q=0\) \(\AtLM\) is not a good measure of the number of defect degrees of freedom --- we should instead use the CST \(C\)-function in equation~\eqref{eq:q0_C}.

\section{Conclusions and outlook}
\label{sec:discussion}

In this paper we computed the contribution of a codimension-one defect or interface to entanglement entropy in four-dimensional ${\mathcal{N}}=4$ SYM theory, realized holographically via probe D5-branes embedded in \(\ads[5] \times \sph[5]\). By introducing a mass deformation, we triggered an RG flow and evaluated the resulting change in entanglement entropy of a ball-shaped region centered on the defect. Our analysis was performed at leading order in the probe approximation, where the number of degrees of freedom dual to the D5-branes is parametrically smaller than those in the SYM theory, $\sqrt{\la} \,  N_5\ll N_3$, corresponding on the gravity side to neglecting the backreaction of the D5-branes. This approximation enabled us to obtain fully analytic results.

We derived an explicit formula \eqref{eq:ee_full} for the defect or interface contribution to the entanglement entropy as a function of $a$, a dimensionless combination of a mass parameter $m$ and the radius $\ell$ of the entangling region, $a=\frac{2\pi}{\sqrt{\lambda}} m\ell$. For the conformal case, $m=0$, our result in \eqref{eq:d5_conformal_entanglement} reproduces the probe limit of a calculation of holographic entanglement entropy from the fully backreacted geometry~\cite{Estes:2014hka}, as well as known results for the universal defect free energy \eqref{eq:free-energy}~\cite{Casini:2011kv,Robinson:2017sup}, providing a nontrivial check of the probe brane prescription. For large $a$, in the interface case, the calculation reproduces half of the entropy of the Coulomb branch vacuum derived in ref.~\cite{Chalabi:2020tlw}.

In addition to entanglement entropy, we defined and computed various entanglement $C$-functions to measure the number of degrees of freedom along the RG flow. For nonzero mass, we constructed a \(C\)-function \eqref{eq:CST_modified_1} from the entanglement entropy, following ref.~\cite{Casini:2023kyj}, which resulted in a monotonic behavior along the defect RG flow for vanishing D3-brane charge $q=0$ on the D5-brane worldvolume. For nonzero $q\ne 0$, however, we found that the same \(C\)-function remains monotonic with increasing $a$ (or mass), but tends to more-and-more negative values as one flows towards the IR, eventually diverging to \(C \to - \infty\) as the radius of the entangling region becomes infinite. This divergence in the IR region is due to the fact that the RG flow dual to $q\neq 0$ is four-dimensional while the \(C\)-function of \cite{Casini:2023kyj} is adapted to three-dimensional defect flows. We attempted to remedy this by using a modified $C$-function \eqref{eq:Ctilde_interface} which is finite in the IR, but turns out to be non-monotonic for sufficiently large $\vert q\vert$.

The challenge we face constructing a $C$-function for \(q \neq 0\) is complementary to that for constructing entropic \(C\)-functions in RG flows across dimensions~\cite{GonzalezLezcano:2022mcd,Jokela:2025cyz}. In the latter case one has an IR that is lower-dimensional than the UV, which is reflected in the behavior of the entanglement entropy in the UV and IR limits. Conversely, our result for \(S_1\) for \(q \neq 0\) has the UV divergence structure of a three-dimensional entanglement entropy, but the leading-order large \(a\) behavior of a four-dimensional entropy. For RG flows across dimensions one can define an entropic \(C\)-function that decreases monotonically, or one that interpolates between effective central charges in the UV and IR, but apparently not one that does both~\cite{GonzalezLezcano:2022mcd}. Similarly, the defect \(C\)-function defined in equation~\eqref{eq:CST_modified_1} is monotonically decreasing with increasing \(a\), but does not saturate to a finite value as \(a\to \infty\), while  the modified \(C\)-function in equation~\eqref{eq:Ctilde_interface} interpolates between finite values, but is not always monotonic.

To overcome these issues for \(q \neq 0\), we also considered three alternative candidate $C$-functions motivated by the works \cite{Liu:2012eea,Liu:2013una,Casini:2017vbe} that are adapted to four-dimensional RG flows, which we refer to as ``\(A\)-functions'' in order to distinguish them from the \(C\)-functions discussed above. In our probe approximation, we find that only one of these functions \eqref{eq:ALM_tilde}, which we denote by $\widetilde{A}_{\text{LM}}$, interpolates between finite values in the UV and IR. In the UV, $\widetilde{A}_{\text{LM}}$ coincides with the average of the type A Weyl anomaly coefficients of the two $\mathcal{N} = 4$ SYM theories on the two sides of the interface, while in the IR, it picks up the expected Weyl anomaly coefficient of $\mathcal{N} = 4$ SYM theory without an interface. It also behaves monotonically for $q \geq 1\slash 2$ but not for smaller values of \(q\).

The computations at the probe level involve certain subtleties. In particular, the role of a boundary term in the entanglement entropy calculations for probe branes has been a point of discussion. This term was omitted in refs.~\cite{Karch:2014ufa,Jokela:2024cxb} whereas it was included in ref.~\cite{Chalabi:2020tlw}. We have clarified the origin of this boundary term, showing that it compensates for a difference between the values of an integral over the worldvolume of the probe brane evaluated in two different coordinate systems --- a difference that arises because of a singularity in the integrand. Thus, all of the end results for entropies calculated in refs.~\cite{Karch:2014ufa,Kumar:2017vjv,Rodgers:2018mvq,Chalabi:2020tlw,Jokela:2024cxb} should be correct, since those \cite{Karch:2014ufa,Kumar:2017vjv,Rodgers:2018mvq,Jokela:2024cxb} that omit the boundary term performed their integrals in Poincar\'e coordinates.

A natural next step would be to move beyond the probe approximation and consider the effects of backreaction~\cite{Gomis:2006cu,DHoker:2007zhm,Conde:2016hbg,Penin:2017lqt,Garbayo:2022pqp}. Such calculations are technically more challenging and typically require numerical methods, or other simplification, such as invoking smearing which breaks the flavor symmetry $\U(N_5)\to \U(1)^{N_5}$~\cite{Nunez:2010sf}. Nevertheless, comparing our analytic probe results to future backreacted analyses would help clarify the regime of validity of the probe approach and further elucidate the interplay between the defect and ambient contributions to entanglement entropy.

One could also consider a different RG flow, localized to the interface, under which \(N_5\) decreases while the ranks of the gauge groups on either side of the interface remain fixed~\cite{Estes:2014hka}. The universal contribution of the defect to the entanglement entropy at the fixed points of this RG flow was computed in ref.~\cite{Estes:2014hka}, and found to always be smaller in the IR than in the UV. We expect that in the probe limit this flow would correspond to two separate stacks of D5-branes of the type we consider. One stack should have an \(\ads[4] \times \sph[2]\) worldvolume of the type studied in section~\ref{sec:conformal} and carry non-zero D3-brane charge \(n_3\), capturing the difference in ranks of the gauge group on either side of the interface. The other stack should be of the type in equation~\eqref{eq:D5_embedding} with \(m \neq 0\), triggering the RG flow, but \(n_3 = 0\). Consequently, we expect that in the probe limit the defect contribution to entanglement entropy under this RG flow would be a linear combination of the entanglement entropies in equations~\eqref{eq:d5_conformal_entanglement} and~\eqref{eq:delta_ee_q0}. Therefore, the defect \(C\)-function would be proportional to the one in equation~\eqref{eq:q0_C}, which is monotonic, confirming the prediction of \cite{Estes:2014hka} in the probe limit. To capture the qualitative difference between this RG flow and the one we consider in this work would require going beyond the probe limit.

Straightforward generalizations to study defect entanglement entropy would be to looking at more complicated entangling regions such as concentric balls~\cite{Fonda:2014cca,Jokela:2019ebz} centered on the defect, enabling a check of the monotonicity of \(C\)-function constructed from mutual information~\cite{Casini:2015woa} or at the very least furnish additional entropic RG diagnostics in defect CFTs. More broadly, our approach could be adapted to other types of defects, including higher-codimension examples or setups in different spacetime dimensions, such as the D3--\(\mathrm{D7}'\) intersection~\cite{Bergman:2010gm,Kristjansen:2012tn,Linardopoulos:2025ypq}.

\acknowledgments

We thank Adam Chalabi, Poul Damgaard, Elias Kiritsis, Charlotte Kristjansen, Javier Subils, and Konstantin Zarembo for useful discussions and comments. N.\,J. has been supported in part by Research Council of Finland grant no. 354533. J.\,K. is supported by the Deutsche Forschungsgemeinschaft (DFG, German Research Foundation) through the German-Israeli Project Cooperation (DIP) grant `Holography and the Swampland', as well as under Germany's Excellence Strategy through the W\"urzburg-Dresden Cluster of Excellence on Complexity and Topology in Quantum Matter - ct.qmat (EXC 2147, project-id 390858490). The work of R.\,R. was supported by the European Union's Horizon Europe research and innovation program under Marie Sk{\l}odowska-Curie Grant Agreement No. 101104286. Nordita is supported in part by Nordforsk. H.\,R. is supported in part by the Magnus Ehrnrooth foundation.

\appendix

\section{Sphere free energy}
\label{app:free_energy}

In this section we use holography to compute \(F_\mathrm{def}\), the free energy contribution of the conformal codimension-one defect on a maximal \(\sph[3] \subset \sph[4]\). To do so, it will be convenient to write \ads[5] in a coordinate system in which it is foliated by four-spheres. Starting from the metric of Euclidean \(\ads[5] \times \sph[5]\) in Poincar\'e coordinates, obtained via a Wick rotation \(t = -i t_\mathrm{E}\) of the coordinates used in section~\ref{sec:d3_d5},
\begin{equation}
    \diff s^2 = \frac{L^2}{z^2} \le[\diff z^2 + \diff t_\mathrm{E}^2 + \diff x^2 + \diff r^2 + r^2 \diff \f^2 \ri] + L^2 \diff s^2_{\sph[5]} \, ,
\end{equation}
we define new coordinates \((u,\a_1,\a_2,\a_3,\a_4)\) through \(\f = \a_4\) and
\begin{equation}\begin{aligned}
    z &= \frac{2u}{1 + u^2 + (1-u^2) \sin \a_1 \cos \a_2} \, ,
    &
    t_\mathrm{E} &= \frac{(1 - u^2) \sin \a_1 \sin \a_2 \cos \a_3}{1 + u^2 + (1-u^2) \sin \a_1 \cos \a_2} \, ,
    \\
    x &= \frac{(1-u^2) \cos \a_1}{1 + u^2 + (1-u^2)\sin \a_1 \cos \a_2}\ , \qquad
    &
    r &= \frac{(1-u^2) \sin \a_1 \sin \a_2 \sin \a_3}{1 + u^2 + (1-u^2) \sin \a_1 \cos \a_2} \, .
\end{aligned}\end{equation}
This transformation puts Euclidean \ads[5] in the desired \sph[4] slicing,
\begin{equation}\begin{aligned} \label{eq:spherical_slicing_metric}
    \diff s^2 &= L^2 \le(\frac{\diff u^2}{u^2} + \frac{(1-u^2)^2}{4u^2} \diff_{\sph[4]}^2 \ri) + L^2 \diff s^2_{\sph[5]} \, ,
    \\
    \diff_{\sph[4]}^2 &= \diff \a_1^2 + \sin^2 \a_1 \diff\a_2^2 + \sin^2 \a_1 \sin^2 \a_2 \diff \a_3^2 + \sin^2 \a_1 \sin^2 \a_2 \sin^2 \a_3 \diff \a_4 \, .
\end{aligned}\end{equation}
The new radial coordinate \(u\) takes values in the range \(u \in [0,1]\), with the boundary at \(u=0\). The defect at \(x=0\) at the boundary is mapped to \(\a_1 = \pi/2\). Thus, if we make a natural choice of defining function, such that the boundary metric is that of \sph[4], the defect will indeed span a maximal \sph[3]. The massless embeddings \(x = q z\) become
\begin{equation} \label{eq:spherical_slicing_embedding}
    \cos \a_1 = \frac{2 q u}{1-u^2} \, .
\end{equation}
Notice that the D5-branes do not span the full range of \(u\), but are instead restricted to \(u \leq \sqrt{1 + q^2} - q\) in order that \(\cos \a_1 \leq 1\). The worldvolume field strength takes the same form as in equation~\eqref{eq:ansatz}.

In Poincar\'e coordinates the five-form field strength is, in Euclidean signature,
\begin{equation}
    F_5 = i\frac{4 L^4}{z^5} \diff z \wedge \diff t_\mathrm{E} \wedge \diff x \wedge \diff r \wedge \diff \f + \ldots \; ,
\end{equation}
where the dots denote an additional term needed to make \(F_5\) self-dual, the explicit form of which will not be needed here. After the coordinate transformation to spherical slicing, this becomes
\begin{equation} \label{eq:F5_spherical}
    F_5 = i\frac{L^4 (1-u^2)^4}{4 u^5} \diff u \wedge \mathrm{vol}(\sph[4]) + \ldots\;,
\end{equation}
where \(\mathrm{vol}(\sph[4])\) is the volume form on the \sph[4] with metric written in equation~\eqref{eq:spherical_slicing_metric}. We employ a gauge in which the corresponding four-form potential is
\begin{equation} \label{eq:C4_spherical}
    C_4 = -i\frac{L^4}{16\r^4} (1 - 8 u^2 + 8 u^6 - u^8 - 24 u^4 \log u) \, \mathrm{vol}(\sph[4]) + \ldots \; ,
\end{equation}
with the dots denoting a term responsible for the dots in equation~\eqref{eq:F5_spherical}. The expression for \(C_4\) in equation~\eqref{eq:C4_spherical} differs from the direct coordinate transformation of equation~\eqref{eq:ads5xs5_c4} by a gauge transformation, chosen to ensure that \(C_4\) vanishes at the center of \ads\ at \(u=1\).

The Euclidean D5-brane action \(I_\mathrm{D5}\) evaluated on the solution in equation~\eqref{eq:spherical_slicing_embedding} diverges due to the behavior of the integrand at small \(u\). Regulating this integral by integrating only over \(u \geq \e\) for some small \(\e \geq 0\), with the gauge choice for \(C_4\) in equation~\eqref{eq:C4_spherical}, we find
\begin{equation} \label{eq:D5_regulated_action}
    I_\mathrm{D5}^\star = \frac{\sqrt{\la} \, N_3 N_5}{3 \pi} \le(\frac{1}{8 \e^3} - \frac{3 (2q^2 + 3)}{8 \e} + (2-q^2)\sqrt{1+q^2} - 3 q \sinh^{-1} q\ri) + O(\e)\, .
 \end{equation}
The divergences must be eliminated through holographic renormalization~\cite{deHaro:2000vlm}, by the addition of appropriate counterterms evaluated on the cutoff surface at \(u=\e\). The necessary counterterms for the D5-brane are~\cite{Karch:2005ms},\footnote{Scheme dependence, in the form of the possible presence of finite counterterms in equation~\eqref{eq:counterterms}, is fixed by supersymmetry~\cite{Karch:2005ms}. This also fixes the gauge ambiguity in the free energy that would otherwise arise due to boundary terms arising when adding an exact form to \(C_4\)~\cite{Jokela:2021knd}.}
\begin{equation}\begin{aligned} \label{eq:counterterms}
    I_\mathrm{ct} &= - \frac{2 N_5 T_5}{3 \pi} \int \diff \a_2 \diff \a_3 \diff \a_4 \, \sqrt{\g} \le(1 - \frac{L^2}{4} R_\g\ri)
    \\
    &= \frac{\sqrt{\la} N_3 N_5}{3 \pi} \le(- \frac{1}{8\e^3} + \frac{3(2q^2+3)}{8\e}\ri) + O(\e) \, ,
\end{aligned}\end{equation}
where \(\g\) is the induced metric of the intersection between the D5-brane and the cutoff surface in \(\ads[5]\) at \(u = \e\), and \(R_\g\) is the Ricci scalar of this surface. The counterterms exactly cancel the divergences in equation~\eqref{eq:D5_regulated_action}, so that the D5-brane contribution to the sphere free energy, \(F_\mathrm{def} = I_\mathrm{D5}^\star + I_\mathrm{ct}\), is given by the finite term in equation~\eqref{eq:D5_regulated_action},
\begin{equation}
    F_\mathrm{def} =\frac{\sqrt{\la} \, N_3 N_5}{3 \pi} \le((2-q^2)\sqrt{1+q^2} - 3 q \sinh^{-1} q\ri)  .
\end{equation}
At \(q=0\), this expression for \(F_\mathrm{def}\) reduces to the earlier result of ref.~\cite{Robinson:2017sup}, which was computed both in holography and with supersymmetric localization.

\section{Evaluation of entropy integrals}
\label{app:integrals}

In this appendix we provide further details for the evaluation of some of the integrals contributing to the entanglement entropy.

\subsection{\texorpdfstring{\(\Sbrane\)}{Sbrane}}
\label{app:integrals_Sbrane}

We begin with \(\Sbrane\), for which we need to evaluate the integral in equation~\eqref{eq:Sbrane_integral}. It will be convenient to split this integral into two pieces
\begin{equation}
    \Sbrane = \Sdbi + \Swz \, ,
\end{equation}
arising from the Dirac--Born--Infeld (DBI) and Wess--Zumino (WZ) terms in the D5-brane action, respectively. Explicitly, these contributions are given by
\begin{equation}\begin{aligned} \label{eq:Sdbi_and_Swz_integrals}
   \Sdbi &=
    \frac{\sqrt{\la} \, N_3 N_5}{3\pi^2 a} \int_{\cR} \diff s \diff \z  \diff \t \, \biggl[
        s^2 \le(q^2 + (1-s^2)^2 \ri) \le(\frac{\cos(2\t)}{\z^2 - 1} - \sin^2 \t  + \frac{2 a \cos \t}{s \sqrt{\z^2 - 1}}\ri)
        \\ & \hspace{7cm}
        + a^2 \le(\frac{q^2}{s^2} + (1-s^2)^2 \ri)
    \biggr] \frac{1}{(1-s^2)^{3/2})} \, ,
    \\
    \Swz &=
    -\frac{4 q^2 \sqrt{\la} \, N_3 N_5}{3\pi^2} \int_{\cR} \diff s \diff \z  \diff \t \, 
        \frac{s^3 \cos\t + s \sqrt{\z^2 -1}}{s^2 \z^3 (1-s^2)^{3/2} \sqrt{\z^2 - 1}}\, .
\end{aligned}\end{equation}
As discussed in section~\ref{sec:flow_ee}, the main challenge in evaluating the integrals in equation~\eqref{eq:Sdbi_and_Swz_integrals} is the complicated form of the integration region \(\cR\). We perform the computation by performing the change of variables in equation~\eqref{eq:to_poincare}, where the integration region simplifies. The details of the evaluation of \(\Sdbi\) and \(\Swz\) are given in the next two subsections.

\subsubsection{\texorpdfstring{\(\Swz\)}{SWZ}}
\label{app:integrals_Swz}

Since the integral for \(\Swz\) converges absolutely, we can safely transform to Poincar\'e coordinates. After the change of variables in equation in equation~\eqref{eq:to_poincare}, the integral for \(\Swz\) in equation~\eqref{eq:Sdbi_and_Swz_integrals} becomes
\begin{equation}
    \Swz = \frac{16 q^2 \sqrt{\la} \, N_3 N_5}{3\pi^2} \int_{\m\e}^1 \diff s \int_0^\infty \diff t \int_0^\infty \diff r  \,  \frac{r}{\sqrt{1-s^2}}\cS_\mathrm{WZ} \, ,
\end{equation}
with integrand
\begin{equation}\begin{aligned}
    \cS_\mathrm{WZ} &= \frac{(1-s^2)(R^2 - a^2) + q^2 s^2}{(1-s^2)^2 \le[(R^2 - a^2)^2 + 4 a^2 t^2\ri] + 2 q^2 s^2 (1-s^2)(R^2 - a^2) + q^4 s^4}
    \\ &\quad\phantom{=}
    -\frac{(1-s^2)(R^2 - a^2) + q^2 s^2}{(1-s^2)^2 \le[(R^2 + a^2)^2 - 4 a^2 r^2\ri] + 2 q^2 s^2 (1-s^2)(R^2 - a^2) + q^4 s^4}
    \\ &\quad\phantom{=}
    -\frac{4 a^2 (1-s^2)^2 \le[s^2 (1-s^2)(R^2- a^2) + 2 a^2 (1-s^2) + q^2 s^4 \ri]}{\le\{(1-s^2)^2 \le[(R^2 + a^2)^2 - 4 a^2 r^2\ri] + 2 q^2 s^2 (1-s^2)(R^2 - a^2) + q^4 s^4\ri\}^2} \, ,
\end{aligned}\end{equation}
where \(R^2 \equiv r^2 + s^2 + t^2\). Mathematica is able to perform the integrals over \(r\) and \(t\),  with the result
\begin{equation}
    \Swz = - \frac{4 q^2 \sqrt{\la} \, N_3 N_5}{3\pi} \int_{\m \e}^1 \diff s \,  \frac{f_1(a,q;s) + f_2(a,q;s) + f_2(-a,q;s)}{(1-s^2)^2} \ ,
\end{equation}
where we have defined
\begin{equation} \label{eq:Swz_piecewise_piece}
    f_1(a,q;s) = \begin{cases}
        2 a \sqrt{1-s^2} \, , & s \leq \s  \, ,
        \\
        2 s \sqrt{1 + q^2 - s^2}\, , & s > \s  \, ,
    \end{cases}
\end{equation}
and
\begin{equation}
    f_2(a,q;s) \! = \! \frac{q a^2 (1-s^2)(1-2s^2) - 2 q s^4 (1 + q^2 - s^2) + a s [q^2 (1-4s^2) + (1-s^2)^2]\sqrt{1-s^2}}{2 q s^2 \sqrt{(q s + a \sqrt{1-s^2})^2 + s^2 - s^4}} .
\end{equation}

The indefinite integral of \(f_1/(1-s^2)^2\) is easy to evaluate, but diverges as \(s \to 1\). We will regulate this divergence by integrating only over \(s \leq 1-\d\). Then
\begin{equation}\begin{aligned} \label{eq:I1_result}
    \cI_1 &\equiv \int_{\m \e}^{1-\d} \diff s \, \frac{f_1(a,q;s)}{(1-s^2)^2}
    \\ &= 
    \frac{q}{2\d} - \frac{\log \d - \log(2 q^2)}{2q}
    + \frac{2+q^2}{4q}
    - \frac{1}{q} \sinh^{-1} \left(\frac{q}{\sqrt{1-\s^2}}\right)
    \\
    &\qquad\qquad\qquad+ \frac{2 a \s}{\sqrt{1-\s^2}}
     - \frac{\sqrt{1 + q^2 - \s^2}}{1-\s^2} \, ,
\end{aligned}\end{equation}
where we have neglected any terms that vanish when \(\e \to 0 \) and \(\d \to 0\). The integral involving \(f_2\) is more challenging. To proceed, we first change integration variables from \(s\) to \(u \equiv  q s/\sqrt{1-s^2}\), in terms of which we find
\begin{align}
    \cI_2 &\equiv \int_{\m\e}^{1-\d} \diff s \frac{f_2(a,q;s) + f_2(-a,q;s)}{(1-s^2)^2}
    \nonumber \\
    &= \int_{u(\m\e)}^{u(1-\d)} \diff u \le[ \cJ_1(a,q;u) + \cJ_1(a,q;-u) + \p_u \cJ_2(a,q;u) - \p_u \cJ_2(a,q;-u) \ri]  ,
\end{align}
where we have defined
\begin{equation}\begin{aligned}
    \cJ_1(a,q;u) &= \frac{a q^2 - u^3}{2 q u \sqrt{q^2 +u^2} \sqrt{q^2 (a+u)^2 + u^2[1 + (a+u)^2]}} \, ,
    \\
    \cJ_2(a,q;u) &=- \frac{1}{2q u} \sqrt{q^2 +u^2} \sqrt{q^2 + (a+u)^2 + u^2 [1+(a+u)^2]} \,.
\end{aligned}\end{equation}
We can straightforwardly do the integral of derivatives of \(\cJ_2\),
\begin{align}
    \cI_3 &\equiv \int_{u(\m\e)}^{u(1-\d)} \diff u \, \p_u \le[\cJ_2(a,q;u) -\cJ_2(a,q;-u) \ri]
    \nonumber \\
    &= - \frac{q}{2\d} + \frac{a}{\m\e} - \frac{2 + q^2}{4q}.
\end{align}
Now we need to evaluate the integral involving \(\cJ_1\),
\begin{equation}
    \int_{u(\m\e)}^{u(1-\d)} \diff u \le[ \cJ_1(a,q;u) + \cJ_1(a,q;-u) \ri] = \frac{\log \d - \log(2q^2)}{2q} + \cI_4 \, ,
\end{equation}
where we have defined
\begin{equation}
    \cI_4 \equiv \int_0^\infty \diff u \le[
        \cJ_1(a,q;u) + \cJ_1(a,q;-u) + \frac{1}{q \sqrt{1+u^2}}
    \ri].
\end{equation}
We have set the integration limits in \(\cI_4\) to \([0,\infty]\) as the integral is finite in the limits \(\e \to 0\) and \(\d \to 0\). It turns out that \(\cI_4\) vanishes. To show this, we first show that \(\cI_4 \) is independent of \(a\) for any non-zero \(a\), since direct calculation shows that for any \(a \neq 0\), we have
\begin{align}
    \p_a \cI_4 &= \int_0^\infty \diff u \, \p_a \le[ \cJ_1(a,q;u) + \cJ_1(a,q;-u) \ri]
    \nonumber \\
    &= \int_0^\infty \diff u \, \p_u \le[\frac{q^2 + u^2}{4 u q^2} \le( \frac{1}{\cJ_2(a,q;u)} + \frac{1}{\cJ_2(a,q;-u)}\ri)
    \ri] = 0 \, .
\end{align}

Since \(\cI_4\) is independent of \(a\), we will evaluate it in the limit \(a \to 0\). To compute this limit we cannot simply set \(a=0\) in \(\cJ_1(a,q;\pm u)\), since for any small but non-zero \(a\), there is part of the integration region close to \(u=0\) where \(u\) is comparable to or smaller than \(a\), which affects the behavior of the terms in \(\cJ_1(a,q;\pm u)\) involving \(u \pm a\). To deal with this, we split the integration region into two pieces, \(u \leq u_*\) and \(u \geq u_*\), where  we choose \(u_* \ll 1\) such that
\begin{equation}
    a \ll u_* \ll q \, .
\end{equation}
In the integral over \(u \geq u_*\) we can simply set \(a=0\). In the integral over \(0 \leq u \leq u_*\) we must keep \(a\) non-zero, but since \(u_*\) is small we can simplify the integrand, enabling us to perform the integral. Concretely, for \(0 \leq u \leq u_*\) we approximate \(\cJ_1\) as
\begin{equation}
    \cJ_1(a,q;u) \approx \tilde{\cJ}_1(a,q;u) \equiv \frac{a q^2 - u^3}{2 q^2 u \sqrt{q^2 (a+u)^2 + u^2}} \, .
\end{equation}
Then we have that
\begin{equation}\begin{aligned}\label{eq:I3_from_limit}
    \cI_4 = \lim_{a \to 0} \cI_3 
    &= \lim_{u_* \to 0} \int_{u_*}^\infty \diff u \le[ \cJ_1(0,q;u) + \cJ_1(0,q;-u) + \frac{1}{q \sqrt{1+u^2}}\ri]
    \\ &\phantom{=}
    +  \lim_{u_* \to 0} \lim_{a \to 0} \int_0^{u_*} \diff u \le[ \tilde{\cJ}_1(a,q;u) + \tilde{\cJ}_1(a,q;-u) \ri]  .
\end{aligned}\end{equation}
The two integrals on the right-hand side are relatively straightforward to evaluate,
\begin{align}
     \int_{u_*}^\infty \diff u \biggl[ \cJ_1(0,q;u) + &\cJ_1(0,q;-u) + \frac{1}{q \sqrt{1+u^2}}\biggr]
     \nonumber \\
     &= \frac{1}{q} \int_{u_*}^\infty \diff u \le[\frac{1}{\sqrt{1+u^2}} -\frac{u}{\sqrt{q^2 + u^2} \sqrt{1 + q^2 + u^2}} \ri]
     \\
     &= \frac{\sinh^{-1} q}{q} + O(u_*) \, ,
     \nonumber
\end{align}
and
\begin{align}
    \int_0^{u_*} \diff u  \biggl[ \tilde{\cJ}_1(a,q;u) + &\tilde{\cJ}_1(a,q;-u) \biggr]
    \nonumber\\
    &= \frac{1}{2q}\int_0^{u_*} \diff u \frac{1}{u}  \le[ \frac{a q^2 - u^3}{\sqrt{q^2 (a+u)^2 + u^2}} -  \frac{a q^2 + u^3}{\sqrt{q^2 (a-u)^2 + u^2}} \ri]
    \\ 
    &= - \frac{\sinh^{-1} q}{q} + O(u_*) + O(a) \,.
    \nonumber
\end{align}
Thus the two contributions to \(\cI_4\) in equation~\eqref{eq:I3_from_limit} exactly cancel, and we obtain \(\cI_4 = 0\), and therefore
\begin{align} \label{eq:I2_result}
    \cI_2 &= \cI_3 + \frac{\log \d - \log(2q^2)}{2q} + \cI_4
    \nonumber \\
    &= - \frac{q}{2\d} + \frac{\log \d - \log(2q^2)}{2q} + \frac{a}{\m\e} - \frac{2 + q^2}{4q}\,.
\end{align}
The contribution to the entanglement entropy from the WZ term is
\begin{equation}
    \Swz = - \frac{4 q^2 \sqrt{\la} \, N_3 N_5}{3\pi} \le[ \cI_1 + \cI_2\ri] .
\end{equation}
Substituting the results in equations~\eqref{eq:I1_result} and~\eqref{eq:I2_result}, we obtain the final result of
\begin{equation} \label{eq:Swz_result}
    \Swz =\frac{4 q^2 \sqrt{\la} \, N_3 N_5}{3\pi} \le[- \frac{\ell}{\e} + \frac{1}{q} \sinh^{-1} \le(\frac{q}{\sqrt{1-\s^2}}\ri) + \frac{\sqrt{1+q^2-\s^2}}{1-\s^2} - \frac{2a\s}{\sqrt{1-\s^2}} \ri] .
\end{equation}

\subsubsection{\texorpdfstring{\(\Sdbi\)}{SDBI}}
\label{app:integrals_Sdbi}

For the evaluation of \(\Sdbi\), it will be convenient to remove the part of the integral in equation~\eqref{eq:Sdbi_and_Swz_integrals} proportional to \(\Swz\), defining
\begin{equation}
    \Sdbi = - \frac{1}{4} \Swz + \Sdbib \, .
\end{equation}
The remaining integral \(\Sdbib\) is UV-finite. It is given by
\begin{equation}\begin{aligned} \label{eq:Sdbib_chm_integral}
    \Sdbib = \frac{\sqrt{\la} \, N_3 N_5}{3\pi^2 a} \int_{\cR}\diff s \diff \z \diff \t \, \biggl[&
        s^2 \le(q^2 + (1-s^2)^2 \ri) \le(\frac{\cos(2\t)}{\z^2 - 1} - \sin^2 \t \ri)
        \\
        & + \le(\frac{q^2}{2} + (1-s^2)^2 \ri) \frac{2 a s \cos \t}{\sqrt{\z^2-1}} + a^2(1-s^2)^2
    \biggr] .
\end{aligned}\end{equation}
Notice that \(\Sdbib\) is not absolutely convergent; near \(\z=1\) the integrand has a singularity of the type discussed in section~\ref{sec:boundary_term}, proportional to \((\z-1)^{-1}\cos(2\t)\). We denote by \(\Sdbib^{(\mathrm{Poinc})}\) the integral resulting from naively performing the coordinate transformation in equation~\eqref{eq:to_poincare}, with the prescription of performing the \(r\) integral first,
\begin{align} \label{eq:Sdbib_poincare_integral}
    \Sdbib^{\text{(Poinc)}} = \frac{4 \sqrt{\la} \, N_3 N_5}{3\pi^2}\int_0^1 \diff s \int_0^\infty \diff t \int_0^\infty \diff r\, r (1-s^2)^{3/2} \biggl[&
        - \frac{\cN_1}{\cD_1^2} - \frac{\cN_2}{(1-s^2)^2 \cD_1} \\\nonumber
        &+ \frac{4 a^2 s^2 \cN_3}{\cD_2^2} + \frac{\cN_2}{(1-s^2)^2 \cD_2}
    \biggr].
\end{align}
where we have defined numerators
\begin{equation}\begin{aligned}
    \cN_1 &= 16 a^2 s^1 (1-s^2) \cQ_1 t^2 \, ,
    \\
    \cN_2 &= (1-s^2) \cQ_2 (R^2- a^2) - 6 (1-s^2) \cQ_1 t^2 + s^2 \le[q^2 \cQ_1 - (1-s^4)\cQ_2 + 2 s^2 (1-s^2)^3 \ri] ,
    \\
    \cN_3 &= \cN_2 + 4(1-s^2) \cQ_1 t^2 + 2 a^2 (1-s^2)^3 \, ,
\end{aligned}\end{equation}
with \(R^2 = r^2 + s^2 + t^2\) and \(\cQ_\a = q^2 + \a (1-s^2)^2\). The denominators are
\begin{equation}\begin{aligned}
    \cD_1 &= (1-s^2)^2 \le[(R^2 - a^2)^2 + 4 a^2 t^2 \ri] + 2 q^2 s^2 (1-s^2) (R^2 - a^2) + q^4 s^4 \, ,
    \\
    \cD_2 &= (1-s^2)^2 \le[(R^2 - a^2)^2 - 4 a^2 r^2\ri] + 2 q^2 s^2 (1-s^2) (R^2 - a^2) + q^4 s^4 \, .
\end{aligned}\end{equation}
As discussed in section~\ref{sec:boundary_term}, we expect \(\Sdbib \neq \Sdbib^{(\mathrm{Poinc})}\).

Mathematica is able to perform the integrals over \(r\) and \(t\), giving
\begin{equation}
    \Sdbib^{\text{(Poinc)}} = - \frac{\sqrt{\la} \, N_3 N_5}{6\pi a} \int_0^1 \diff s \frac{f_3(a,q;s) + q f_4(a,q;s) - q f_4(-a,q;s)}{(1-s^2)^{5/2}} \, ,
\end{equation}
where we have defined
\begin{equation}
    f_3(a,q;s) = \begin{cases}
        2 a^2 (1-s^2) \le[ q^2 - (1-s^2)^2 \ri] + 6 s^2 (1+q^2 - s^2) \le[q^2 + (1-s^2)^2\ri] \, , & s \leq \s \, ,
        \\[0.5em]
        4 a q^2 s \sqrt{\dfrac{1-s^2}{1+q^2 - s^2}} \, , & s > \s \, ,
    \end{cases} 
\end{equation}
and
\begin{equation}\begin{aligned}
    f_4(a,q;s) = &\frac{1}{\sqrt{(q s + a \sqrt{1-s^2})^2 + s^2 - s^4}} \biggl[
        -3 s^3 (1+q^2 - s^2)^2  + a^2 s (3+2s^2)(1-s^2)^2
        \\
        &\qquad - a q \le[7 s^2 (1+q^2 - s^2) + (a^2 -5 s^4)(1-s^2)\ri] \sqrt{1-s^2}
        \\
        &\qquad + s (3 s^4 - 5 a^2) (1-s^2)(1+q^2 - s^2)\biggr] .
\end{aligned}\end{equation}
The integral involving \(f_3\) is straightforward. The integrand diverges near \(s=1\), so we regulate by integrating only over \(s \leq 1- \d\), with the result
\begin{align} \label{eq:I5}
    \cI_5 &\equiv \int_0^{1-\d} \diff s \frac{f_3(a,q;s)}{(1-s^2)^{5/2}}
    \nonumber \\
    &= \frac{2 a q^3}{\d} + \frac{3 - 4 a^2 - 12 q^3}{4} \sin^{-1}\s + a q (2-3q^2) + \frac{a(12 q^2 -3 +5 \s^2 - 2 \s^4)}{4 \sqrt{1 + q^2 - \s^2}} \, ,
\end{align}
where we neglect terms that vanish when \(\d \to 0\). We can simplify the integral involving \(f_4\) by changing variables to \(u = q s/\sqrt{1-s^2}\), to obtain
\begin{align}
    \cI_6 &\equiv q \int_0^{1-\d} \diff s \frac{f_4(a,q;s) - f_4(-a,q;s)}{(1-s^2)^{5/2}}
    \nonumber \\
    &= q \int_0^{u(\d)} \diff u \, \p_u \le[f_5(a,q;u) - f_5(-a,q;u) \ri]  ,
\end{align}
where
\begin{equation}
    f_5(u,a;q) = u\frac{(u+a)(u^2 + q^2) + u}{(u^2 + q^2)^{3/2}} \sqrt{(u+a)^2(u^2 + q^2)^2+u^2} \, .
\end{equation}
We can then straightforwardly evaluate \(\cI_6\), to find
\begin{equation} \label{eq:I6}
    \cI_6 = - \frac{2 a q^3}{\d} + a q (3q^2 - 2) \, .
\end{equation}
Combining the results in equation~\eqref{eq:I5} and~\eqref{eq:I6}, we find
\begin{align}\label{eq:app_Sdbib_Poincare}
    \Sdbib^{\text{(Poinc)}} &= - \frac{\sqrt{\la} \, N_3 N_5}{6 \pi a} (\cI_5 + \cI_6)
    \nonumber \\
    &= \frac{\sqrt{\la} \, N_3 N_5}{24\pi} \le[\frac{3 -12 q^2 - 5 \s^2 +2 \s^4}{\sqrt{1 + q^2 - \s^2}}- \frac{3 + 4 a^2 + 12 q^2}{a} \sin^{-1} \s\ri].
\end{align}
As discussed in section~\ref{sec:boundary_term}, we expect the difference between \(\Sdbib\) and \(\Sdbib\) to be accounted for by the boundary term \(\Svar\), so that
\begin{align} \label{eq:app_Sdbib}
    \Sdbib &= \Sdbib^{\text{(Poinc)}} - \Svar
    \nonumber \\
    &= \frac{\sqrt{\la} \, N_3 N_5}{12 \pi} \le(\frac{2a^2 + 2q^2 - 1}{a} \sin^{-1}\s + \frac{1 - 2 q^2 - \s^2}{\sqrt{1+ q^2-\s^2}} \ri).
\end{align}
Combining equations~\eqref{eq:Swz_result} and~\eqref{eq:app_Sdbib}, we obtain the result for
\begin{equation}
    \Sbrane = \Sdbi + \Swz = \Sdbib + \frac{3}{4} \Swz
\end{equation}
given in equation~\eqref{eq:Sbrane_result}.

\subsection{\texorpdfstring{\(\Svar\)}{Svar}}
\label{sec:Svar}

In this subsection we describe the evaluation of the boundary term in the entanglement entropy \(\Svar\), defined in equation~\eqref{eq:Svar_def}. In order to obtain an explicit integral to evaluate, we follow the method outlined in ref.~\cite{Chalabi:2020tlw}.

The equations of motion following from the action in equation~\eqref{eq:d5_chm_action} admit expansions near \(\z=\z_H\) of the form
\begin{equation}\begin{aligned} \label{eq:near_horizon_expansion}
    \q &= \q_0(\z_H;v) + \Q_1(\z_H;\t,v)  \sqrt{\z - \z_H} + O(\z - \z_H) \, , 
    \\
    \xi &= \xi_0(\z_H;v) + \Xi_1(\z_H;\t,v)  \sqrt{\z - \z_H} + O(\z - \z_H) \, .
\end{aligned}\end{equation}
The equations of motion imply that the coefficients of the \(O(\sqrt{\z - \z_H})\) terms take the form
\begin{equation}\begin{aligned} \label{eq:near_horizon_expansion_coefficients}
    \Q_1(\z_H;\t,v) &= \q_{1}(\z_H;v) \cos(2\pi T\t) + \tilde{\q}_{1}(\z_H;v) \sin(2\pi T\t) \, ,
    \\
    \Xi_1(\z_H;\t,v) &= \xi_{1}(\z_H;v) \cos(2\pi T\t) + \tilde{\xi}_{1}(\z_H;v) \sin(2\pi T\t) \, ,
\end{aligned}\end{equation}
where \(T\) depends on \(\z_H\) through equation~\eqref{eq:chm_temperature}.  Substituting equations~\eqref{eq:near_horizon_expansion} and~\eqref{eq:near_horizon_expansion_coefficients} into the expression for \(\Svar\) in equation~\eqref{eq:Svar_def}, one finds that the boundary term is
\begin{equation}\begin{aligned} \label{eq:Svar_partial_subbed}
    \Svar &= \frac{\sqrt{\la} \, N_3 N_5}{6 \pi^2} \int \diff v  \diff \t \sinh^2 v \sin \xi_0 \sqrt{q^2 + \cos^4 \q_0} \, \le. \frac{A}{\sqrt{B}} \ri|_{\z_H = 1}\,,
    \\
    A &\equiv \frac{2 \Q_1^2}{\sinh^2 v} + 2 \Xi_1^2 + (\q_1 \tilde{\xi_1}- \tilde{\q}_1 \xi_1)^2 + 2 \le(\Q_1 \p_v \xi_0 - \Xi_1 \p_v \q_0 \ri)^2 \, ,
    \\
    B &\equiv 4 \le(\frac{1 + (\p_v \q_0)^2}{\sinh^2 v} + (\p_v \xi_0)^2 \ri)
    + 2 \le| (\q_1 + i \tilde{\q}_1)\p_v \xi_0  - (\x_1 + i \tilde{\xi}_1) \p_v \q_0\ri|^2
    \\ &\phantom{=}
    + 2 \le(\frac{\q_1^2 + \tilde{\q}_1^2}{\sinh^2 v} + (\xi_1^2 + \tilde{\xi}_1^2) \ri)
    +(\q_1 \tilde{\xi}_1 - \tilde{\q}_1 \xi_1)^2.
\end{aligned}\end{equation}

Crucially for our purposes, since there are no derivatives with respect to \(\z_H\) appearing in equation~\eqref{eq:Svar_partial_subbed}, \(\Svar\) may be evaluated knowing only the form of the expansion coefficients at \(\z_H=1\) (corresponding to \(T=1/(2\pi)\)). These coefficients may be obtained by expanding the solution in equation~\eqref{eq:massive_solution_chm} for \(\z_H=1\). The result is that \(\tilde{\q}_1(1;v) = \tilde{\xi}_1(1;v) = 0\) and
\begin{equation}\begin{aligned}
    \q_0(1;v) &= \sin^{-1} \le(\frac{a}{\cosh v}\ri),
    &
    \q_1(1;v) &= - \frac{\sqrt{2}\, a}{\cosh v \sqrt{\cosh^2 v - a^2}}\, ,
    \\
    \xi_0(1;v) &= \cos^{-1} \le(\frac{q}{\sqrt{\cosh^2 v - a^2}}\ri),
    &
    \xi_1(1;v) &= \frac{\sqrt{2} \, a^2 q \sech v }{(\cosh^2 v - a^2) \sqrt{\sinh^2 v - a^2 \tanh^2 v - q^2 }} \, .
\end{aligned}\end{equation}
Substituting these into equation~\eqref{eq:Svar_partial_subbed} we arrive at the integral to be solved for \(\Svar\),
\begin{equation}
    \Svar = -\frac{\sqrt{\la} \, N_3 N_5}{3\pi^2} a^2 \int_{v_\mathrm{min}}^{v_\mathrm{max}} \diff v\int_0^{2\pi} \diff \t \, \cos^2 \t \, \tanh v \, \frac{q^2 + (1-a^2 \sech^2 v)^2}{(\cosh^2 v - a^2)^{3/2}} \, ,
\end{equation}
the integral quoted in equation~\eqref{eq:Svar_integral}.

\section{Coulomb branch entanglement entropy}
\label{sec:coulomb}

The entanglement entropy contribution \(P^{(N_3,n_3)}_\mathrm{Coul}\) of a probe D3-brane spanning \((t,x,r,\f)\) in \(\ads[5] \times \sph[5]\), dual to \(\cN=4\) SYM theory on the Coulomb branch --- as defined in equation~\eqref{eq:Coulomb_branch_entropy_expansion} --- was computed in ref.~\cite{Chalabi:2020tlw}. In this appendix we provide a translation of their results to our notation, to facilitate comparison with our results. We also provide an evaluation of \(\Sbrane\) for the D3-brane in Poincar\'e coordinates, showing that it differs from the correct result for \(\Sbrane\) by \(\Svar\), as discussed in section~\ref{sec:boundary_term}.

Ref.~\cite{Chalabi:2020tlw} considered a single D3-brane at constant radial coordinate \(z = \m^{-1}\) and a ball-shaped entangling region of radius \(\ell\).\footnote{Ref.~\cite{Chalabi:2020tlw} uses \(v\) and \(R\) where we use \(\m\) and \(\ell\), respectively.} Defining a dimensionless radius \(a = \m\ell\) and multiplying their results by \(n_3\) to account for \(n_3\) coincident D3-branes, their results for the contributions to the entanglement entropy defined in section~\ref{sec:flow_ee} are\footnote{In ref.~\cite{Chalabi:2020tlw}, \(P^{(N_3,1)}_\mathrm{Coul}\), \(\Sbrane\), \(\Shor\), and \(\Svar\) are denoted by \(S_\mathrm{Coul}\), \(\cS\), \(\cS_1^{\mathrm{(bdy)}}\), and \(\cS_2^{\mathrm{(bdy)}}\), respectively.}
\begin{subequations} \label{eq:S_coulomb_contributions}
\begin{align}
    \Sbrane &= \Q(a-1) \frac{2}{3} n_3 N_3 \le[\cosh^{-1} a - \le(a + \frac{1}{a}\ri) \sqrt{a^2 - 1} \, \ri] \, ,
    \label{eq:Sbrane_D3_result}
    \\
    \Shor &= \Q(a-1)\frac{4}{3} n_3 N_3 \cosh^{-1} a \, ,
    \\
    \Svar &= -\Q(a-1) \frac{2}{3} n_3 N_3 \frac{\sqrt{a^2 - 1}}{a} \, ,
\end{align}
\end{subequations}
where \(T_\mathrm{D3}\) is the D3-brane tension and \(\Q\) is the Heaviside step function. Adding up these results one obtains \(P_\mathrm{Coul} \equiv \Sbrane + \Shor  +\Svar\), the leading contribution of the probe D3-branes to the holographic entanglement entropy in the probe limit~\cite{Chalabi:2020tlw},
\begin{equation} \label{eq:S_coul}
    P^{(N_3,n_3)}_\mathrm{Coul} = \Theta(a-1) \frac{2}{3} n_3 N_3 \le[3 \cosh^{-1} a - \le(a + \frac{2}{a} \ri) \sqrt{a^2 - 1} \, \ri].
\end{equation}

The result for \(\Sbrane\) in equation~\eqref{eq:Sbrane_D3_result} was obtained in ref.~\cite{Chalabi:2020tlw} by direct evaluation of the integral
\begin{equation}\begin{aligned} \label{eq:coulomb_branch_Sbrane_integral}
    \Sbrane &= - \frac{N_3}{3\pi} \int_\mathcal{R} \diff \z \diff \t \, \cI,
    \\
    \cI &= \frac{1}{a \z^3} \le[ \le(\frac{2}{\z^2 - 1} + 1\ri) \cos(2\t) - 1 - 2 a^2 \ri] \sqrt{\le( \cos \t \, \sqrt{\z^2 - 1} - a\ri)^2 - \z^2} \, .
\end{aligned}\end{equation}
The integration region is the worldvolume of the D3-branes in the \((\z,\t)\) coordinates, specified by the inequalities
\begin{equation}
    0 \leq \t \leq 2 \pi \, ,
    \qquad
    \z \geq 1 \, ,
    \qquad
    a - \cos\t\,  \sqrt{\z^2 - 1} \geq \z \, .
\end{equation}
Notice that the integrand in equation~\eqref{eq:coulomb_branch_Sbrane_integral} has the expected \(\cI \sim \cos(2\t)/(\z^2 - 1)\) divergence near \(\z=1\) described in section~\ref{sec:boundary_term}, and so to obtain a finite entanglement entropy one must peform the integral over \(\t\) before the integral over \(\z\). As a check that this prescription indeed produces the correct entanglement entropy, ref.~\cite{Chalabi:2020tlw} showed that equation~\eqref{eq:S_coul} matches the result for \(P^{(N_3,n_3)}_\mathrm{Coul}\) obtained using the RT prescription in the back-reacted multi-center D3-brane geometry, at linear order in a small \(n_3/N_3\) expansion.

Here we will show that if one tries to evaluate \(\Sbrane\) by changing variables in the integral to Poincar\'e coordinates, one obtains a result \(\SbranePoinc\) satisfying equation~\eqref{eq:S_brane_poincare_difference}, i.e.
\[
    \SbranePoinc = \Sbrane + \Svar \, ,
\]
in line with the discussion in section~\ref{sec:boundary_term}. We adopt Poincar\'e coordinates by inverting the transformation in equation~\eqref{eq:chm_map}, restricted to the D3-branes' worldvolume at \(z = \m^{-1}\). A trivial integral over the angular coordinate \(\xi\) was performed in the derivation of~\eqref{eq:coulomb_branch_Sbrane_integral}, so the necessary transformation depends \(r\) and \(x\) only through the \(\xi\)-invariant combination \(\varrho= \sqrt{r^2 + x^2}\). To simplify notation, we will rescale \(t\) and \(\varrho\) by factors of \(\ell\) to make them dimensionless. In sum, the resulting change of variables is
\begin{equation} \label{eq:chm_to_poincare_coulomb}
    \tan \t = \frac{2 t}{1  - t^2 - \varrho^2 - a^{-2}} \, ,
    \qquad
    \z = \frac{a}{2} \sqrt{(1 + t^2 + \varrho^2 + a^{-2})^2 - 4  \varrho^2} \, .
\end{equation}
Applied to equation~\eqref{eq:coulomb_branch_Sbrane_integral} this gives the integral
\begin{equation}\begin{aligned}
\label{eq:coulomb_branch_S_brane_poinc_integral}
   \SbranePoinc = \frac{32 n_3 N_3}{3\pi}  \int_{-\infty}^\infty \diff t \int_0^\infty \diff \varrho\, \varrho^2 \biggl[&
        \frac{\varrho^2 - 1}{\S^2} - \frac{2t^2}{(\S-4 a^{-2})^2}
        \\
        &+ \frac{2t^2-\varrho^2}{\S (\S-4a^{-2})} + 4\frac{1 + a^2(\varrho^2 + t^2)}{a^4 \S^2(\S-4a^{-2})} \biggr] \, ,
\end{aligned}\end{equation}
where \(\S =(1 + t^2 + \varrho^2 + a^{-2})^2 - 4  \varrho^2\).

The integrand in equation~\eqref{eq:coulomb_branch_S_brane_poinc_integral} is an even function of \(\varrho\), so we may extend the domain of integration to \(\varrho\in (-\infty,\infty)\), multiplying by a compensating factor of one half. Then, thinking of the integral as a contour integral in complex \(\varrho\) plane, we close the contour with a large semicircle in the upper half-plane. The semicircle provides no-contribution to the integral since the integrand decays as \(\varrho^{-6}\) at large \(\varrho\). The integral is then a sum of residues at the zeroes of \(\S\) and \(\S-4a^{-2}\) in the upper half-plane, which are located at
\begin{equation}\begin{aligned}
    \S &= 0: && \qquad \varrho= \pm 1  + i \sqrt{t^2 + a^{-2}}\,,
    \\
    \S-4a^{-2} &= 0: && \qquad \varrho= i\sqrt{a^{-2} + (t \pm i)^2} \, .
\end{aligned}\end{equation}
Evaluating the sum of residues, one finds
\begin{equation}
    \SbranePoinc = \frac{a}{3} n_3 N_3 \int_{-\infty}^\infty \diff t \biggl[
        - \frac{2 + 6 a^4 t^4 + a^2 + 8 a^2 t^2}{(1 + a^2 t^2)^{3/2}} +  2 \Re \le(\frac{1 + 3 i t \le[1 + a^2 (t+i)^2 \ri]}{\sqrt{1 + a^2 (t+i)^2}} \ri)
    \biggr] \, .
\end{equation}
The integral over \(t\) then yields
\begin{equation} \label{eq:Sbrane_poinc_coulomb_result}
    \SbranePoinc = \Q(a-1) \frac{2}{3} n_3 N_3 \le[\cosh^{-1} a - \le(a + \frac{2}{a} \ri) \sqrt{a^2 - 1} \ri],
\end{equation}
which is indeed equal to \(\Sbrane + \Svar\).

We note that it is not possible to commute the integrals in equation~\eqref{eq:coulomb_branch_S_brane_poinc_integral}, similar to how the integral in equation~\eqref{eq:model_integral_cartesian} depends on the order of integration over \(x\) and \(y\). Let us define \(\cS\) as the iterated integral obtained by swapping the order of the integrals in equation~\eqref{eq:coulomb_branch_S_brane_poinc_integral},
\begin{equation}\begin{aligned}
\label{eq:coulomb_branch_S_brane_poinc_integral_2}
   \cS = \frac{32 n_3 N_3}{3\pi}  \int_0^\infty \diff \varrho\int_{-\infty}^\infty \diff t  \, \varrho^2 \biggl[&
        \frac{\varrho^2 - 1}{\S^2} - \frac{2t^2}{(\S-4 a^{-2})^2}
        \\
        &+ \frac{2t^2-\varrho^2}{\S (\S-4a^{-2})} + 4\frac{1 + a^2(\varrho^2 + t^2)}{a^4 \S^2(\S-4a^{-2})} \biggr] \, .
\end{aligned}\end{equation}
This may be evaluated in a similar manner to \(\SbranePoinc\), treating the integral over \(t\) as a contour integral in the complex \(t\) plane, which may be closed by a large semicircle in the upper half-plane. The contour integral then picks up contributions from the poles at \(\Sigma =0\) and \(\Sigma = 4a^{-2}\) in the upper half-plane, which are located at
\begin{equation}\begin{aligned}
    \S &= 0: && \qquad t = i \sqrt{a^{-2} + (\varrho\pm 1)^2}\,,
    \\
    \S-4a^{-2} &= 0: && \qquad t = i \le|1 \pm \sqrt{a^{-2} + \r^2} \ri| .
\end{aligned}\end{equation}
The sum of the residues of these poles
\begin{equation}
    \cS = \frac{n_3 N_3}{3} \int_0^\infty \diff \varrho\, \varrho\le[\cS_1(a;\r) - \cS_1 (a;-\r) + \cS_2(a;\r)\ri] \, ,
\end{equation}
where we have defined
\begin{equation}
    \cS_1(a;\varrho) = \frac{3 + 6 a^4(\varrho-1)^4 +a^2 (7 - 17 \varrho+ 9 \varrho^2)}{a^2\le[ a^{-2} + (\varrho-1)^2 \ri]^{3/2}}\, ,
\end{equation}
and
\begin{equation}
    \cS_2(a;\varrho) = \begin{cases}
         \dfrac{3 a^3 \varrho(4 + 3 a^2 \varrho^2)}{(1 + a^2 \varrho^2)^{3/2}},
         & \sqrt{a^{-2} + \varrho^2} > 1 \,,
         \\
         12 a^2 \varrho,
         &
         \text{otherwise} \,.
    \end{cases}
\end{equation}
Then performing the integral over \(\varrho\) we obtain
\begin{equation}
    \cS = \Q(a-1) \frac{2}{3}n_3 N_3 \le[\cosh^{-1} a - a \sqrt{a^2 - 1}\ri].
\end{equation}
Comparing to equation~\eqref{eq:Sbrane_poinc_coulomb_result} we see that \(\cS \neq \SbranePoinc\), and rather than equation~\eqref{eq:S_brane_poincare_difference} satisfies \(\cS = \Sbrane - \Svar\).

\bibliographystyle{JHEP}
\bibliography{probe_entanglement}

\end{document}